\documentclass{aa}  

\usepackage{graphicx}
\usepackage{txfonts}
\usepackage{natbib}
\usepackage{xcolor}
\bibpunct{(}{)}{;}{a}{}{,}
\usepackage{xcolor}

\begin{document} 

% Personal definitions
\def\hi {H\,{\sc i}}
\def\hii {H\,{\sc ii}}
\def\water {H$_2$O}
\def\meth {CH$_{3}$OH}
\def\dg{$^{\circ}$}
\def\kms{km\,s$^{-1}$}
\def\ms{m\,s$^{-1}$}
\def\jyb{Jy\,beam$^{-1}$}
\def\mjyb{mJy\,beam$^{-1}$}
\def\ujyb{$\mu$Jy\,beam$^{-1}$}
\def\ujy{$\mu$Jy}
\def\solmass {\hbox{M$_{\odot}$}}
\def\solum {\hbox{L$_{\odot}$}} 
\def\d {$^{\circ}$}
\def\n {$n_{\rm{H_{2}}}$}
\def\kmsg{km\,s$^{-1}$\,G$^{-1}$}
\def\tbo {$T_{\rm{b}}\Delta\Omega$}
\def\mtbo {$\langle T_{\rm{b}}\Delta\Omega \rangle$}
\def\tb {$T_{\rm{b}}$}
\def\om{$\Delta\Omega$}
\def\dvi {$\Delta V_{\rm{i}}$}
\def\mdvi{$\langle\Delta V_{\rm{i}}\rangle$}
\def\dvz {$\Delta V_{\rm{Z}}$}
\def\masyr {$\rm{mas\,yr^{-1}}$}
\def\code {FRTM code}
\def\NW {Nedoluha \& Watson}
\defcitealias{sur142}{S+14}
\defcitealias{sur112}{S+11a}
\title{Monitoring of the polarized \water ~maser emission around the massive protostars W75N(B)-VLA\,1 and W75N(B)-VLA\,2.}

\author{G.\ Surcis  \inst{1}
  \and 
  W.H.T. \ Vlemmings \inst{2}
 \and
  C. \ Goddi \inst{1,3,4,5}
  \and
  J.M. \ Torrelles \inst{6,7}
  \and
  J.F. \ G\'{o}mez \inst{8}
  \and
  A. \ Rodr\'{i}guez-Kamenetzky \inst{9}
  \and
  C. \ Carrasco-Gonz\'{a}lez \inst{10}
  \and
  S. \ Curiel \inst{11}
  \and
  S.-W. \ Kim  \inst{12}
  \and
  J.-S. \ Kim \inst{12}
%  \and
% J. \ Cant\'{o} \inst{11} 
  \and
  H.J.~van Langevelde \inst{13,14} 
  }

\institute{INAF - Osservatorio Astronomico di Cagliari, Via della Scienza 5, I-09047, Selargius, Italy\\
 \email{gabriele.surcis@inaf.it}
 \and
 Department of Space, Earth and Environment, Chalmers University of Technology, Onsala Space Observatory, SE-439 92 Onsala, Sweden
 \and
 Dipartimento di Fisica, Universit\`{a} degli Studi di Cagliari, SP Monserrato-Sestu km 0.7, I-09042 Monserrato, Italy
 \and
INFN, Sezione di Cagliari, Cittadella Univ., I-09042 Monserrato (CA), Italy
 \and
Universidade de S\~{a}o Paulo, Instituto de Astronomia, Geof\'{i}sica e Ci\^{e}ncias Atmosf\'{e}ricas, Departamento de Astronomia, S\~{a}o Paulo, SP 05508-090, Brazil
\and 
 Institut de Ci\`{e}ncies de l’Espai (ICE, CSIC), Can Magrans s/n, E-08193, Cerdanyola del Vall\`{e}s, Barcelona, Spain
 \and
 Institut d’Estudis Espacials de Catalunya (IEEC), Barcelona, Spain
 \and
 Instituto de Astrof\'{i}sica de Andaluc\'{i}a, CSIC, Glorieta de la Astronom\'{i}a s/n, E-18008 Granada, Spain
 \and
Instituto de Astronomía Teórica y Experimental (IATE, CONICET-UNC), Laprida 854, Córdoba, X5000BGR, Argentina
 \and
 Instituto de Radioastronom\'{i}a y Astrof\'{i}sica (IRyA-UNAM), Morelia, Mexico
 \and
 Instituto de Astronom\'{i}a, Universidad Nacional Autónoma de M\'{e}xico (UNAM), Apdo Postal 70-264, M\'{e}xico, D.F., Mexico
 \and
 Korea Astronomy and Space Science Institute, 776 Daedeokdaero, Yuseong, Daejeon 305-348, Republic of Korea
 \and
 Joint Institute for VLBI ERIC, Oude Hoogeveensedijk 4, 7991 PD Dwingeloo, The Netherlands
 \and
 Sterrewacht Leiden, Leiden University, Postbus 9513, 2300 RA Leiden, The Netherlands
 }

\date{Received ; accepted }
\abstract
%context heading (optional)
{Several radio sources have been detected in the high-mass star-forming region W75N(B), 
among them the massive young stellar objects VLA\,1 and VLA\,2 are of great interest. 
These are thought to be in different evolutionary stages. In particular,
VLA\,1 is at the early stage of the photoionization and it is driving a thermal radio jet,
while VLA\,2 is a thermal, collimated ionized wind surrounded by a dusty disk or envelope.
In both sources 22~GHz \water ~masers have been detected in the past. Those around VLA\,1
show a persistent linear distribution along the thermal radio jet
and those around VLA\,2 have instead traced the evolution from a non-collimated to a
collimated outflow over a period of $\sim$20 years. The magnetic field inferred
from the \water ~masers showed an orientation rotation following the direction
of the major-axis of the shell around VLA\,2, while it is immutable around VLA\,1.
}
{By monitoring the polarized emission of the 22~GHz \water ~masers around both VLA\,1 and 
VLA\,2 over a period of six years, we aim to determine whether the \water ~maser distributions
show any variation over time and whether the magnetic field behaves accordingly.} 
{The European VLBI Network was used in full polarization and phase-reference mode in order
to determine the absolute positions of the 22~GHz \water ~masers with a beam size of 
$\sim1$~mas and to determine both the orientation and the strength of the magnetic field. We 
observed four epochs separated by two years from 2014 to 2020.}
{We detected polarized emission from the \water ~masers around both VLA\,1 and VLA\,2  
in all the epochs. By comparing the \water ~masers detected in the four epochs,
we find that the masers around VLA\,1 are tracing a nondissociative shock originating from
the expansion of the thermal radio jet, while the masers around VLA\,2 are tracing an
asymmetric expansion of the gas that is halted in the northeast where the gas likely 
encounters a very dense medium. We also found that the magnetic field inferred from
the \water ~masers in each epoch can be considered as a portion of a quasi-static magnetic
field estimated in that location rather than in that time. This allowed us to study locally the 
morphology of the magnetic field around both VLA\,1 and VLA\,2 in a larger area by considering 
the vectors estimated in all the epochs as a whole. We find that the magnetic field in VLA\,1
is along the jet axis and bends toward north and south at the northeast and southwest ends of 
the jet, respectively, reconnecting with the large-scale magnetic field.
The magnetic field in VLA\,2 is perpendicular to the expansion directions till it encounters
the denser matter in the northeast, here the magnetic field is parallel to the expansion 
direction and agrees with the large-scale magnetic field. We also measured the magnetic field 
strength along the line of sight in three of the four epochs, whose values are
$-764~\rm{mG}<B_{\rm{||}}^{\rm{VLA\,1}}<-676$~mG and $-355~\rm{mG}<B_{\rm{||}}^{\rm{VLA\,2}}<-2426$~mG. }
{}
\titlerunning{.}
\authorrunning{Surcis et al.}
\maketitle
%________________________________________________________________
\section{Introduction}
\label{intro}
All stars in the sky are generally divided into two main groups according to their mass:
low-mass stars ($M<8$~\solmass) and high-mass stars ($M>8$~\solmass). While for the first
group the formation process is quite well established, for the high-mass stars there are
still several open questions that need to be addressed \cite[e.g.,][]{tan14, mot18}. 
However, based on observational constraints, an evolutionary scenario of high-mass star 
formation (HMSF) was proposed by \cite{mot18}. This is summarized in the following.
High-mass stars
form in molecular complexes, in particular in parsec-scale massive clumps/clouds that first 
undergo a global controlled collapse forming low-mass prestellar cores. This first phase
is known as starless massive dense cores (MDCs). After about 10$^4$~years, low-mass 
prestellar cores become protostars with growing mass. This phase is called protostellar 
MDC phase. Only after $\sim10^5$~years more we have the high-mass protostellar phase where, 
thanks to the gas flow streams generated by the global collapse, the protostars become 
high-mass protostars, even though they still harbor low-mass stellar embryos. At this
stage of the evolution the high-mass protostars are still quiet in the infrared (IR) and 
if their accretion rates are efficient and strong they drive outflows. In this phase the
accretion disks are already formed. As soon as the stellar embryos reach more than 
8~\solmass ~the high-mass protostars become IR-bright and they develop ultra-compact 
\hii ~region (UC\hii) that are quenched by infalling gas or confined to the 
photoevaporating disks. Finally, we have the \hii ~region phase that lasts about
$10^{5}-10^{6}$~years. In this phase, the ultraviolet radiation from the stellar embryos 
produces the \hii ~region and the gas accretion toward the newborn star first slows 
down and then stops by terminating the main accretion phase. The high-mass star is 
formed. \\
\indent The great importance that the magnetic field has in several phases of 
HMSF has been showed through magnetohydrodynamical (MHD) simulations
\cite[e.g.,][]{mye14,kui16,mat18,mac20,ros20,oli22}. Despite the observational difficulties, 
which are due to the low number of high-mass protostars that are usually found densely 
clustered in molecular clouds and to their long distances to the Sun, measurements of the
morphology and strength of magnetic fields close to high-mass protostars are possible. These
can be obtained by observing the polarized emission of  
dust and molecular lines with the Atacama Large Millimeter Array (ALMA; e.g.,
\citealt{dal19,sanh21}), and by observing the polarized maser emission with the very long 
baseline interferometry (VLBI) technique \cite[e.g.,][]{sur112,sur111,sur142,sur22}. 
A relevant star-forming region where it is possible to measure the magnetic field close to 
high-mass young stellar objects (YSOs) in different evolutionary phases is W75N(B).\\
\indent The active high-mass star-forming region (HMSFR) W75N is part of the Cygnus X complex
\citep{wes58, har77, hab79} at a distance of 1.30$\pm$0.07~kpc \citep{ryg12}.
\cite{has81} identified three compact radio regions within W75N at a spatial resolution
of $\sim1''\!\!.5$: W75N(A), W75N(B), and W75N(C). In \citeyear{hun94},
\citeauthor{hun94} further mapped the continuum emission of W75N(B) at a resolution of
$\sim0''\!\!.5$ revealing the presence of three very compact subregions (Ba, Bb,
and Bc). For the first time \cite{hun94} also underlined the great importance of the
region for understanding the star formation process thanks to its significant activity. 
The subregions Ba and Bb were renamed as VLA\,1 and VLA\,3 in 1997
when \cite{tor97} imaged them at a resolution of $\sim0''\!\!.1$ with the Very Large 
Array (VLA). \cite{tor97} also imaged for the first time another weaker and more
compact radio source between VLA\,1 and VLA\,3, called VLA\,2. While VLA\,1 and VLA\,3 
showed a well elongated radio continuum emission along northeast-southwest and
northwest-southeast directions, respectively, that were consistent with the 
morphology of thermal radio jets, VLA\,2 showed a quasi-circular morphology that was 
supposed to be an UC\hii. Despite their small separation (VLA\,2 
is only $\sim0''\!\!.8$, $\sim 1000$~au at 1.3~kpc, from VLA\,1), the three
radio sources are thought to be massive YSOs \citep{she03} at three different 
evolutionary stages, with VLA\,1 the most evolved and VLA\,2 the least evolved 
\citep{tor97}. In addition, \cite{she03} report a systemic velocity for W75N(B) 
of $V_{\rm{lsr}}=+10.0$~\kms, which can also be considered as the systemic velocity for
VLA\,1, VLA\,2, and VLA\,3.
Five more new radio sources were identified in W75N(B) thanks to
high sensitive observations made with the upgraded VLA. They are VLA\,4, that is
about $\sim1''\!\!.5$ south of Bc \citep{car10}, VLA\,[NE] ($\sim8''$ northeast 
of VLA\,2), VLA\,[SW] ($\sim6''$ southwest of VLA\,2), and Bd ($\sim0.5''$
northeast of VLA\,4; \citealt{rod20}). 
Thanks to ALMA observations at 1.3~mm (spatial
resolution of $\sim1.2''$), \cite{rod20} were able to associate VLA\,[NE] and VLA\,[SW]
with the millimeter cores MM3 and MM2, respectively, indicating that they are also
embedded YSOs. Furthermore, VLA\,1, VLA\,2, and VLA\,3 are also associated with a
millimeter core (MM1) but VLA\,4, Bc, and Bd are not, which suggests that they are not 
embedded YSOs but shock-ionized gas \citep{rod20}. In addition, the multi wavelength 
observations made by \cite{car15} with the VLA showed that VLA\,2 changed its morphology between
1996 and 2014 from a compact roundish source ($\leq 160$~au; \citealt{tor97}) to an
extended source that is elongated in the northeast-southwest direction ($220 \times
\leq 160$~au, position angle of PA=65\d), while VLA\,1 still shows an unchanged  
morphology since 1996 \citep{rod20}. The spectral index analysis indicates that VLA\,2 
is a thermal, collimated ionized wind surrounded by a dusty disk or envelope 
\citep{car15}, VLA\,1 is at the early stage of the photoionization and it is driving a 
thermal radio jet, and VLA\,3 is also driving a thermal radio jet whose shocks are 
traced by the obscured Herbig-Haro (HH) objects Bc and VLA\,4 \citep{rod20}. \\
\indent A large-scale high-velocity CO-outflow, with an extension greater than 3~pc
(PA=66\d) and with a total molecular mass greater than 255~\solmass, was detected
from W75N(B) \citep[e.g.,][]{hun94,dav98,she03,mak18}. \cite{she03} found that the
entire CO emission of W75N uncovers a complex morphology of multiple, overlapping
outflows. In particular, they suggest that VLA\,2 may drive the large-scale CO-outflow,
while VLA\,1 and VLA\,3 are instead the centers of two additional, more compact outflows
(extension of $\sim$0.21~pc and $\sim$0.15~pc, respectively).
However, this is not clear yet and the main powering source of the large-scale
CO-outflow remains undetermined \citep[e.g.,][]{qiu08}. In addition, \cite{she03} 
also determined that more than 10\% of the molecular gas in W75N is outflowing material,
and the combined outflow energy is roughly half the gravitational binding energy of the 
cloud thus preventing its further collapse.\\
\begin {table*}[t]
\caption []{Observational details.
%: observation date, EVN antennas that participated in the observations, observed bandwidth, number of spectral channels used during correlations, observed target and calibrators, size and position angle of the restoring beam, peak intensity of the calibrated radio emission, rms after calibration, self-noise produced by the maser emission, mean linear polarization fraction, and polarization angle.
} 
\begin{center}
\scriptsize
\begin{tabular}{ l c c c c c c c c c c c }
\hline
\hline
\,\,\,\,\,\,\,\,\,\,(1)   &(2)                       & (3)        &  (4)       & (5)        & (6)                 & (7)      & (8)         & (9)                   & (10)   & (11)  & (12) \\
Observation     & Antennas                           & Bandwidth  & Spectral   & Source     & Restoring           & Position & Peak   & rms\tablefootmark{a}  & $\sigma_{\rm{s.-n.}}$\tablefootmark{b} & $P_{\rm{l}}$\tablefootmark{c}  & Polarization\\ 
\,\,\,\,\,\,\,\,\,date  &                            &            & channels   & name       & Beam size           & Angle    & intensity (I) &                       &        &       & angle\\ 
                &                                    & (MHz)      &            &            & (mas~$\times$~mas)  & (\d)     & ($\frac{\rm{Jy}}{\rm{beam}}$) &($\frac{\rm{mJy}}{\rm{beam}}$) & ($\frac{\rm{mJy}}{\rm{beam}}$)  & (\%) & (\d) \\ 
\hline
17 June 2014    & Ef, On, Nt, Tr, Ys, Mh             &  4         & 2048       & W75N(B)       & $1.1 \times 0.8$    & $-60.02$ & -\tablefootmark{d}            &  \textit{13}     &  25  & -\tablefootmark{d} & -\tablefootmark{d}  \\
                &                                    &            &            & J2040+4527\tablefootmark{e}  & $1.1 \times 0.7$ & $-62.40$ & 0.029 & \textbf{0.6} & -    & - & -  \\
                &                                    &            &            & J2202+4216\tablefootmark{f}  & $1.5 \times 0.8$ & $-52.68$ & 1.868 & \textbf{2.5} & -    & 7.0 & $-64\pm8$\tablefootmark{g}  \\
12 June 2016    & Ef, Jb, Mc, Nt, Sr, Ys,            &  4         & 2048       & W75N(B)       & $1.9 \times 1.2$    & $-86.02$ & -\tablefootmark{h}            & \textit{11}     &  73  & -\tablefootmark{h} & -\tablefootmark{h}  \\
                &  Mh                                &            &            & J2040+4527\tablefootmark{e}  & $1.0 \times 0.7$ & $-86.84$ & 0.026 & \textbf{0.7} & -    & - & -  \\
                &                                    &            &            & J2202+4216\tablefootmark{f}  & $1.5 \times 1.0$ & $+74.56$ & 0.685 & \textbf{1.5} & -    & 3.1 & $-61\pm12$\tablefootmark{i} \\
                &                                    &            &            & 3C48\tablefootmark{j}  & $1.5 \times 0.9$ & $-59.93$ & 0.016 & \textbf{1.1} & -    & <21\tablefootmark{k} & $-$  \\
09 June 2018    & Ef, Jb, Mc, On, Nt, Tr,            &  8         & 4096       & W75N(B)       & $1.2 \times 1.0$    & $+86.92$ & -\tablefootmark{l}            & \textit{14}     &  28  & -\tablefootmark{l} & -\tablefootmark{l}  \\
                & Sr, Ys, Mh                         &            &            & J2040+4527\tablefootmark{e}  & $1.1 \times 0.8$ & $-79.82$ & 0.021 & \textbf{0.4} & -    & - & -  \\
                &                                    &            &            & J2202+4216\tablefootmark{f}  & $1.2 \times 0.8$ & $-61.91$ & 0.765 & \textbf{1.7} & -    & 4.0 & $+5\pm1$\tablefootmark{m} \\
                &                                    &            &            & 3C48\tablefootmark{j}  & $1.8 \times 0.7$ & $-42.47$ & 0.125 & \textbf{0.5} & -    & 4.2 & $-70\pm1$  \\  
25 Oct. 2020    & Ef, Jb, Mc, On, Tr, Sr,            &  8         & 4096       & W75N(B)       & $1.4 \times 1.0$    & $-53.96$ & -\tablefootmark{n}            & \textit{13}     &  12  & -\tablefootmark{n} & -\tablefootmark{n}  \\
                & Ys, Mh                             &            &            & J2040+4527\tablefootmark{e}  & $1.3 \times 0.9$ & $-56.63$ & 0.034 & \textbf{0.6} & -    & - & -  \\
                &                                    &            &            & J2202+4216\tablefootmark{f}  & $1.6 \times 1.0$ & $-75.93$ & 1.011 & \textbf{2.2} & -    & 3.8 & $+15\pm1$\tablefootmark{m} \\
                &                                    &            &            & 3C48\tablefootmark{j}  & $2.2 \times 1.1$ & $-38.84$ & 0.027 & \textbf{0.3} & -    & 6.4 & $-70\pm1$  \\       
\hline
\end{tabular}
\end{center}
\tablefoot{
\tablefoottext{a}{The spectral rms (in italics) is measured in channels with no line emission. The rms of the radio continuum (in boldface) is obtained by averaging all the channels.}
\tablefoottext{b}{Self-noise in the maser emission channels \citep[e.g.,][]{sau12}. When more than one maser feature shows circularly polarized emission, we present here the self-noise of the weakest feature.
When no circularly polarized emission is detected, we consider the self-noise of the brightest maser feature.}
\tablefoottext{c}{Linear polarization fraction.}
\tablefoottext{d}{See Tables~\ref{VLA1.1_tab} and \ref{VLA2.1_tab}.}
\tablefoottext{e}{Phase-reference calibrator at 2.856\d ~from W75N(B). The errors of $\alpha_{2000}$ and $\delta_{2000}$ are 0.68~mas and 0.80~mas, respectively \citep{petr11}. }
\tablefoottext{f}{Primary polarization calibrator. }
\tablefoottext{g}{Calibrated using the maser feature VLA1.1.05, see Sect.~\ref{obssect}. }
\tablefoottext{h}{See Tables~\ref{VLA1.2_tab} and \ref{VLA2.2_tab}.}
\tablefoottext{i}{Calibrated using the value measured on 1$^{\rm{st}}$ June 2016 by one of the Korean VLBI Network (KVN) antennas and calibrated by using 3C286 (\textit{private communication}). }
\tablefoottext{j}{Secondary polarization calibrator.}
\tablefoottext{k}{Considering a 3$\sigma$ detection threshold.}
\tablefoottext{l}{See Tables~\ref{VLA1.3_tab} and \ref{VLA2.3_tab}.}
\tablefoottext{m}{Calibrated by using 3C48.}
\tablefoottext{n}{See Tables~\ref{VLA1.4_tab} and \ref{VLA2.4_tab}.}
}
\label{Obs}
\end{table*}
\indent
The great interest that W75N(B) has aroused in the past is also due to the presence 
of several maser species (OH, \meth, and \water) around the two sources VLA\,1 and 
VLA\,2 \citep[e.g.,][]{has81,hun94,tor97,lek00,min00,sur09,fis11,kan16,col18,col21}. OH masers
are distributed throughout the region with the majority of the maser spots associated 
with VLA\,1 \citep{hut02,fis05}. Nevertheless, VLA\,2 is the site of the most intensive 
OH flare ever registered in a star-forming region \citep[1000~Jy,][]{ala05,sly10}, while other 
OH maser emission 
sites are situated on a ring structure around VLA\,1, VLA\,2, and VLA\,3 \citep{hut02}. 
In \citeyear{fis11}, \citeauthor{fis11} measured the proper motions of the OH masers 
showing that most of those near VLA\,1 (located $\sim1''$ northwest) are moving 
northward ($V<+5$~\kms) and those associated with VLA\,2, and located southwest, 
are moving both toward southwest and southeast ($V<+5$~\kms). The detection of OH maser 
Zeeman-pairs provided measurements of magnetic field strength between 6 and 8~mG close to
VLA\,1, where the 22~GHz \water ~maser are detected, and up to about 17~mG around VLA\,2
\citep{fis11}. However, higher values \citep[40-70~mG,][]{sly06, sly10} were measured
during the strong OH maser flare. 
The 6.7 GHz \meth ~masers are only associated with VLA\,1 and they 
are distributed parallel to the thermal radio jet of VLA\,1 \citep{min00,min01,sur09}. 
No \meth ~masers were detected around VLA\,2 \citep{sur09, ryg12} till 2014, when three
maser spots were detected in the southwest of the source \citep{car15}.
%\cite{ryg12} measured the proper motion of the 6.7~GHz \meth ~masers around VLA\,1. 
Thanks to the polarized emission
of the \meth ~masers, \cite{sur09} were able to measure a magnetic field oriented 
southwest - northeast, that is perfectly aligned with VLA\,1, and whose 
strength on the line of the sight was found to be $|B_{||}|>10$~mG \citep{sur19}.\\
\indent
\indent The 22~GHz \water ~masers have been widely studied and monitored both with
single dishes and interferometers, revealing a high intensity variations, with extreme 
maser flares (up to 10$^3$~Jy), and important variation in the maser distribution
\citep[e.g.,][]{lek84,lek94,lek00,tor03,sur112,kim13,sur142,kra15,kim18}. The \water 
~masers are associated with VLA\,1 and VLA\,2, and only one maser was associated with 
VLA\,3 in 1996 \citep{tor97}, but it has never been detected again 
\citep[e.g.,][]{tor03, sur142}. Over a period of 16~years, 
VLBI observations have showed that the evolution of
the \water ~masers around VLA\,1 and VLA\,2, despite their close separation, is completely
different. Whereas the \water ~masers around VLA\,1 are always linearly distributed 
($\rm{PA}\approx43$\d) along the thermal radio jet, those detected around 
VLA\,2 are instead tracing an expanding shell \citep[expanding velocity of 
$\sim$30~\kms,][]{sur142} that 
evolved from a quasi-circular \citep{tor03,sur112} to an elliptical structure
\citep{kim13,sur142} following the morphology change in the continuum emission observed 
by \cite{car15}. Therefore, in VLA\,2 the \water ~masers might be tracing the evolution 
from a non-collimated to a collimated outflow \citep{sur142}. 
Furthermore, \citet{sur142} also showed that the magnetic field around
VLA\,1 has not changed from 2005 to 2012 and it is always oriented along the direction
of the thermal radio jet. Whereas, the orientation of the magnetic field around VLA\,2 
changed in a way that is consistent with the new direction of the major-axis of the 
shell-like structure that is now aligned with the thermal radio jet of VLA\,1.\\
\begin{figure*}[h!]
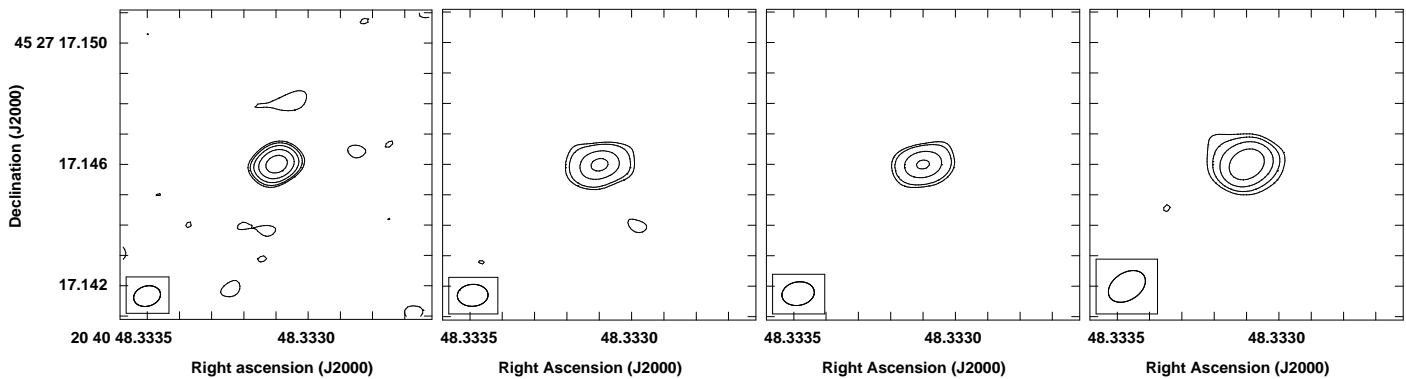

\centering
\includegraphics[width = 5.6 cm]{J2040+4527_ep1.eps}
\includegraphics[width = 4.17 cm]{J2040+4527_ep2.eps}
\includegraphics[width = 4.17 cm]{J2040+4527_ep3.eps}
\includegraphics[width = 4.17 cm]{J2040+4527_ep4.eps}
\caption{Contour maps of the phase-reference calibrator J2040+4527 in the EVN epochs 2014.46, 
2016.45, 2018.44, and 2020.82 from left to right, respectively. Contours are 4, 8, 16, and 32 
times 0.6~\mjyb. The restoring beam is overplotted on the bottom left corner of each panel (see 
Table~\ref{Obs}).}
\label{phcal}
\end{figure*}
\begin {table*}[t]
\caption []{Absolute position of the reference maser feature.} 
\begin{center}
\scriptsize
\begin{tabular}{ l c c c c c c c c c c c }
\hline
\hline
\,\,\,\,\,(1) &(2)   &   (3)           &    (4)                    & (6) & (7) & \multicolumn{2}{c}{(8)}            &  (9)                        &   (10) &(11) \\
epoch   & Reference  & $V_{\rm{lsr}}$  & $\alpha_{2000}$           & $\varepsilon_{\rm{rms}}$&  $\varepsilon_{\rm{gauss}}^{\alpha}$ & \multicolumn{2}{c}{$\Delta\alpha$\tablefootmark{a}}  & $\delta_{2000}$ & $\varepsilon_{\rm{gauss}}^{\delta}$            & $\Delta\delta$\tablefootmark{a}  \\ 
        & maser      &                 &                           & &   &                                &                             &      &\\
        &            &  (km/s)         & ($\rm{^{h}:~^{m}:~^{s}}$) & (mas)  &  (mas) &  (mas)  &  ($\rm{^{s}}$)     & ($\rm{^{\circ}:\,':\,''}$)  & (mas) & ($''$) \\
\hline
2014.46 & VLA1.1.15  & 10.45           & 20:38:36.43399            & 0.003 & 0.006& $\pm$0.9 &  $\pm$0.00008        & +42:37:34.8710              & 0.04  & $\pm$0.0009              \\
2016.45 & VLA1.2.07  & 10.92           & 20:38:36.43403            & 0.029 & 0.05 & $\pm$1.2 &  $\pm$0.00011        & +42:37:34.8667              & 0.04  & $\pm$0.0010              \\
2018.44 & VLA1.3.04  & 8.63            & 20:38:36.43201            & 0.018 & 0.04 & $\pm$0.9 &  $\pm$0.00008        & +42:37:34.8588              & 0.03  & $\pm$0.0009              \\
2020.82 & VLA1.4.05  & 9.87            & 20:38:36.43111            & 0.029 & 0.06 & $\pm$1.0 &  $\pm$0.00009        & +42:37:34.8794              & 0.05  & $\pm$0.0009              \\
\hline
\end{tabular}
\end{center}
\tablefoot{
\tablefoottext{a}{The uncertainties of the absolute positions ($\Delta\alpha$ and $\Delta\delta$)
are obtained by adding quadratically the systematic errors ($\varepsilon_{sys}=0.028$~mas), the 
errors due to the thermal noise ($\varepsilon_{\rm{rms}}$), the Gaussian fit errors 
($\varepsilon_{\rm{gauss}}$), the position errors of the phase-reference source J2040+4527 
($\Delta\alpha^{\rm{J2040}}=0.68$~mas and $\Delta\delta^{\rm{J2040}}=0.80$~mas), and half of the 
restoring beam of W75N(B) to account for the maser spots scatter of each maser features (see 
Table~\ref{Obs} and Sect.~\ref{obssect}).}
}
\label{abspos}
\end{table*}
\indent The peculiarity of the \water ~maser shell expansion with the 
contemporary variation of the magnetic field around VLA\,2, together with the
presence of a close-by immutable VLA\,1 source, made W75N(B) one of the most interesting 
case where to investigate the evolution of early massive YSOs. For this reason that we performed 
every two years VLBI monitoring observations of 22~GHz \water ~maser emission in full
polarization mode from 2014 to 2020 for a total of four epochs. Here, we report in
Sect.~\ref{res} the results of the monitoring observations, that are described in
Sect.~\ref{obssect}. We discuss the magnetic fields around VLA\,1 and VLA\,2 in Sect.~\ref{magn}, 
and we finally present a full picture of the two massive YSOs in Sect.~\ref{discussion}.
\begin {table*}[t!]
\caption []{Comparison of 22~GHz \water ~maser parameters between epochs~2014.46, 2016.45, 2018.44, and 2020.82.} 
\begin{center}
\scriptsize
\begin{tabular}{ l c c c c c c c c}
\hline
\hline
                                         & \multicolumn{4}{c}{VLA\,1} & \multicolumn{4}{c}{VLA\,2}     \\ 
                                         &  2014.46            &     2016.45         &  2018.44            &     2020.82      &  2014.46            &     2016.45         &  2018.44         &     2020.82 \\
\hline                                         
Number of maser features                     &  28                 &    20               &  20                 &    10            &    43               &  37                 &    44            &    39  \\
$V_{\rm{lsr}}$ range (\kms)              & $[+7.9; +19.7]$     & $[+2.7; +16.4]$     & $[+7.8; +15.0]$     & $[+8.8; +26.4]$  & $[-11.6; +21.2]$    & $[-15.6; +20.7]$    & $[-15.9; +28.2]$ & $[-1.0; +27.2]$\\
$I$ range (\jyb)                         & $[0.43; 1205.52]$   & $[0.24; 1619.87]$   & $[2.08; 446.25]$    & $[0.14; 302.47]$ & $[0.21; 31.70]$     & $[0.23; 90.99]$     & $[0.04; 128.02]$ & $[0.05; 285.16]$ \\
$\langle\Delta v_{\rm{L}}\rangle$ (\kms)  & $1.3$               & $1.1$               & $1.0$               & $1.7$            & $0.7$                 & $0.7$                  & $0.7$            & $0.7$ \\
\hline
                                         & \multicolumn{8}{c}{Polarization} \\
\hline
 $P_{\rm{l}}$ range (\%)                 & $[0.4; 15.6]$       & $[0.7; 2.2]$        & $[0.7; 10.6]$       & $[0.1; 0.3]$     & $[0.9; 4.6]$        & $[0.6; 2.7]$        & $[1.0; 2.3]$     & $[0.2; 1.9]$ \\
 $P_{\rm{V}}$ range (\%)                 & $-$                 & $3.5$               & $1.6$               &  $-$             & $-$                 & $[4.9; 7.8]$        & $[1.5; 2.9]$     & $1.3$ \\
  \hline
                                         & \multicolumn{8}{c}{Intrinsic characteristics} \\
\hline
 $\Delta V_{\rm{i}}$ range (\kms)        & $[0.4; 4.2]$        & $[3.8; 4.3]$        & $[0.4; 3.9]$        &   $2.6$          & $[1.6; 1.9]$        & $[0.4; 4.2]$        & $[0.8; 3.6]$     & $[2.6; 3.8]$  \\
$T_{\rm{b}}\Delta\Omega$ range (log K sr)& $[7.0; 9.4]$        & $[6.8; 9.3]$        & $[8.9; 10.5]$       &    $10.6$        & $[8.9; 9.2]$        & $[6.1; 9.4]$        & $[6.0; 10.5]$    & $[8.2; 8.4]$  \\
 $\langle\Delta V_{\rm{i}}\rangle$\tablefootmark{a} (\kms)& $3.0^{+0.2}_{-0.3}$& $2.8^{+0.3}_{-0.2}$ & $2.7^{+0.5}_{-1.4}$ & $2.6^{+0.1}_{-0.3}$ & $1.6^{+0.2}_{-0.2}$& $2.9^{+0.2}_{-0.2}$ & $2.1^{+0.8}_{-0.4}$ & $2.9^{+0.1}_{-0.4}$\\
$\langle T_{\rm{b}}\Delta\Omega\rangle$\tablefootmark{a} (log K sr)&$8.7^{+0.8}_{-0.9}$&$8.7^{+1.2}_{-0.9}$  &$9.1^{+0.6}_{-0.3}$& $10.5^{+0.1}_{-0.4}$ & $9.2^{+0.3}_{-1.2}$& $9.1^{+0.4}_{-1.9}$ & $11.0^{+0.1}_{-4.2}$ & $8.3^{+0.8}_{-0.8}$\\
$T$\tablefootmark{b} (K)                 & $3600^{+496}_{-684}$ & $3136^{+708}_{-432}$ & $2916^{+1180}_{-2240}$  & $2704^{+212}_{-588}$   & $1024^{+272}_{-240}$  & $3364^{+480}_{-448}$ & $1764^{+1600}_{-608}$              & $3364^{+236}_{-864}$  \\
$\Gamma+\Gamma_{\nu}$\tablefootmark{c}   & $20$                 & $18$                & $16$                 & $15$      & $7$               & $18$              & $10$              & $18$         \\
\hline
                                         & \multicolumn{8}{c}{Magnetic field} \\
\hline
 $\chi$ range (\d)                       &  $[-62; +48]$       & $[-76; -11]$        & $[-88; +88]$     & $[-67; -5]$    &  $[-47; -31]$       & $[-77; +88]$        & $[-89; +90]$     & $[-54; +40]$    \\
 $\theta$ range	(\d)			 &  $[+75; +90]$       & $+90$        & $[+62; +90]$     & $[+5; +54] $           &  $[+79; +86]$       & $[+76; +90]$        & $[+14; +90]$     & $[+66; +90]$   \\
 $\Phi_{\rm{B}}$ range	(\d)		 &  $[-42; +68]$         & $[+14; +18]$           & $[-4; +88]$     & $[-5; -67]$    &  $[+43; +51]$       & $[-3; +45]$        & $[-25; +85]$     & $[-64; +80]$   \\
 $|B_{||}|$ range (mG)                   &  $-$        & $676$         & $[733; 764]$      & $-$    &  $-$       & $[1498; 2426]$        & $[355; 439]$     & $452$   \\
 $\langle \chi \rangle$\tablefootmark{d} (\d) &  $-47\pm30$         & $-60\pm16$          & $-87\pm5$       & $-32\pm44$    &  $-40\pm7$         & $-74\pm19$          & $+76\pm12$       & $-73\pm62$      \\
 $\langle \theta \rangle$\tablefootmark{d} (\d) & $90^{+21}_{-21}$& $90^{+19}_{-19}$ & $+78^{+11}_{-11}$ & $+55^{+2}_{-48}$ & $86^{+3}_{-13}$& $90^{+18}_{-18}$ & $90^{+37}_{-37}$ & $90^{+49}_{-49}$\\
 $\langle\Phi_{\rm{B}}\rangle$\tablefootmark{d} (\d)      &  $+43\pm30$         & $+30\pm16$          & $-10\pm36$       & $-32\pm44$    &  $+50\pm7$         & $+16\pm19$          & $-29\pm36$       & $+61\pm31$      \\
 $\langle |B_{||}| \rangle$\tablefootmark{e} (mG) &  $-$ & $676\pm102$         & $748\pm22$       & $-$        &  $-$         & $1755\pm656$          & $399\pm59$       & $452\pm68$  \\
 $\langle |B| \rangle$\tablefootmark{e,f} (mG) &  $-$  & $>2076\tablefootmark{g}$       & $3598\pm106$ & $-$    &  $-$         & $>5679\tablefootmark{h}$          & $>663\tablefootmark{i}$       & $>598$\tablefootmark{l}    \\
Arithmetic mean of $|B_{||}|$ (mG)                 &  $-$              & $676$               & $748$            &  $-$     &  $-$              & $1962$               & $397$            &  $452$         \\
\hline
\end{tabular}
\end{center}
\tablefoot{
\tablefoottext{a}{The averaged values are determined by analyzing the total full probability distribution function.}
\tablefoottext{b}{$T\approx100 \cdot (\langle\Delta V_{\rm{i}}\rangle/0.5)^2$ is the gas temperature of the region where the \water ~masers arise, with \dvi ~the intrinsic maser
linewidth \citep[see][]{ned92}, in case turbulence is not present.}
\tablefoottext{c}{Here $\Gamma$ is the decay rate and $\Gamma_{\nu}$ is the cross-relaxation rate \citep[e.g.,][]{ned92}. The values of \tbo ~have to be adjusted according to the gas
temperature by adding +1.3 ($\Gamma+\Gamma_{\nu}=20$), +1.3 ($\Gamma+\Gamma_{\nu}=18$),
+1.2 ($\Gamma+\Gamma_{\nu}=16$), +1.2 ($\Gamma+\Gamma_{\nu}=15$), +0.8 ($\Gamma+\Gamma_{\nu}=7$), +1.3 ($\Gamma+\Gamma_{\nu}=18$),
+1.0 ($\Gamma+\Gamma_{\nu}=10$), and +1.3 ($\Gamma+\Gamma_{\nu}=18$) as described in \cite{and93}.}
\tablefoottext{d}{Error-weighted values, where the weights are $w_{\rm{i}}=1/e_{\rm{i}}$ and $e_{\rm{i}}$ is the error
of the i\textit{th} measurements.}
\tablefoottext{e}{Error-weighted values, where we assumed weights of $w_{\rm{i}}=1/e_{\rm{i}}^{2}$, with $e_{\rm{i}}$ being the error of the \textit{i}th measurement, to take into more consideration the less uncertain measures.}
\tablefoottext{f}{$|B|=\langle |B_{||}| \rangle/\rm{cos}\langle \theta \rangle$ if $\theta\neq\pm90$\d.}
\tablefoottext{g}{We report the lower limit estimated by considering $\theta=\theta+\varepsilon^{-}=71$\d, where $\varepsilon^{-}$ is one of the associated
errors to $\theta$ (i.e., $\theta^{\varepsilon^{+}}_{\varepsilon^{-}}$).}
\tablefoottext{h}{We report the lower limit estimated by considering $\theta=72$\d.}
\tablefoottext{i}{We report the lower limit estimated by considering $\theta=53$\d.}
\tablefoottext{l}{We report the lower limit estimated by considering $\theta=41$\d.}
}
\label{para}
\end{table*}
\section{Observations and analysis}
\label{obssect}
W75N(B) was observed at 22 GHz in full polarization spectral mode with several European VLBI
Network\footnote{The European VLBI Network is a joint facility of European, Chinese, South 
African and other radio astronomy institutes funded by their national research councils.} (EVN) 
antennas on four epochs separated by two years (see Table~\ref{Obs}). The observations were
carried out in June 2014 (epoch 2014.46), 2016 (epoch 2016.45), and 2018 (epoch 2018.44) and in
October 2020 (epoch 2020.82), for a total observation time per epoch of 12~h. The bandwidth was
4~MHz in epochs 2014.46 and 2016.45, providing a local standard of rest velocity 
($V_{\rm{lsr}}$) range of $\sim55$~\kms ~(after calibration ranging from $-15.7$~\kms ~to 
$+31.7$~\kms), and 8~MHz in epochs 2018.44 and 2020.82 (after calibration ranging from 
$-30.3$~\kms ~to $+48.1$~\kms). We observed with a bandwidth two times wider in the last
two epochs to search for maser emission at velocities $<-15.7$~\kms, as indicated by the results
obtained in epoch 2016.45 (see Sec.~\ref{res16} and Table~\ref{VLA2.2_tab}). To measure the
absolute positions of the \water ~masers, the observations were conducted in phase-reference 
mode (with cycles phase-calibrator – target of 45 sec – 45 sec). The phase-reference calibrator 
was J2040+4527 (separation $=2^{\circ}\!\!.856$). The data were correlated with the EVN software
correlator \citep[SFXC;][]{kei15} at the Joint Institute for VLBI ERIC (JIVE) using 2048 
channels in epochs 2014.46 and 2016.45, and 4096 channels in epochs 2018.44 and 2020.82, 
generating all 4 polarization combinations (RR, LL, RL, and LR) with a spectral resolution in all 
epochs of $\sim$2~kHz ($\sim$0.03~\kms). \\
\indent The data were calibrated using the Astronomical
Image Processing Software package (AIPS) by following the standard calibration procedure 
\citep[e.g.,][]{sur112}. Specifically, the bandpass, the delay, the phase, and the polarization 
calibration were performed in all epochs on the calibrator J2202+4216. Then we performed the
fringe-fitting and the self-calibration on the brightest maser feature of each epoch (reference
maser features VLA1.1.15, VLA1.2.07, VLA1.3.04, and VLA1.4.05 in Tables~\ref{VLA1.1_tab},
\ref{VLA1.2_tab}, \ref{VLA1.3_tab}, and \ref{VLA1.4_tab}, respectively; for the 
notation definition see Sect.~\ref{res}). The \textit{I},
\textit{Q}, \textit{U}, and \textit{V} Stokes cubes were then imaged using the AIPS task IMAGR.
Afterward, the \textit{Q} and \textit{U} cubes were combined to produce cubes of linearly polarized
intensity ($POLC=\sqrt{Q^{2}+U^{2}-\sigma_{P}^2}$) and polarization angle
($POLA=1/2\times~atan(U/Q)$). The polarized intensity cubes were corrected according to the noise 
$\sigma_{P}=\sqrt{[(Q \times \sigma_{Q})^2+(U \times \sigma_{U})^2]/(Q^{2}+U^{2})}$, where
$\sigma_{Q}$ and $\sigma_{U}$ are the noise of the \textit{Q} and \textit{U} Stokes cubes,
respectively. The formal error on $POLA$ due to the thermal noise is given by 
$\sigma_{POLA}=0.5 ~(\sigma_{P}/POLC) \times (180^{\circ}/\pi)$  \citep{war74}.\\
%
%\begin{figure}[h!]
%\centering
%\includegraphics[width = 9 cm]{3C48_ep3.eps}
%\includegraphics[width = 9 cm]{3C48_ep4.eps}
%\caption{Maps of the polarization calibrator 3C48 in epochs 2018.44 (\textit{top}) and
%2020.82 (\textit{bottom}). The contours represent the total intensity and they correspond to 2, 4,
%8, 16, 32, and 64 times 1.6~\mjyb. The black segments are the linear polarization vectors drawn inside
%the total intensity contours and with linearly polarized intensity greater than three times the rms
%of the map. The restoring beam is overplotted on the bottom left corner of each panel (see
%Table~\ref{Obs}).}
%\label{3c48}
%\end{figure}
%
\indent To measure the absolute positions of the \water ~maser features, we self-calibrated, in all
four epochs, the phase-reference source J2040+4527 and the amplitude and phase solutions were
applied only to the uncalibrated peak channel of the brightest maser feature of each epoch. 
We show the contours
maps of J2040+4527 in Fig.\ref{phcal} and the absolute positions, with their uncertainties, of the
reference maser feature of each epoch are listed in Table~\ref{abspos}. The uncertainties of the
absolute positions of the reference maser features are estimated by adding quadratically the 
systematic errors ($\varepsilon_{sys}=0.028$~mas) due to the source elevation limit and the 
separation between the calibrator and the target \citep[][maximum and minimum elevation of W75N(B) 
at each station and in all epochs were $\sim80$\d ~and $\sim35$\d, respectively]{reid14},
the errors due to the thermal noise ($\varepsilon_{\rm{rms}}$), the errors of the Gaussian fit 
of the peak spot of the reference maser feature ($\varepsilon_{\rm{gauss}}$), and the errors 
of $\alpha_{2000}$ and $\delta_{2000}$ of J2040+4527 
\citep[0.68~mas and 0.80~mas, respectively,][]{petr11}. The later is usually unnecessary in 
relative astrometry studies, as the one presented here, however due to the large time interval 
between the EVN epochs we include it in our analysis to account for possible variation of the 
calibrator position from one epoch to the others. Another source of uncertainty that is
taken into consideration here is that due to the identification of the maser features, that is to 
account for the maser spots scatter of each maser features \cite[see][hereafter S+11a]{sur112},
and it is equal to half of the restoring beam of W75N(B).\\
\begin{figure*}[th!]
\centering
\includegraphics[width = 9.1 cm]{VLA1_over_Q_UN_final.eps}
\includegraphics[width = 9.1 cm]{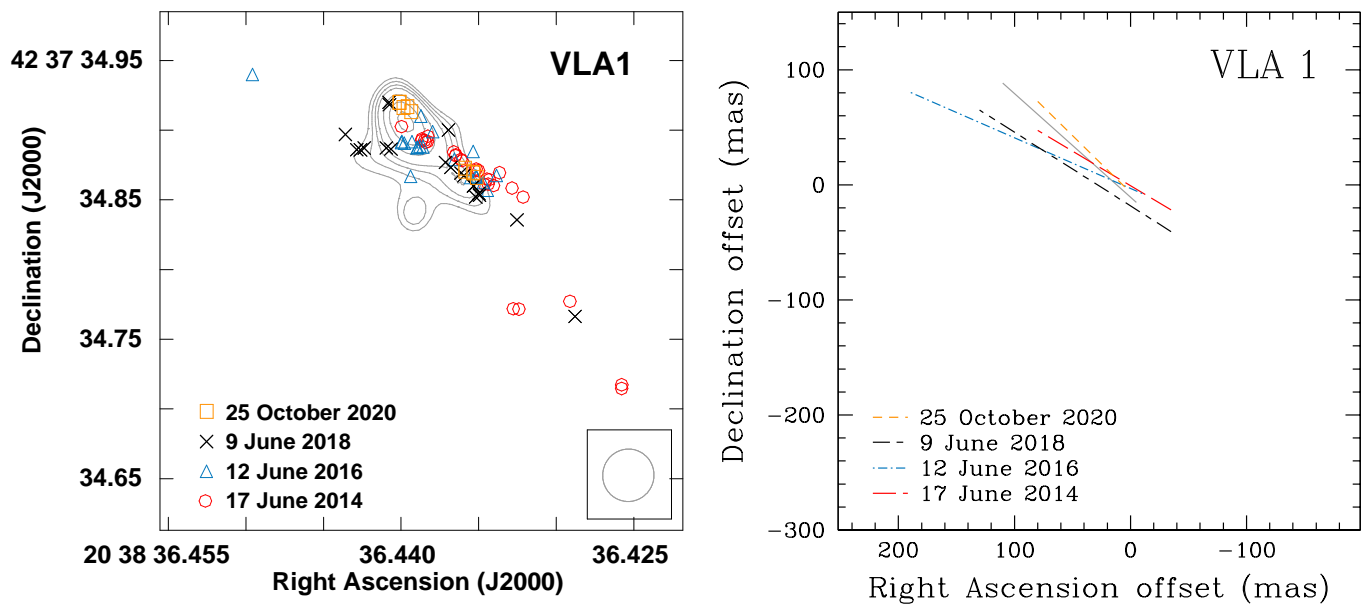}
\caption{Comparison of the \water ~maser features detected toward VLA\,1 (\textit{left panel}) in the
four EVN epochs (2014.46, 2016.45, 2018.44, and 2020.82) and superimposed to the uniform-weighted 
continuum map at Q-band (central frequency 44~GHz) of the thermal radio jet driven by VLA\,1 obtained 
with the VLA by \cite{rod20}. The light gray contours are 5, 6, 7, 8, 10, 12, 14 times 
$\sigma=100~\rm{\mu Jy~beam^{-1}}$ and the VLA beam is shown on the bottom right corner.
The positions of the maser features are corrected assuming the proper motion of the region W75N(B) 
equal to the median proper motion measured for the 6.7~GHz \meth ~maser features by \cite{ryg12}, 
$\langle\mu_{\rm{\alpha}}\rangle=(-1.97\pm0.10)~\rm{mas~yr^{-1}}$ and
$\langle\mu_{\rm\delta}\rangle=(-4.16\pm0.15)~\rm{mas~yr^{-1}}$. In addition, the positions of the maser
features associated with VLA\,1 and detected in the last three epochs have been further corrected 
considering the proper motion of VLA\,1 within W75N(B). This proper motion is 
$\langle\mu_{\rm{\alpha}}\rangle^{\rm{VLA\,1}}=(-6.3\pm0.4)~\rm{mas~yr^{-1}}$ and
$\langle\mu_{\rm\delta}\rangle^{\rm{VLA\,1}}=(+5.0\pm0.4)~\rm{mas~yr^{-1}}$ (see Appendix~\ref{appB}). 
The size of the symbols are ten times the uncertainties of the absolute positions of the maser features 
(see Table~\ref{abspos}).
For clarity we show in the \textit{right panel} the linear fit of the maser features for each epoch, the
length of the segments correspond to the position of the maser features used in the fitting. The 
reference position is $\alpha_{2000}=20^{\rm{h}}38^{\rm{m}}36^{\rm{s}}\!.43399$ and 
$\delta_{2000}=+42^{\circ}37'34''\!\!.8710$.
The parameters of the linear fit are reported in Table~\ref{linearfit}. The solid light-gray line 
indicates the orientation of the thermal radio jet \citep[PA=+42\d$\pm$5\d;][]{rod20}.
}
\label{over_vla1}
\end{figure*}
\indent We had to use different approaches than we did in the past to calibrate the linear
polarization angles of the \water ~maser features. This is because the last National Radio Astronomy Observatory (NRAO) POLCAL 
observations\footnote{http://www.aoc.nrao.edu/$\sim$smyers/evlapolcal/polcal\_master.html} of
J2202+4216 were made in May/June 2012. 
For epoch 2014.46, we assumed that the magnetic field orientation on the plane of the sky
around VLA\,1 has not changed from the last VLBI epoch (i.e., 2012.54, \citealt{sur142} hereafter S+14), 
as it was
the case between epochs 2005.89 and 2012.54 \citepalias{sur142}. We aligned the \water ~maser features
detected toward W75N(B) in epoch 2012.54, for which the absolute positions were unknown, with those
in epoch 2014.46 by associating the maser feature VLA1.1.21 ($V_{\rm{lsr}}=11.37$~\kms;
Table~\ref{VLA1.1_tab}) with the maser feature VLA1.28 ($V_{\rm{lsr}}=11.35$~\kms;
\citetalias{sur142}), because both are spatially coincident and have similar radial velocities.
As a consequence we found that the maser feature VLA1.1.05 (Table~\ref{VLA1.1_tab}) 
can be considered to be part of the same maser clump gas of VLA1.07 \citepalias{sur142}, although 
they are not 
exactly the same maser feature. Therefore, we can assume that VLA1.1.05 has
a mean linear polarization angle ($\chi$) equal to -25\d, which is the linear polarization
angle measured in the maser clump gas of VLA1.07 by \citetalias{sur142}. We were thus able 
to estimate the polarization angles of the \water ~maser features in epoch 2014.46 with a systemic error 
of no more than $\sim$~8\d.
For epoch 2016.46, we calibrated the linear polarization angles of the \water ~maser
features by rotating the linear polarization angle measured for J2202+4216 from our EVN data to the
one measured on 1$^{\rm{st}}$ June 2016 by one of the Korean VLBI Network (KVN) antennas 
($\chi_{\rm{J2202+4216,KVN}}=-61^{\circ}\pm12^{\circ}$ calibrated by using 3C286, \textit{private
communication}). In this case the systemic error was
of no more than $\sim$~12\d. In this epoch, we also tried to calibrate the linear polarization 
angles by observing the well known polarization calibrator 3C48, but the rms was not sufficient to
detect the linear polarization intensity above $3\sigma$ (see Table~\ref{Obs}).
Thanks to the wider bandwidth in the last two epochs (2018.44 and 2020.82), we were instead able 
to calibrate the linear polarization angles by using the polarization
calibrator 3C48. 
%(Fig.~\ref{3c48})
We assumed for 3C48 a polarization angle at K-band equal to $-70$\d ~\citep{per13}.\\
\indent Similarly to \citetalias{sur142}, we analyzed the polarimetric data following the procedure 
reported in \citetalias{sur112}. Therefore, we first identified the \water ~maser features and then
we measured the mean linear polarization fraction ($P_{\rm{l}}$) and the mean linear polarization
angle ($\chi$) for each identified \water ~maser feature considering only the consecutive channels
(more than two) across the total intensity spectrum for which the polarized intensity is
$\geq10~\rm{rms}$. Afterward, by using the full radiative transfer method (FRTM) code for 22~GHz 
\water ~masers \citep{vle06}, which is based on the model for unsaturated 22~GHz \water 
~masers of \cite{ned92}, we modeled the observed total intensity and linear polarization 
spectra of the linearly polarized maser features by gridding the intrinsic maser
linewidth (\dvi) between 0.4 and 4.5~\kms, in steps of 0.025~\kms, using a least square fitting routine
($\chi^2$-model) with an upper limit of the emerging brightness temperature (\tbo, 
where $\Delta \Omega$ is the maser beaming) of $10^{11}$~K~sr ~\citep[for more details see 
Appendix~A of][]{vle06}. In this
way we were able to obtain as outputs of the \code ~the values of \tbo ~and \dvi ~that 
produce the best fit models for our linearly polarized maser features. Because the \code ~
is based on a model for unsaturated \water ~maser, it cannot properly disentangle the values of \tbo 
~and \dvi ~in the case of saturated maser features and therefore it provides only a lower limit for 
\tbo ~and an upper limit for \dvi. An upper limit for \tbo ~below which the maser features can be
considered unsaturated is \tbo$< 6.7 \times 10^9$~K~sr \citep{sur111}. Then, from \tbo ~and 
$P_{\rm{l}}$ we could estimate the angle between the maser propagation direction and the
magnetic field ($\theta$) from which the 90\d ~ambiguity of the 
magnetic field orientation with respect to the linear polarization vectors can be solved. 
\begin {table*}[t!]
\caption []{Linear fit parameters of
\water ~maser features along VLA\,1 and around VLA\,2.} 
\begin{center}
\scriptsize
\begin{tabular}{ l c c c c c c c c}
\hline
\hline
\,\,\,\,\,(1)&(2)            & (3)             & (4)       & (5)     & \multicolumn{2}{c}{(6)}     & (7) & (8)      \\
 Epoch       & $m$\tablefootmark{a}  & $q$\tablefootmark{a} & PA & $\rho$\tablefootmark{b}     & \multicolumn{2}{c}{Proper Motion\tablefootmark{c}} &  $\langle V \rangle$\tablefootmark{d} & $V_{\rm{max}}$\tablefootmark{d}\\
             &               &                 & (\d)      &                &  ($\rm{mas~yr^{-1}}$) & (\kms)  & (\kms) & (\kms) \\ 
\hline
\multicolumn{9}{c}{VLA\,1}\\
\hline
 2014.46\tablefootmark{e} & $0.60\pm0.03$  & $-0.65\pm0.72$ & $+59\pm2$   & $+0.98$    & $-$ & $-$& +11.7 & +16.8  \\
 2016.45     & $0.44\pm0.06$ & $-3.11\pm2.95$ & $+66\pm3$ & $+0.87$ & $-$           & $-$ & +11.2 & +16.4 \\
 2018.44     & $0.64\pm0.10$ & $-18.49\pm5.10$ & $+57\pm6$ & $+0.84$ & $-$  & $-$ & +11.3 & +15.0 \\
 2020.82     & $0.99\pm0.02$ & $-6.76\pm0.90$ & $+45\pm1$ & $+1.00$ & $-$ & $-$ & +18.5 & +26.3\\
\hline
\multicolumn{9}{c}{VLA\,2 - zone 2}\\
\hline
 2014.46     & $-$            & $-$         & $-$       & $-$     & $-$        & $-$       & $-$  & $-$ \\
 2016.45     & $0.04\pm0.01$  & $-628\pm10$ & $+87\pm1$ & $+1.00$ & $-$        & $-$       & +2.0 & +2.3 \\
 2018.44     & $-0.08\pm0.01$ & $-554\pm10$ & $-85\pm1$ & $-1.00$ & $3.8^{+15.6}_{-2.1}$   & $23.4^{+96.2}_{-12.9}$ & +2.7 & +3.5\\
 2020.82     & $0.30\pm0.04$  & $-750\pm19$ & $+73\pm2$ & $+0.95$ & $4.3^{+22.3}_{-1.2}$ & $26.5^{+137.5}_{-7.4}$ & +4.5 & +9.3\\
\hline     
\multicolumn{9}{c}{VLA\,2 - zone 4}\\
\hline
 2014.46     & $-$            & $-$            & $-$        & $-$     & $-$    & $-$    & $-$   & $-$ \\
 2016.45     & $1.26\pm0.32$  & $-1358\pm138$  & $+38\pm18$ & $+0.91$ & $-$    & $-$    & +18.3 & +20.7 \\
 2018.44     & $-7.33\pm6.50$ & $+2219\pm2686$ & $-8\pm12$  & $-0.62$ & $-$    & $-$    & +23.1 & +28.2 \\
 2020.82     & $-6.15\pm4.40$ & $+1545\pm1677$ & $-9\pm10$  & $-0.57$ & $12.5^{+0.9}_{-5.8}$ & $77.1^{+5.5}_{-35.8}$ & +23.3 & +27.2\\
\hline         
\end{tabular} 
\end{center}
\tablefoot{
\tablefoottext{a}{Considering the equation for a line $y=mx+q$.}
\tablefoottext{b}{Pearson product-moment correlation coefficient $-1\leq\rho\leq+1$; $\rho=+1$ 
($\rho=-1$) is total positive (negative) correlation, $\rho=0$ is no correlation.}
\tablefoottext{c}{For the proper motion on the plane of the sky we
considered the distance of the median point ($\Delta\alpha$, $\Delta\delta$) of
the line of one epoch from the line of the next epoch. The velocity reported for an epoch is 
always calculated with respect to the previous epoch. The errors are estimated
considering the uncertainties of $m$ and $q$ of the two epochs between which the velocity is measured.}
\tablefoottext{d}{$\langle V \rangle$ and $V_{\rm{max}}$ are the mean and maximum velocities observed along the line of sight, respectively.}
\tablefoottext{e}{We did not consider in the fit the features VLA1.1.01, VLA1.1.02, VLA1.1.03, 
VLA1.1.05, and VLA1.1.06.}
}
\label{linearfit}
\end{table*}
Indeed, if $\theta>\theta_{\rm{crit}}=55$\d ~the magnetic field appears to be perpendicular to the 
linear polarization vectors; otherwise, it is parallel \citep{gol73}. 
\citet{ned92} found that \tbo ~scaled linearly with $\Gamma+\Gamma_{\nu}$, which are the 
maser decay rate and cross-relaxation rate, respectively. As explained in \citet{sur112}, 
$\Gamma_{\nu}$ varies with the temperature and $\Gamma=1~\rm{s^{-1}}$ for the \water ~maser emission,
therefore in our fit
we consider a value of $\Gamma+\Gamma_{\nu}=1~\rm{s^{-1}}$ that allows us to adjust the fitted \tbo
~values by simply scaling it according to the real $\Gamma+\Gamma_{\nu}$ value 
as described in \citet{and93}. Note that \dvi ~and $\theta$ ~do not need to be adjusted.
The errors of \tbo, \dvi, and $\theta$ were determined by analyzing the
probability distribution function of the full radiative transfer $\chi^2$-model fits.
In case a maser feature is also circularly polarized, we can use the best estimates of \tbo ~and
\dvi ~in the \code ~to produce $I$ and $V$ models that we used to fit the $V$ spectra of
the circularly polarized maser features from which we can measure the Zeeman splitting. 
Due to the typical weak circularly polarized emission of \water ~masers ($<1\%$), it is important
to consider the self noise ($\sigma_{\rm{s.-n.}}$) produced by the maser features in their channels 
to determine whether
the circularly polarized emission is real. The self-noise becomes important when the power
contributed by the astronomical maser is a significant portion of the total received power
\citep{sau12}. Therefore, a detection of circularly polarized emission has been considered real 
only when the $V$ peak intensity of a maser feature is both $>3~\rm{rms}$ and $>3~\sigma_{\rm{s.-n.}}$
(see Table~\ref{Obs}). 
\section{Results}
\label{res}
We report in Table~\ref{para} the number of \water ~maser features detected around VLA\,1 and
VLA\,2 in the four EVN epochs with their corresponding local standard of rest velocity 
($V_{\rm{lsr}}$) and peak intensity ($I$) ranges, the mean linewidth of the 
maser features ($\langle\Delta v_{\rm{L}}\rangle$), the ranges of $P_{\rm{l}}$ and of the circular 
polarization fraction ($P_{\rm{V}}$), the ranges of $\chi$, and all the outputs of the \code 
~(\tbo, \dvi, and $\theta$) with the derived magnetic field parameters. These are the ranges of 
the estimated orientation of the magnetic field on the plane of the sky ($\Phi_{\rm{B}}$) and 
its error-weighted value ($\langle\Phi_{\rm{B}}\rangle$), the range of the magnetic field 
strength along the line of sight in absolute values ($|B_{||}|$) and their error-weighted 
values ($\langle |B_{||}| \rangle$), and the error-weighted values of the estimated 3D magnetic 
field strength ($\langle |B| \rangle$).
In addition, we report the detailed results obtained from the four EVN epochs with their plots and 
Tables in Appendix~\ref{appA}. We should mention here that the detected 
\water ~maser features are called throughtout the paper as VLA1.x.yy and VLA2.x.yy, where x is a 
number from 1 to 4 indicating the EVN epoch from the first (2014.46) to the fourth (2020.82) and 
yy is the number of the \water ~maser feature counted from west to east in each epoch. Maser features
with the same yy value but with different x value are not necessarily related to each others, that 
is they are not necessarily the same maser feature detected in different epochs.\\
\indent The main objective of our monitoring project is to determine whether the 22~GHz \water ~maser
distributions (Sect.~\ref{sv}), and the maser features characteristics such intensity 
(Sect.~\ref{fiv}) and polarization (Sect.~\ref{mp}), 
around VLA\,1 and VLA\,2 show any variation over time and how the magnetic field behaves 
accordingly. To do this we have to correct the positions of the detected \water ~maser features 
from epoch 2016.45 to epoch 2020.82 by considering the proper motion of the entire region with 
respect to the Earth and assuming epoch 2014.46 as the reference epoch. 
\cite{ryg12} measured the median proper motion of the 6.7~GHz \meth ~maser features associated with 
VLA\,1 and VLA\,2. They assumed that this motion represents the proper motion of the entire region 
W75N(B). The components of this proper motion along right ascension and declination are  
$\langle\mu_{\rm{\alpha}}\rangle=(-1.97\pm0.10)~\rm{mas~yr^{-1}}$ and
$\langle\mu_{\rm\delta}\rangle=(-4.16\pm0.15)~\rm{mas~yr^{-1}}$. We therefore corrected the 
positions of the \water ~maser features of the last three EVN epochs by assuming that both VLA\,1 and
VLA\,2 moved from epoch 2014.45 with a proper motion equal to that measured by \cite{ryg12}.
Furthermore, a comparison of the absolute positions of the continuum emission of VLA\,1
and VLA\,2 at K-band, as measured by \citet[][epoch 1996.96]{tor97}, \cite{car15} and 
\citet[][epoch 2014.20]{rod20} with the VLA, has showed that, while VLA\,2 does not show any further
motion within the region W75N(B), VLA\,1 does actually move. The proper motion of VLA\,1 within 
W75N(B) is $\langle\mu_{\rm{\alpha}}\rangle^{\rm{VLA\,1}}=(-6.3\pm0.4)~\rm{mas~yr^{-1}}$ and
$\langle\mu_{\rm\delta}\rangle^{\rm{VLA\,1}}=(+5.0\pm0.4)~\rm{mas~yr^{-1}}$ (see Appendix~\ref{appB}). 
Therefore, before comparing the maser features around VLA\,1 we must apply a further 
correction to their positions.\\
\indent Unfortunately, we cannot compare the maser 
distributions of the EVN epochs with those observed previously with the VLBA in epochs 2005.89 and 
2012.54, because we do not have any information on the absolute positions of the maser features 
in those epochs \citepalias{sur112, sur142}. Nevertheless, when necessary refer to Table~A.3 of 
\citetalias{sur142} for the parameters of the VLBA epochs 2005.89 and 2012.54.
\subsection{Spatial and velocity distribution of the \water ~masers}
\label{sv}
\subsubsection{VLA\,1}
\label{svVLA1}
\begin{figure*}[th!]
\centering
\includegraphics[width = 8 cm]{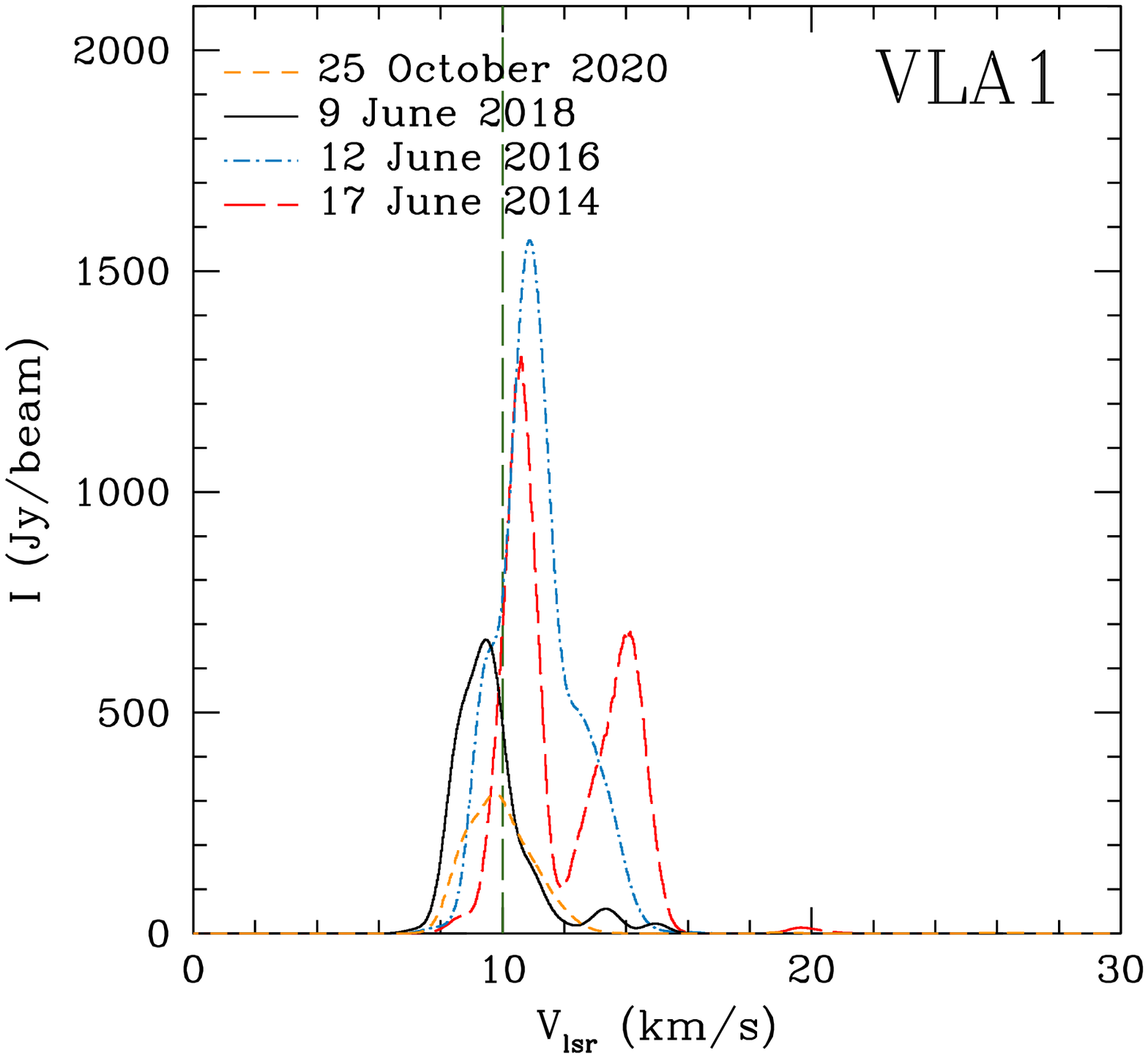}
\includegraphics[width = 8 cm]{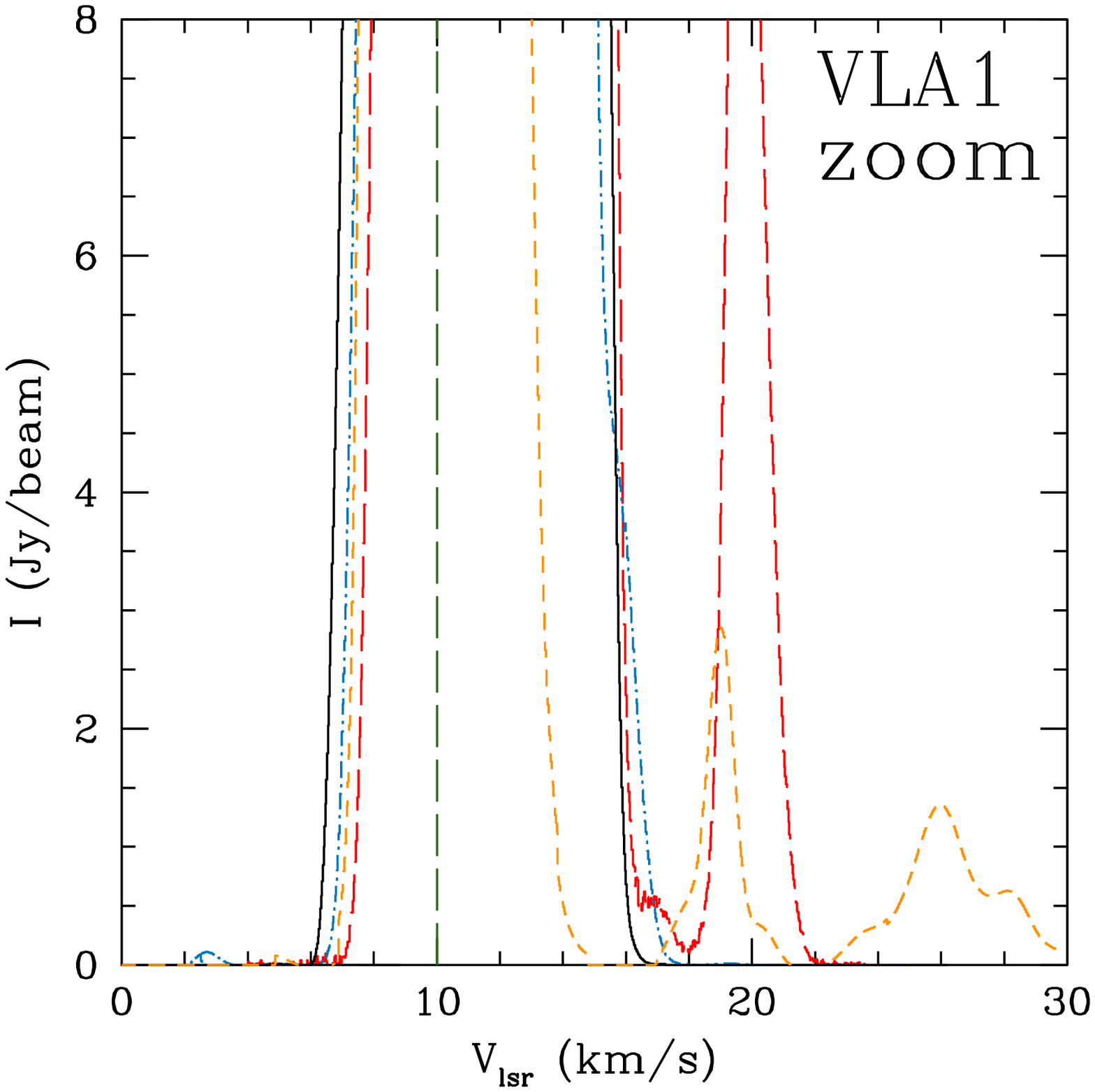}
\caption{Comparison of the total sum of the \water ~maser spectra detected toward VLA\,1
with the EVN in epochs 2014.46, 2016.45, 2018.44, and 2020.82. The vertical dashed dark-green line 
indicates the assumed systemic velocity of the region ($V_{\rm{lsr}}=+10.0$~\kms, \citealt{she03}). }
\label{totspectravla1}
\end{figure*}
The number of 22~GHz \water ~maser features detected around VLA\,1 has decreased from the EVN epoch
2014.46 to the EVN epoch 2020.82 (see Table~\ref{para}). If we plot all the \water ~maser features
detected in all the four EVN epochs, after correcting their positions as reported above, we see that
these are always distributed along the radio continuum emission of VLA\,1 presented by \cite{rod20}.
This continuum emission was obtained at a resolution of tens of milliarcseconds by observing with 
the VLA a wide range of frequencies ($4-48$~GHz) in 2014. \cite{rod20} concluded that VLA\,1 
is at the early stage of photoionization and it is driving a thermal radio jet at scale of about 
0.1~arcsec ($\approx130$~au). In
Fig.~\ref{over_vla1} we overplot the \water ~maser features detected in the four EVN epochs to the
continuum emission measured at Q-band by \cite{rod20}. The velocities of all the \water 
~maser features are not showed in this figure, however, they are shown in the left panels of 
Fig.~\ref{posplot}, where every EVN epoch is reported in a different panel.
In order to better visualize the accordance of the maser features distribution with the position angle of 
the thermal radio jet, we make a linear fit of 
the maser features in each epoch. The results of these fits are shown in the right panel
of Fig.~\ref{over_vla1} and their parameters are reported in Table~\ref{linearfit}. For the linear fit
of epoch 2014.46, we do not consider the group of five maser features located south because their
positions would strongly affect the position angle of the fitting line. This is not the case for 
the northeast and southeast maser features of epochs 2016.45 and 2018.44, respectively, which we indeed 
include in our linear fits. We see that the position angle of the fitted lines (see Fig.~\ref{over_vla1}
and Table~\ref{linearfit} for comparison) are consistent with the position angle of the thermal radio 
jet \citep[solid gray line in Fig.~\ref{over_vla1}, PA=+42\d$\pm$5\d;][]{rod20}.\\
\begin{figure*}[th!]
\centering
\includegraphics[width = 9 cm]{VLA2_over_K_final.eps}
\includegraphics[width = 9.2 cm]{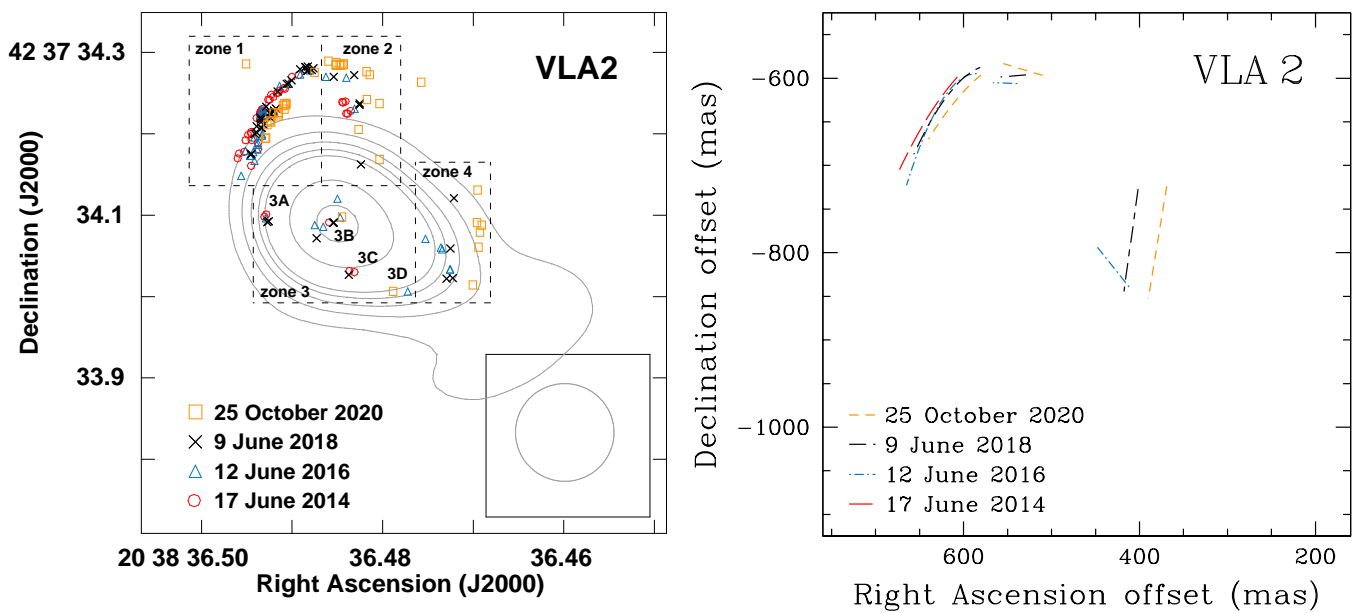}
\caption{Comparison of the \water ~maser features detected toward VLA\,2 (\textit{left panel}) in the
four EVN epochs (2014.46, 2016.45, 2018.44, and 2020.82) and superimposed to the natural-weighted 
continuum map at K-band (central frequency 22~GHz) of the thermal, collimated ionized wind emitted by 
VLA\,2 obtained with the VLA by \cite{car15}. The light gray contours are 5, 10, 15, 20, 25, 50, and 
75 times $\sigma=10~\rm{\mu Jy beam^{-1}}$ and the VLA beam is shown on the bottom right corner. 
The positions of the maser features are corrected assuming the proper motion of the subregion W75N(B)
equal to the median proper motion measured for the 6.7~GHz \meth ~maser features by \cite{ryg12}, 
$\langle\mu_{\rm{\alpha}}\rangle=(-1.97\pm0.10)~\rm{mas~yr^{-1}}$ and
$\langle\mu_{\rm\delta}\rangle=(-4.16\pm0.15)~\rm{mas~yr^{-1}}$. The size of the symbols are ten times 
the uncertainties of the absolute positions of the maser features (see Table~\ref{abspos}). For clarity
we show in the \textit{right panel} the parabolic fit of the maser features of zone 1 and the linear 
fit of the maser features of zone 2 (north features) and zone 4 for each epoch. The reference position
is $\alpha_{2000}=20^{\rm{h}}38^{\rm{m}}36^{\rm{s}}\!.43399$ and 
$\delta_{2000}=+42^{\circ}37'34''\!\!.8710$.
The parameters of the linear and parabolic fits are reported in Tables~\ref{linearfit} and \ref{polyfit}, 
respectively.}
\label{over_vla2}
\end{figure*}
\indent We note that the mean velocities along the line of sight of the maser features
($\langle V \rangle$, Col.~7 of Table~\ref{linearfit}) in the first three epochs, which are all around 
+11.5~\kms, and their similar maximum line of sight velocities ($V_{\rm{max}}\approx+16$~\kms, Col.~8 
of Table~\ref{linearfit}) are largely different than those of epoch~2020.82 
($\langle V \rangle=+18.5$~\kms ~and  $V_{\rm{max}}=+26.3$~\kms). These differences might be explained
either with an acceleration of the motion of VLA\,1 along the line of sight and away from us, or with
the variation of the masing conditions. Even though a combination of the two seems to be the 
case (see Fig.~\ref{totspectravla1}). In addition, we also note that the maser velocities are spatially 
mixed on the plane of the sky in each epoch.
\subsubsection{VLA\,2}
The number of 22~GHz \water ~maser features detected around VLA\,2 in the four EVN epochs ranges
between 37 and 44 (see Table~\ref{para}),
which is roughly half of the number of maser features detected in the two previous VLBA epochs (88 and
68 in 2005.89 and 2012.54, respectively; \citetalias{sur142}). These differences might not be because of 
the sensitivity of the EVN observations, which was better than that of the VLBA epochs
(see Table~\ref{Obs} and \citetalias{sur112,sur142}), but of the maser activity of the region. 
%The total
%flux of the \water ~maser emission almost constantly increased between the VLBA epoch 2012.54 and the
%EVN epoch 2020.82 (see Fig.~\ref{totflux}). This reflects the fact that the maser features in the four
%EVN epochs, although half in number, are actually brighter than those detected in the two VLBA epochs
%(see Table~\ref{para} and \citetalias{sur112,sur142}). Nevertheless,
The maser distribution in the four EVN epochs is consistent with the elliptical shell observed for the 
first time by \cite{kim13} and confirmed by \citetalias{sur142}. \citetalias{sur142} also measured a mean 
expansion velocity of the maser shell on the plane of the sky, which begun to expand in 1999 when the shell 
was quasi-circular and continued by becoming elliptical in 2007 (\citealt{tor03}; \citetalias{sur112}; 
\citealt{kim13}), of 4.9~\masyr ~(30~\kms ~at a distance of 1.3~kpc). However, the previous observations 
could not be properly compared because only those presented in \cite{kim13} were made in phase-reference 
mode and therefore the absolute positions of most of the maser features in the other epochs were unknown. 
For estimating the expansion velocity, \citetalias{sur142} assumed that the center of the shells coincides 
in the different epochs and consequently the measured expansion is radial. 
Since we were able to measure the absolute positions of the maser features in the 
four EVN epochs, we are now able to properly measure the expansion velocity. Nevertheless, we can 
compare our expansion velocities only in magnitude and not in direction with that measured
by \citetalias{sur142}. We plot all the \water ~maser features detected with the EVN
\label{svVLA2}
\begin {table}[t!]
\caption []{Polynomial fit parameters of
the \water ~maser features of zones 1 of VLA\,2.} 
\begin{center}
\scriptsize
\begin{tabular}{ l c c c c c c }
\hline
\hline
\,\,\,\,\,(1)&(2)            & (3)             & (4)       & (5)     & \multicolumn{2}{c}{(6)}           \\
 Epoch       & $a$\tablefootmark{a}  & $b$\tablefootmark{a} & $c$\tablefootmark{a} & $R^2$\textbf{\tablefootmark{~b}}  & \multicolumn{2}{c}{Proper Motion} \\
             &               &                 &       &                &  ($\rm{mas~yr^{-1}}$) & (\kms)  \\ 
\hline
 2014.46     & $-0.00757$    & $8.08$ & $-2714$   & $0.85$  & $-$ & $-$  \\
 2016.45     & $-0.0204$     & $23.9$ & $-7595$   & $0.89$  & $-$ & $-$  \\
 2018.44     & $-0.0118$     & $13.3$ & $-4332$   & $0.96$  & $-$ & $-$ \\
 2020.82     & $-0.00669$    & $6.93$ & $-2365$   & $0.94$  & $-$ & $-$ \\
             &               &                 &           &        & $-2.5$\tablefootmark{c} & $-15.4$\tablefootmark{c} \\
\hline
\end{tabular} 
\end{center}
\tablefoot{
\tablefoottext{a}{Considering the equation of a polynomial of second order $y=ax^2+bx+c$.}
\tablefoottext{b}{$R^2$ is the coefficient of determination of the polynomial fit.}
\tablefoottext{c}{Between epoch 2014.46 and epoch 2020.82. The minus sign indicates that the motion
is opposite to the expansion velocity measured by \citetalias{sur142}.}
}
\label{polyfit}
\end{table}
and overplotted to the continuum emission at K-band \citep{car15} 
in the left panel of Fig.~\ref{over_vla2}. Here, we group the maser features in four different zones.
As for VLA\,1, the velocities of all the \water ~maser features detected in VLA\,2 are showed in 
the right panels of Fig.~\ref{posplot}. \\

\begin{figure}[t!]
\centering
\includegraphics[width = 8.2 cm]{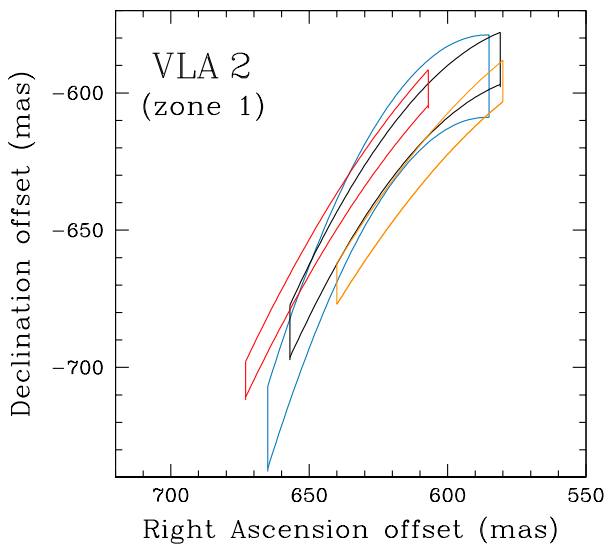}
\caption{Each area represents the area in each epoch where the maser features considered in the polynomial fit 
(see Fig.~\ref{over_vla2} and Table~\ref{polyfit}) are located. The colors correspond to those used in Fig.~\ref{over_vla2}. }
\label{para_area}
\end{figure}

\noindent {\bf Zone~1}. This is located northeast. Interestingly, while \citetalias{sur142} found 
expanding motions, our data now shows the opposite trend, with apparent motions toward the central source.
Indeed, the maser features seem to not trace an expansion, but they actually seem to "bounce". The 
maser features of epoch 2014.46 are generally slightly at northeast of those of epochs 2016.45 and 
2018.44, which are more northeast than those of epoch 2020.82. This apparent "bouncing" could be 
interpreted
as evidence that the outflowing gas, where the masers arise, encounters an obstacle, as already 
proposed by \cite{kim18}. This obstacle can be either a much denser medium that
stops the expansion of the gas (case~A) or the absence of physical conditions, such as density and
temperature, for producing the \water ~maser emission (case~B). In  case~A, the impact of the gas
with a denser medium might have produced an additional slow inward shock that pumped the maser features
in the four EVN epochs. To estimate the proper motion of the gas on the plane of the sky due to this inward 
shock we fit the maser features of zone~1
with polynomials of second order, one per EVN epoch, and the results are shown in the right panel of
Fig.~\ref{over_vla2} and the parameters are listed in Table~\ref{polyfit}. In addition in 
Fig.~\ref{para_area} we show the areas where the maser features considered in the polynomial fit are 
located. We note that only the areas of epochs 2014.46 and 2020.82 do not 
overlap and therefore their polynomial fits can be used to estimate the proper motion.
Hence, we estimate the proper motion by measuring a mean distance between the curves of epochs 
2014.46 and 2020.82 and the resulting velocity is equal to 
-2.5~\masyr ~that corresponds to $\sim-15.4$~\kms ~at a distance of 1.3 kpc, the 
minus sign indicates that the motion is opposite to the expansion measured by \citetalias{sur142}.
Although case~A can be plausible, we should note that the curvature of the maser distribution in zone~1 is
opposite than one would expect (see right panel of Fig.~\ref{over_vla2}). In case~B, the physical
conditions of the gas in the northeast
are not suitable for the maser emission and what we observe are maser features pumped by different
outward shocks in each epoch. However, both cases are questioned by the presence of the broad maser
feature VLA2.4.39 ($I=1.94$~\jyb, $\Delta v_{\rm{L}}=1.65$~\kms; see Table~\ref{VLA2.4_tab}) that is 
located farther northeast of all the other maser features of zone 1. This maser feature shows physical 
parameters, such as the velocity ($V_{\rm{lsr}}=+16.35$~\kms), consistent with those of the other maser 
features of the zone. However, the presence of
this isolated maser feature can also be justified by the presence of a belt of gas, between VLA2.4.39
and the other maser features, where the maser conditions are not met. Therefore, if the shock that
pumped VLA2.4.39 is the same that pumped the maser features in the EVN epoch 2014.46, we can estimate
the proper motion due to this shock. VLA2.4.39 is at about 53~mas from the front of the maser features
detected in epoch 2014.46, therefore the proper motion is $7.9$~\masyr ~that corresponds to
$\sim49$~\kms ~at a distance of 1.3~kpc. This is higher than the expansion velocity measured
by \citetalias{sur142} and the proper motions measured by \cite{kim13}, $\sim20-30$~\kms.\\
\begin {table}[t!]
\caption []{Individual \water ~maser features around VLA\,2 
for which the velocity of the pumping shock is measured.} 
\begin{center}
\scriptsize
\begin{tabular}{ l c c c c c }
\hline
\hline
\,\,\,\,\,(1)&(2)     & (3)            & (4)           & \multicolumn{2}{c}{(5)}          \\
 Epoch    & maser     & $V_{\rm{lsr}}$\tablefootmark{a} & Peak      &  \multicolumn{2}{c}{Proper  motion} \\
          & feature   &                & Intensity (I)   &                        &  \\ 
          &           &  (\kms)        & (\jyb)    &   (\masyr) &  (\kms) \\ 
\hline
\multicolumn{6}{c}{VLA\,2 - zone 2 - below linear fit}\\
\hline
 2014.46  & VLA2.1.04 & $+4.26$        & $5.78\pm0.02$ &  $-$                   & $-$     \\
 2016.45  & VLA2.2.07 & $+5.63$        & $0.67\pm0.02$ &  $4.8$                 & $29.6$  \\
 2018.44  & VLA2.3.06 & $+6.27$        & $5.86\pm0.03$ &  $4.5$                 & $27.8$  \\
 2020.82  & VLA2.4.13 & $+0.65$        & $0.38\pm0.05$ &  $4.7$                 & $29.0$  \\
          &           &                &               &  $4.6$\tablefootmark{b} & $28.4$\tablefootmark{b}  \\
\hline
\multicolumn{6}{c}{VLA\,2 - zone 3 - group 3A}\\
\hline
 2014.46  & VLA2.1.22 & $+5.61$        & $2.77\pm0.01$ &  $-$                   & $-$     \\
 2016.45  & VLA2.2.18 & $+5.87$        & $1.58\pm0.02$ &  $2.0$                 & $12.3$  \\
 2018.44  & VLA2.3.32 & $+5.90$        & $0.90\pm0.02$ &  $2.0$                 & $12.3$  \\
          &           &                &               &  $2.0$\tablefootmark{c} & $12.3$\tablefootmark{c}  \\
\hline
\multicolumn{6}{c}{VLA\,2 - zone 3 - group 3B}\\
\hline
 2014.46  & VLA2.1.09 & $+21.19$       & $0.27\pm0.01$ &  $-$                   & $-$     \\
 2016.45  & VLA2.2.12 & $+13.45$       & $10.42\pm0.52$&  $5.0$                 & $30.8$  \\
 2018.44  & VLA2.3.13 & $+12.80$       & $9.10\pm0.10$ &  $7.5$                 & $46.2$  \\
          &           &                &               &  $6.2$\tablefootmark{c} & $38.3$\tablefootmark{c}  \\
\hline     
\multicolumn{6}{c}{VLA\,2 - zone 3 - group 3C}\\
\hline
 2014.46  & VLA2.1.03 & $+14.90$       & $20.80\pm0.25$&  $-$                   & $-$     \\
 2018.44  & VLA2.3.09 & $+14.69$       & $4.77\pm0.13$ &  $1.3$                 & $8.0$  \\
\hline         
\end{tabular} 
\end{center}
\tablefoot{
\tablefoottext{a}{The assumed systemic velocity of the region is $V_{\rm{lsr}}=+10.0$~\kms ~\citep{she03}}
\tablefoottext{b}{Between epoch 2014.46 and epoch 2020.82.}
\tablefoottext{c}{Between epoch 2014.46 and epoch 2018.44.}
}
\label{vzone23}
\end{table}
\indent In \citeyear{car15}, \citeauthor{car15} presented the most recent continuum maps of VLA\,2, which were
obtained by observing four different frequency bands (C, U, K, and Q) with the VLA in 2014. From these new
maps it was possible to verify the variation of collimation of the outflow emitted from VLA\,2 as
traced by the \water ~maser features from 1999 to 2012 \citepalias{sur142}. We can therefore compare our
maser distributions with the VLA continuum emission at K-band (see Fig.~\ref{over_vla2}). The
asymmetric morphology of the K-band continuum emission, which shows a weaker emission toward southwest
and none toward northeast, suggests the presence of an obstacle toward northeast as we supposed above.
This obstacle might be an inhomogeneity within the dusty disk or envelope supposed by \cite{car15}.
In particular, the distribution of the maser features of zone 1 seems to be the continuation toward 
northwest of the last external contour of the continuum emission at 50~\ujyb (see Fig.~\ref{over_vla2}), 
even though no continuum emission is detected where these maser features arise. \\
\indent We note that the velocity range of the maser features of zone~1 in the four EVN epochs 
($-11$~\kms$<V_{\rm{lsr}}^{\rm{2014.46}}<+17$~\kms, 
$-16$~\kms$<V_{\rm{lsr}}^{\rm{2016.45}}<+17$~\kms, 
$-16$~\kms$<V_{\rm{lsr}}^{\rm{2018.44}}<+16$~\kms, and $+6$~\kms$<V_{\rm{lsr}}^{\rm{2020.82}}<+17$~\kms) 
is similar to that covered in the VLBA epoch 2012.54
($-10$~\kms$<V_{\rm{lsr}}^{\rm{2012.54}}<+18$~\kms; \citetalias{sur142}), even though the range observed in
the EVN epoch~2020.82 shows only redshifted velocities.
This might suggest that the
maser features of zone~1 traced the same shock, that moved outward, from 2012.54 to 2018.44, and
then they quenched any time between 2018.44 to 2020.82. We also note that, in the first three
EVN epochs, the blue- and redshifted maser features do not follow any particular spatial 
distribution, that is blue- and redshifted maser features are spatially coincident and do not show
any velocity gradient. This might further indicate that the gas expands along the walls of the 
denser medium when encountering it. Consequently, the maser features in the last EVN epoch
(2020.82) might trace a different shock that still
move outward rather than inward, according to the morphology of the maser distribution. This might
exclude the possible inward shock supposed above when we discussed case~A.\\

\noindent {\bf Zone~2}. The low number of \water ~maser features of zone~2 (northwest, see 
Fig.~\ref{over_vla2}) detected from the VLBA epoch 2012.54 to the EVN epoch 2018.44 and their total
\water ~maser intensity (see Sect.~\ref{fivVLA2}) indicate a low maser activity in this zone
since the appearance of the elliptical maser distribution. However, the maser features detected in 
the last EVN epoch 2020.82 are in number about three times more and their total \water ~maser 
intensity is almost ten times higher than previously detected (see Sect.~\ref{fivVLA2}), suggesting a 
sudden increment of the maser activity as never observed before in the northeast part of the maser 
distribution, neither when the distribution was quasi-circular (\citealt{tor97,tor03};
\citetalias{sur112}) nor afterward (\citealt{kim13}; \citetalias{sur112,sur142}). Comparing only 
the EVN epochs, we note that all the maser features in zone~2 north of declination 42\d37'34.\!\!''2
in Fig.~\ref{over_vla2} have a velocity range between $-1$~\kms ~and +9~\kms, that is, they are all
blueshifted,
with the exception of VLA2.4.12 that shows a redshifted velocity of $V_{\rm{lsr}}=+12.77$~\kms. 
The rest of the 
maser features of zone~2 (south of declination 42\d37'34.\!\!''2 in Fig. \ref{over_vla2}), which are
the only ones detected on the continuum emission at K-band (see Fig.~\ref{over_vla2}), show much 
higher redshifted velocities (+15~\kms$<V_{\rm{lsr}}<+23$~\kms). Differently from zone~1, we see that 
the maser features in zone~2 detected in one epoch are always detected outward with respect the previous 
epoch. To better estimate the apparent motion between different epochs, we make a linear fit of the 
maser features in each epoch. In particular, we are able to measure the expansion velocity of the 
north maser features in zone~2 by considering the distance of the median point of the line of one 
epoch from the line of the next epoch. The results are reported in Table~\ref{linearfit} and in the
right panel of Fig.~\ref{over_vla2}. The expansion velocities on the plane of the sky measured between 
the epochs 2016.45 and 2018.44 (3.8~\masyr) and between epochs 2018.44 and 2020.82 (4.3~\masyr) are 
consistent with the expansion velocity of 30~\kms ~(4.9~\masyr) measured by \citetalias{sur142}. In 
zone~2 it is also possible to identify four maser features, each detected in a different EVN epoch, 
located below the linear fits and at the center of the dashed rectangle that highlights zone~2
in Fig.~\ref{over_vla2}, that apparently seem to trace an outward motion. These maser features 
are VLA2.1.04, VLA2.2.07, VLA2.3.06, and VLA2.4.13 (see Table~\ref{vzone23}). Although these maser 
features have different line of sight velocities (see Col.3 of Table~\ref{vzone23}), we can estimate 
the expansion velocity between the four EVN epochs by assuming that the shock that pumped them is the 
same. We find a proper motion of 4.8~\masyr ~($\sim30$~\kms ~at 1.3~kpc) between the epochs 2014.46 
and 2016.45, 4.5~\masyr ~($\sim28$~\kms) between epochs 2016.45 and 2018.44, and 4.7~\masyr 
~($\sim29$~\kms) between the epochs 2018.44 and 2020.82 (see Table~\ref{vzone23}). The mean proper 
motion between epochs 2014.46 and 2020.82 is equal to 4.6~\masyr ~($\sim28$~\kms) that is again 
consistent with the expansion velocity measured by \citetalias{sur142} on the plane of the sky, that is 
30~\kms. The consistency of the expansion velocities measured from the maser features of zone~2 with 
that measured previously by \citetalias{sur142} is a further clue of the presence, since epoch 2014.46, 
of an obstacle in front of the expanding gas in zone~1 that is absent in front of the gas in zone~2 
that can freely expand.\\

\noindent {\bf Zone~3}. The \water ~maser features of zone~3 are all located toward the bright core of 
the continuum emission observed by \citet[][see Fig.~\ref{over_vla2}]{car15} and show velocities in
the range between +1~\kms ~and +22~\kms. We note that the blueshifted maser features are always 
located outward than the redshifted in all the epochs. Looking at 
Fig.~\ref{over_vla2} we can divide the maser features in zone~3 in four groups: one in the east (group 
3A), one in the center (group 3B), one in the south (group 3C), and the fourth in the southwest (group
3D). Group 3B coincides with the peak of the continuum emission at K-band ($\sim800$~\ujyb), groups 3A 
and 3C are located between the contours at 25$\sigma$ (250~\ujyb) and 50$\sigma$ (500~\ujyb) of 
Fig.~\ref{over_vla2}, while group 3D between the contours at 20$\sigma$ (200~\ujyb) and 25$\sigma$ 
(250~\ujyb). For the first three groups (3A-3C), we can identify maser features with similar velocity
in at least two epochs and we are therefore able to estimate their proper motions. In
particular, we identify in group~3A a maser feature in three consecutive epochs (from 2014.46 to 
2018.44) corresponding to VLA2.1.22, VLA2.2.18, and VLA2.3.32 (see Table~\ref{vzone23}). We measured a 
constant velocity of $\sim$12~\kms ~(2.0~\masyr) pointing slightly toward southwest between epochs 
2014.46 and 2016.45 and between epochs 2016.45 and 2018.44 (see Table~\ref{vzone23}). The maser 
features VLA2.1.03 and VLA2.3.09 of group~3C can be considered tracing the same gas and therefore the 
estimated proper motion is $\sim$8~\kms ~(1.3~\masyr; see Table~\ref{vzone23}). Both proper motions
of groups~3A and 3C are much lower than that measured by \citetalias{sur142}. The identification
of common maser features with similar $V_{\rm{lsr}}$ in group~3B is very difficult. However, we can 
identify three maser features by considering their relative position in Fig~\ref{over_vla2}. These are 
VLA2.1.09, VLA2.2.12, and VLA2.3.13 (see Table~\ref{vzone23}). The proper motion is then $\sim$31~\kms 
~(5.0~\masyr) between epochs 2014.46 and 2016.45, and $\sim$46~\kms ~(7.5~\masyr) between epochs 
2016.45 and 2018.44 (see Table~\ref{vzone23}). The mean proper motion between epochs 2014.46 and 
2018.44 is about 38~\kms ~(6.2~\masyr) that is slightly higher than the expansion velocity of 30~\kms 
~measured by \citetalias{sur142}.\\

\noindent {\bf Zone~4}. This zone is located southwest and the \water ~maser features distribution
is roughly aligned north-south as observed previously (\citealt{tor03,kim13};\citetalias{sur142}). No 
\water ~maser emission was detected toward this zone in the EVN epoch~2014.46. A comparison with the 
continuum emission at K-band reported by \cite{car15} reveals that the maser features of zone~4, 
which are all redshifted (+16~\kms$<V_{\rm{lsr}}<+29$~\kms), are all associated with the weak 
continuum emission at around 200-300~\ujyb ~(see Fig.~\ref{over_vla2}). The results of 
the linear fit of the maser features are reported in Table~\ref{linearfit} and displayed in the right 
panel of Fig.~\ref{over_vla2}. Because the linear fit of epoch~2016.45 crosses that of epoch~2018.44 
(see right panel of Fig.~\ref{over_vla2}), we can estimate an expansion velocity on the plane of the 
sky only between epochs 2018.44 and 2020.82. This is the largest ever measured toward the maser 
features around VLA\,2 and its value is equal to 12.7~\masyr ~($\sim$78~\kms). This high velocity might 
suggest that the gas does not encounter any dense matter toward southwest that could slow it down.
\begin{figure}[t!]
\centering
\includegraphics[width = 8 cm]{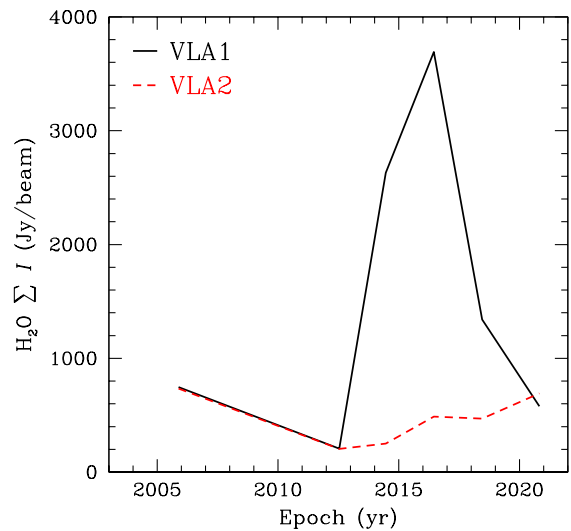}
\caption{Comparison of the total sum of the \water ~maser intensity ($I$) measured toward VLA\,1
and VLA\,2 with the VLBI in
epochs 2005.89 \citepalias[VLBA,][]{sur112}, 2012.54 \citepalias[VLBA,][]{sur142}, 2014.46 (EVN), 2016.45 (EVN),
2018.44 (EVN), and 2020.82 (EVN). The lowest intensities coincide with the minimum of activity
registered toward W75N(B) by the 22-m Pushchino telescope in May-July 2012 \citep{kra15}.}
\label{totflux}
\end{figure}
\begin{figure*}[t!]
\centering
\includegraphics[width = 8 cm]{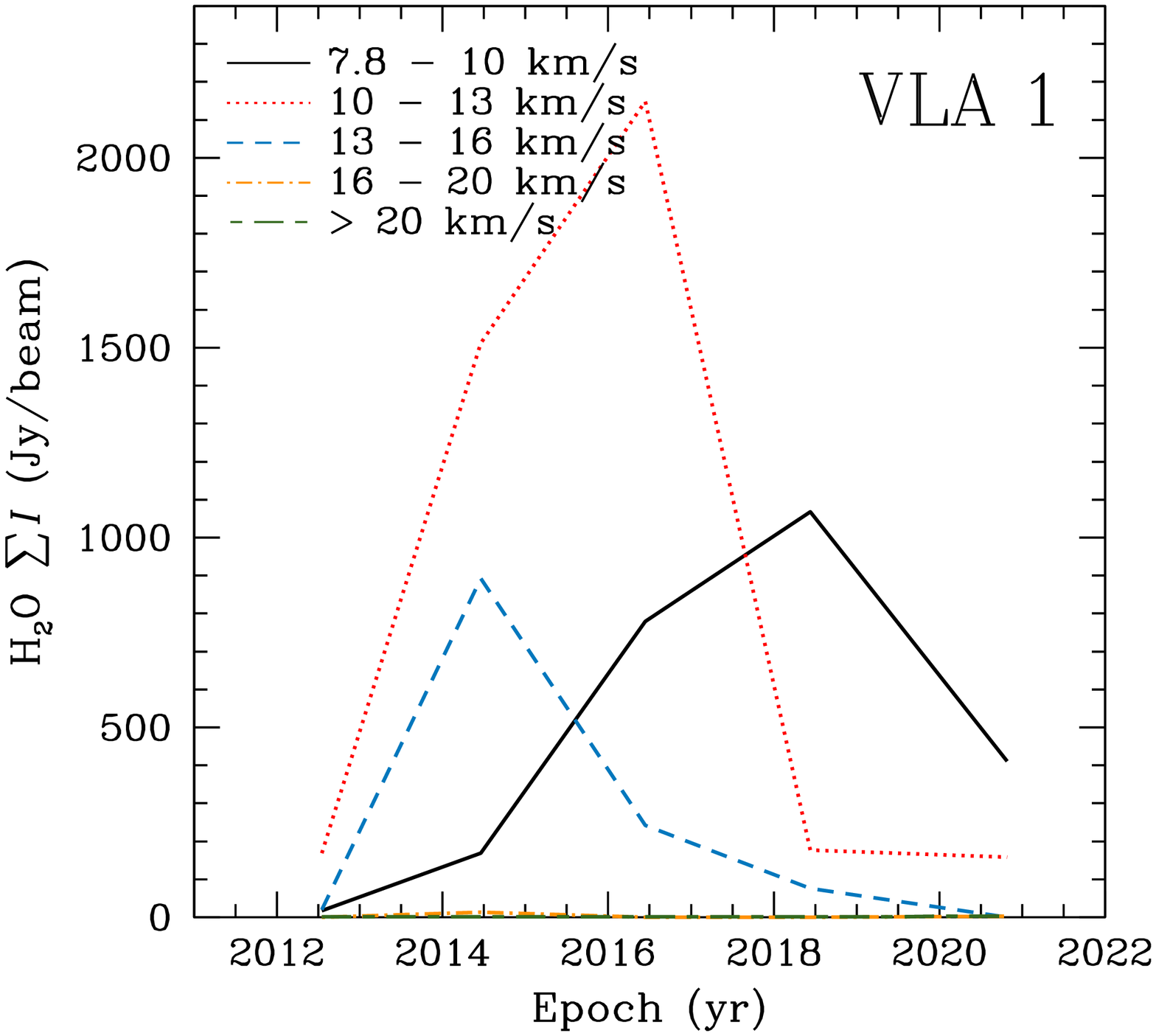}
\includegraphics[width = 8 cm]{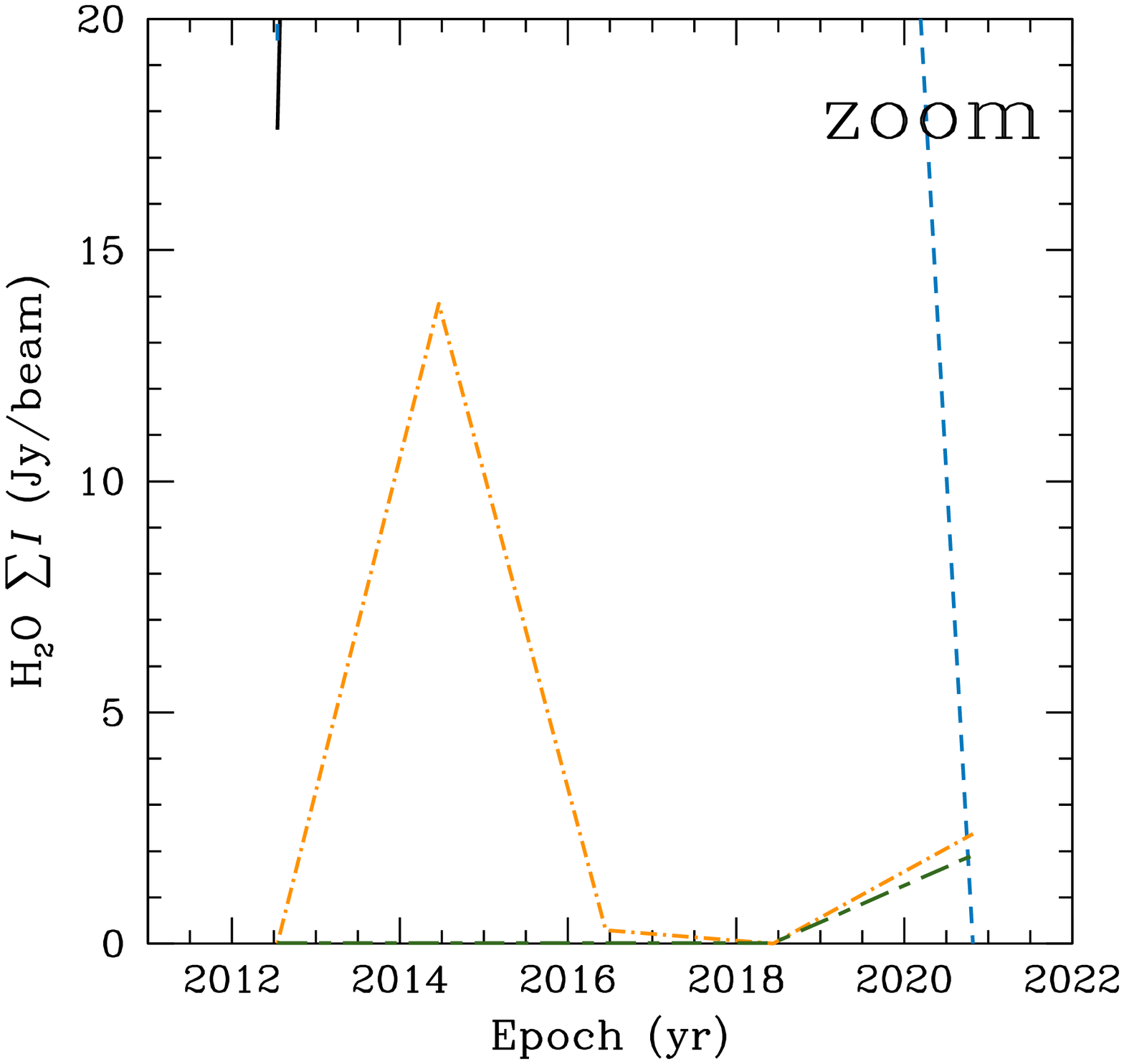}
\caption{Comparison of the total sum of the \water ~maser intensity ($I$) for five different ranges of velocities in VLA\,1
as measured in epochs 2012.54 \citepalias[VLBA,][]{sur142}, 2014.46 (EVN), 2016.45 (EVN), 2018.44 (EVN), and
2020.82 (EVN). }
\label{VLA1flux}
\end{figure*}
\begin{figure*}[th!]
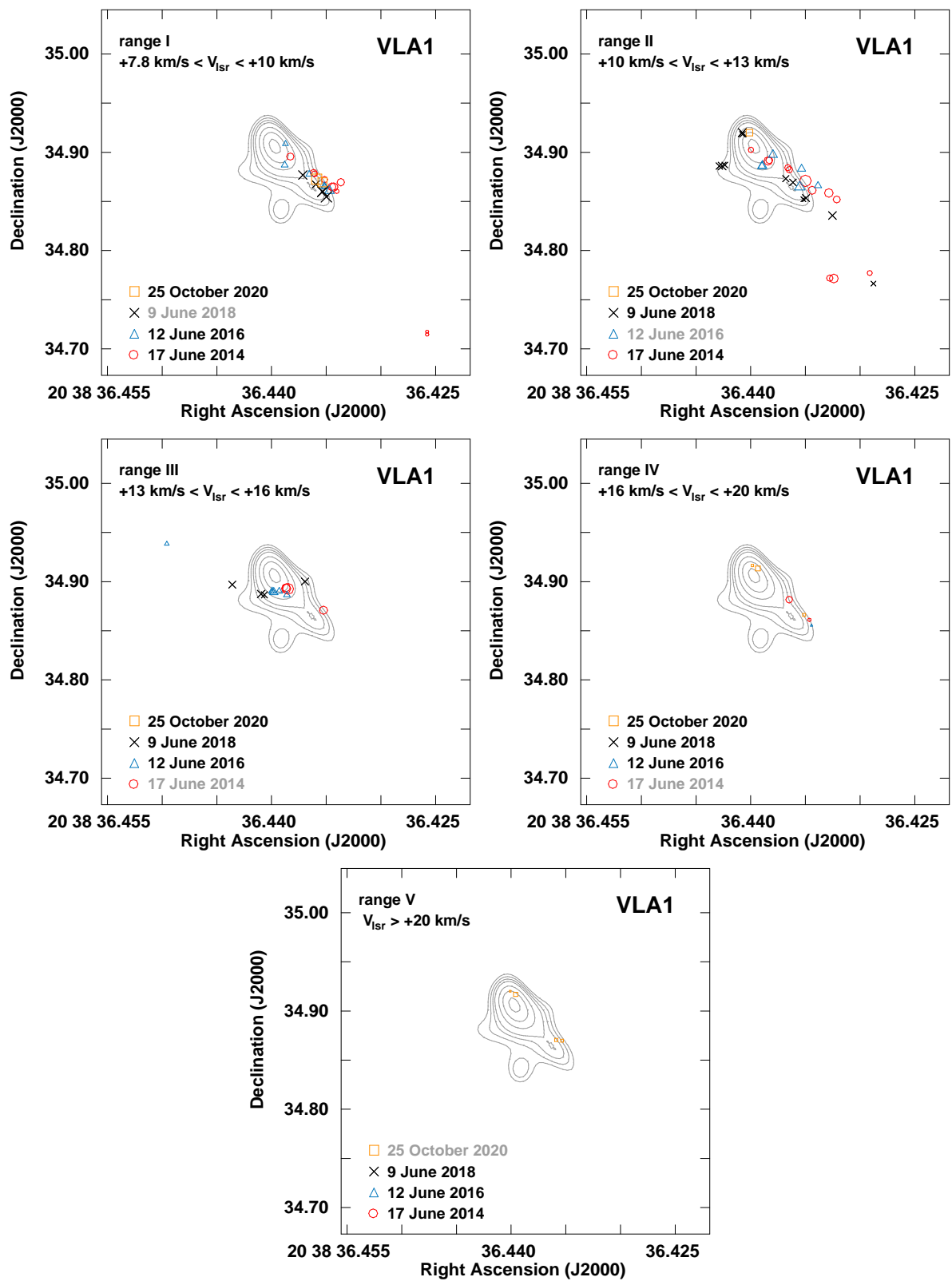

\centering
\includegraphics[width = 7.8 cm]{7.8-10_vla1.eps}
\includegraphics[width = 7.8 cm]{10-13_vla1.eps}
\includegraphics[width = 7.8 cm]{13-16_vla1.eps}
\includegraphics[width = 7.8 cm]{16-20_vla1.eps}
\includegraphics[width = 7.8 cm]{greater_20_vla1.eps}
\caption{Comparison of the \water ~maser features detected for five different ranges of velocities in
VLA\,1 in the four EVN epochs 2014.46, 2016.45, 2018.44, and 2020.82. The size of the symbols are scaled 
logarithmically according to their peak intensities. The epoch for which the maximum of the
intensities is registered for each velocities range is colored in light gray. The assumed velocity of the
region is $V_{\rm{lsr}}=+10.0$~\kms ~\citep{she03}.}
\label{VLA1ranges}
\end{figure*}
\subsection{Intensity variability}
\label{fiv}
\subsubsection{VLA\,1}
\label{fivVLA1}
\begin{figure*}[th!]
\centering
\includegraphics[width = 8 cm]{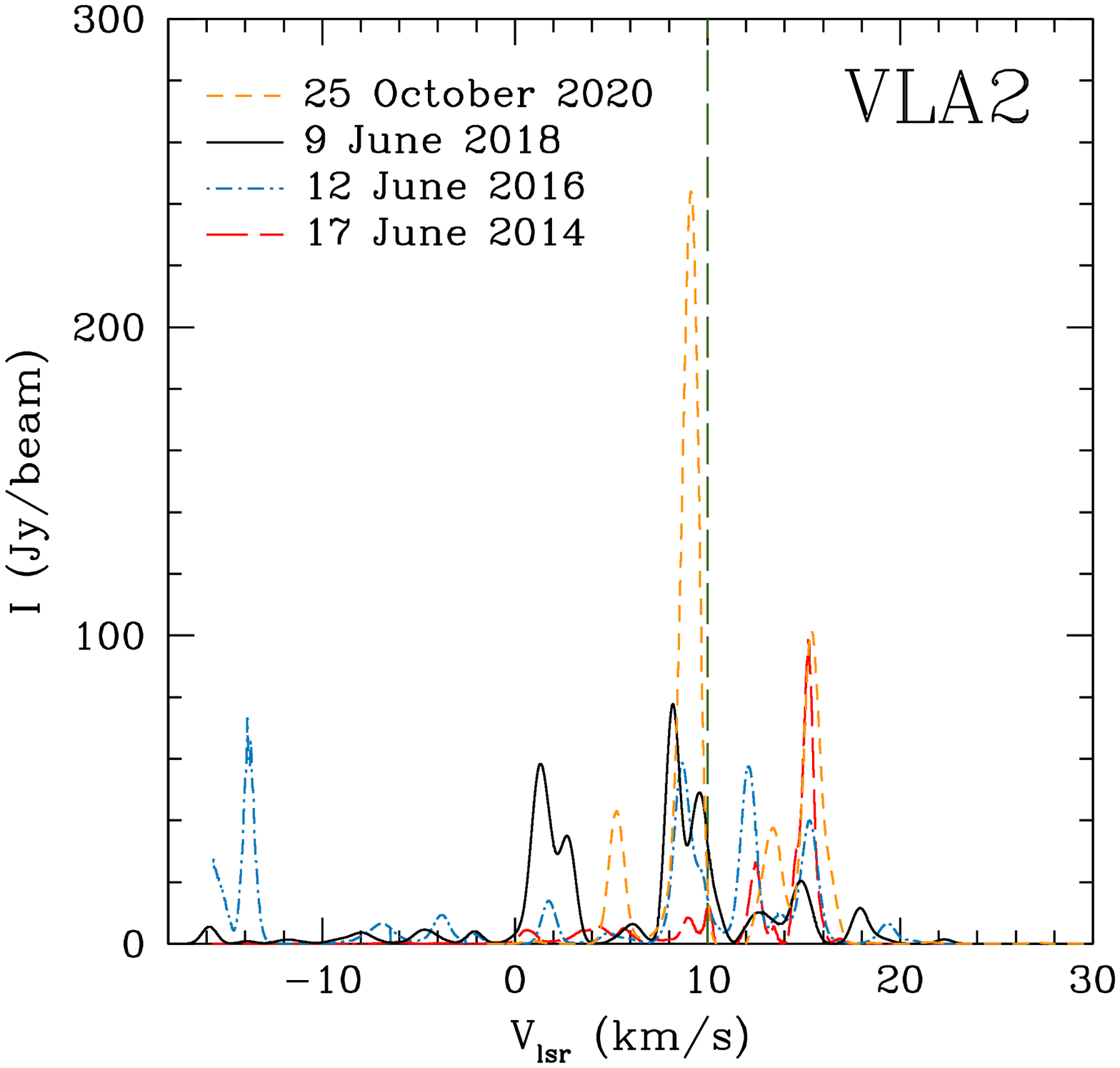}
\includegraphics[width = 8 cm]{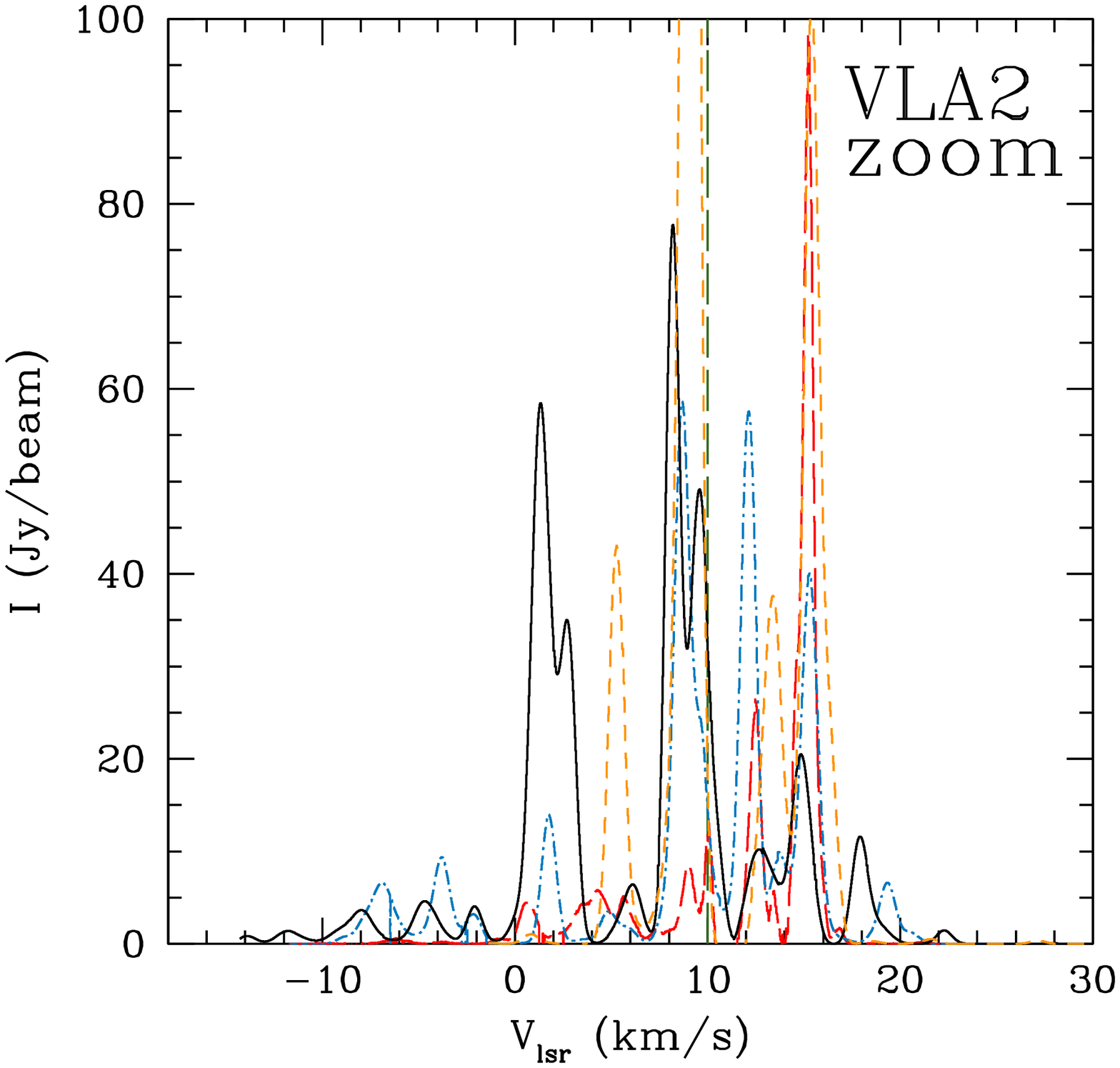}
\caption{Comparison of the total sum of the \water ~maser spectra detected toward VLA\,2
with the EVN in epochs 2014.46, 2016.45, 2018.44, and 2020.82. The vertical dashed dark-green line 
indicates the assumed systemic velocity of the region ($V_{\rm{lsr}}=+10.0$~\kms, \citealt{she03}). }
\label{totspectravla2}
\end{figure*}
\indent The total \water ~maser intensity ($\sum{I}$) of the maser features first had an increment
between the VLBA epoch 2012.54 \citepalias{sur142} and the EVN epoch 2016.45 and then it decreased 
again till the last EVN epoch 2020.82. This variation of the intensity can be seen in 
Fig.~\ref{totflux}, where we also show the intensities measured in epochs 2005.89 and 2012.54 
with the VLBA \citepalias{sur112, sur142}. The lowest intensity was measured in 2012.54 
($\sum I=206.31$~\jyb) by \citetalias{sur142} and it coincides with the minimum of activity
registered toward W75N(B) by the 22-m Pushchino telescope in May-July 2012 \citep{kra15}. The \water
~maser features cover a large line of sight velocity range from +7.8~\kms ~to +26.4~\kms ~over the 
four EVN epochs and, as mentioned before, with maser features with line of sight velocities 
$\geq+20$~\kms ~detected only in epoch 2020.82 (see Table~\ref{para} and 
Fig.~\ref{totspectravla1}). We therefore compare the
 intensities for five different ranges of velocities along the line of sight (ranges I-V) in
Fig.~\ref{VLA1flux} and we plot the corresponding maser features superimposed to the continuum 
emission at Q-band \citep{rod20} in the five panels of Fig.~\ref{VLA1ranges}. We note that each 
range of velocities has the maximum of total intensity at different epochs (the corresponding 
epoch is colored in light gray in Fig.~\ref{VLA1ranges}). In particular the maximum is reached
earlier for the ranges with the highest velocities, except for the maser features with velocities 
$\geq+20$~\kms ~(range V) that are detected only in epoch 2020.82. Indeed, we see that $\sum{I}$ of 
the maser features with +7.8~\kms$<V_{\rm{lsr}}<+10$~\kms (range I), which are the only blueshifted 
features with respect to the systemic velocity of the region ($V_{\rm{lsr}}=+10.0$~\kms; 
\citealt{she03}), and with +10~\kms$<V_{\rm{lsr}}<+13$~\kms ~(range II) show the maximum in epochs
2018.44 and 2016.45, respectively. Whereas the ranges +13~\kms$<V_{\rm{lsr}}<+16$~\kms ~(range III) 
and +16~\kms$<V_{\rm{lsr}}<+20$~\kms ~(range IV) show their maximum in epoch 2014.46.
Furthermore, we note that the maser features of ranges I, II, which are respectively within $\mp$3~\kms
~from the systemic velocity, and IV are mainly located along the southwestern tail of the thermal radio 
jet at Q-band, and in particular the intensity of ranges I and II reaches its maximum when the 
maser features are along this tail (see Fig.~\ref{VLA1ranges}). A few maser features of ranges II and 
IV are close to the position of the strong core of the continuum emission at Q-band and these all but one
are detected in the last two epochs, the one is detected in epoch 2014.46 (range II). Most of the
maser features of range III are instead aligned east-west below the core of the continuum emission at Q-band 
and its  
intensity reaches the maximum in epoch 2014.46. The most redshifted maser features (range V) are the 
weakest ones among all of those detected in the VLBA and EVN epochs, this can be due to the fact that 
they have arisen only recently. These maser features are located in the very west edge of the southern 
tail and slightly north of the core. The spatial coincidence of the blue- and redshifted maser features 
of ranges I and II may suggest that the maser features are tracing the central part of one of the lobes 
of the thermal jet, and that this has such an inclination that the maser features located on the surface 
closer to us appear slightly blueshifted, while those located on the opposite surface appear slightly 
redshifted.
\subsubsection{VLA\,2}
\label{fivVLA2}
\begin{figure}[t!]
\centering
\includegraphics[width = 8 cm]{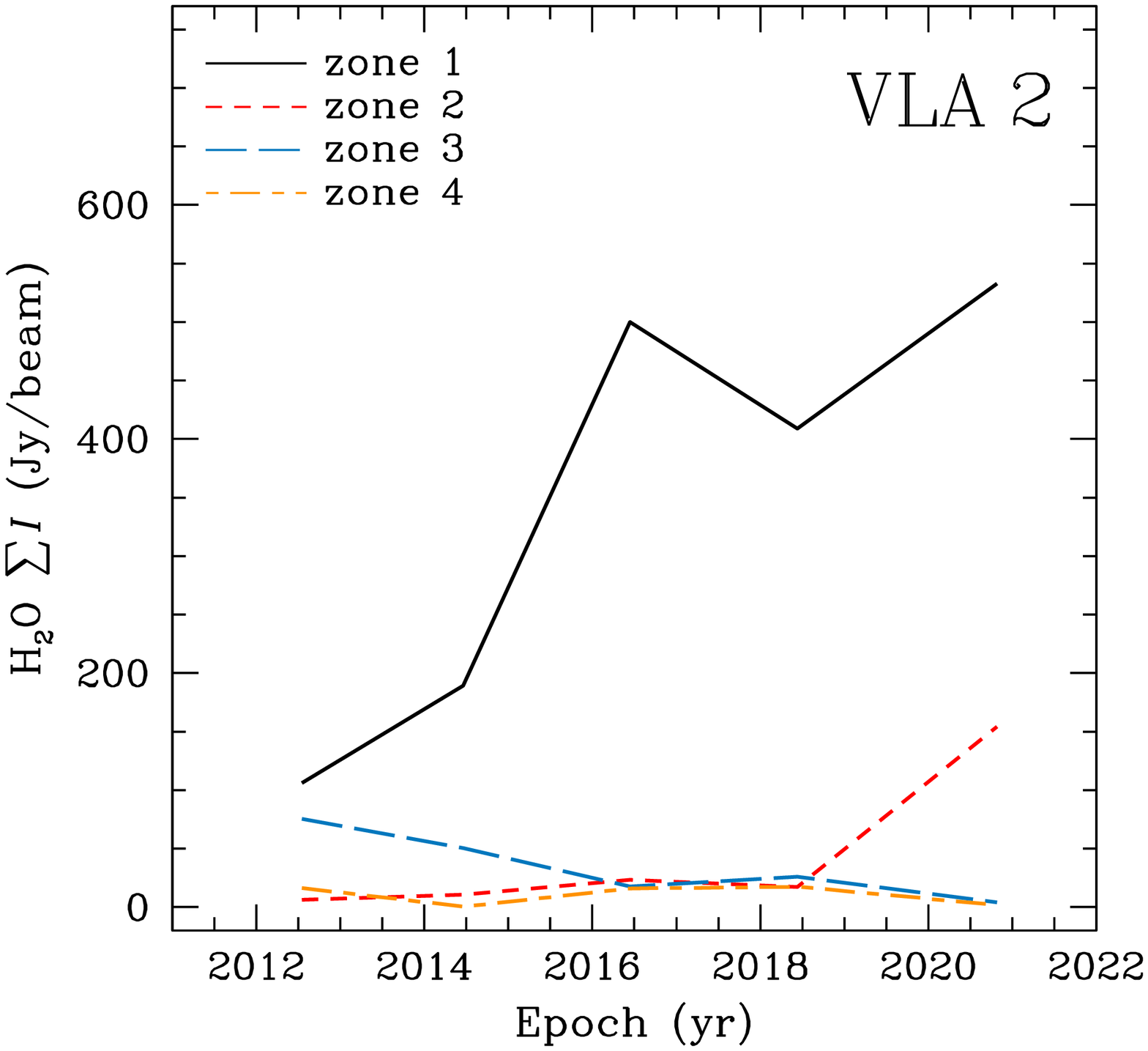}
\caption{Comparison of the total \water ~maser intensity in the four zones of VLA\,2 (see
Fig.\ref{over_vla2}) measured in epochs 2012.54 \citepalias[VLBA,][]{sur142}, 2014.46
(EVN), 2016.45 (EVN), 2018.44 (EVN), and 2020.82 (EVN). }
\label{zonesflux}
\end{figure}

The total intensity of the \water ~maser emission almost constantly increased between the VLBA 
epoch 2012.54 and the EVN epoch 2020.82 (see Fig.~\ref{totflux}). This reflects the fact that the maser 
features in the four EVN epochs, although half in number, are actually brighter than those detected in 
the two previous VLBA epochs (see Table~\ref{para} and \citetalias{sur112,sur142}).
We show the total sum of the \water ~maser spectra detected towards VLA\,2 in the four EVN epochs 
in Fig.~\ref{totspectravla2}, from where the complexity of the \water ~maser emission around VLA\,2 is
further confirmed.\\
\indent We compare the $\sum{I}$ for the four zones (see Sect.~\ref{fivVLA2}) in Fig.~\ref{zonesflux}. 
Here we see that zone~1 in the four EVN epochs shows increments of intensity between epochs
2014.46 and 2016.45, and between epochs 2018.44 and 2020.82, with a slight decrement between epochs 
2016.45 and 2018.44. The increments of total intensity are due to the high intensity of 
individual maser features rather than to their number (see Tables~\ref{VLA2.1_tab}, \ref{VLA2.2_tab},
\ref{VLA2.3_tab}, and \ref{VLA2.4_tab}), indicating that the masing conditions are more favorable 
inward than outward. Also the maser features of zone~2 show an increment of $\sum{I}$ between epochs 
2018.44 and 2020.82 similar to that of zone~1 between the same two epochs, this is 127~\jyb ~for zone~1 
and 137~\jyb ~for zone~2.\\
\indent In zone~3, we instead observe a decrement of $\sum{I}$ between the VLBA epoch~2012.54 and the
four EVN epochs. We detected the brightest maser features in the EVN epoch~2014.46 toward south 
(VLA2.1.01 and VLA2.1.03, each with $\rm{I}\approx21$~\jyb) while the rest of the maser features of 
zone~3 show a maser intensity  $\rm{I}<13$~\jyb ~in all four EVN epochs, with the brightest of them 
detected toward the center of the continuum emission. The trend of $\sum{I}$ in zone~4 between epochs 
2016.45 and 2020.82 is identical to that observed for the maser features of zone~3.
\subsection{Maser polarization}
\label{mp}
The analysis of the polarized emission from the \water ~maser features detected in the four EVN
epochs allow us to compare, in addition to the magnetic field that we will present in Section~\ref{magn}, 
some physical parameters of the maser features and of the gas where they arise.
\begin{figure*}[t!]
\centering
\includegraphics[width = 8.2 cm]{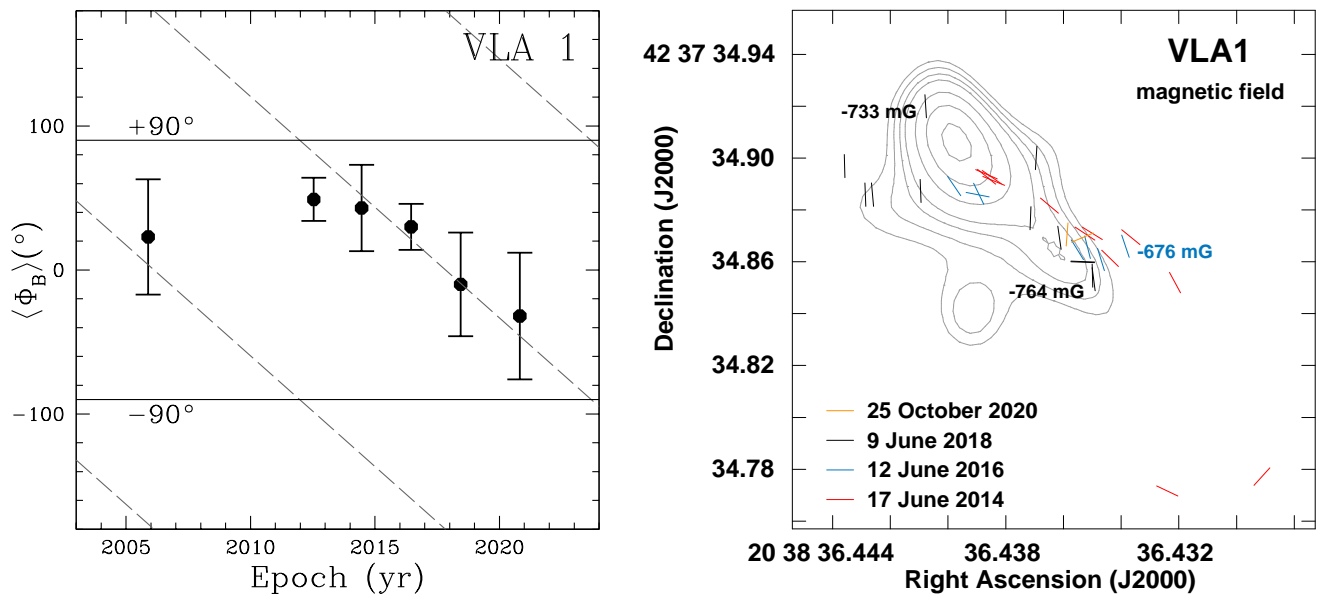}
\includegraphics[width = 9.1 cm]{VLA1_B_over_Q_UN.eps}
\caption{Multi-epoch comparison of $\langle\Phi_{\rm{B}}\rangle$ angles for VLA\,1 (\textit{left panel}). 
The epochs are 2005.89 \citepalias[VLBA,][]{sur112}, 2012.54 \citepalias[VLBA,][]{sur142}, 2014.46
(EVN), 2016.45 (EVN), 2018.44 (EVN), and 2020.82 (EVN). The two horizontal solid black lines indicates 
the $\pm 90$\d ~angles. The dashed gray lines represents the best linear fit results of the data 
considering that the position angle of a magnetic field vector has contemporary multiple values equal to
$\langle\Phi_{\rm{B}}\rangle=\langle\Phi_{\rm{B}}\rangle\pm180$\d. The slopes of these lines are 
$m_{\rm{VLA\,1}}=(-15^{\circ}\!.38\pm1^{\circ}\!.80)~\rm{yr^{-1}}$. \textit{Right panel}: 
Magnetic field vectors and strength along the line of sight (next to the corresponding maser feature) 
as estimated from all the linearly and circularly polarized maser features detected around VLA\,1 
during the four EVN epochs (2014.46, 2016.45, 2018.44, and 2020.82). The vectors are superimposed to 
the uniform-weighted continuum map at Q-band (central frequency 44~GHz) of the thermal jet driven by 
VLA\,1 \citep{rod20}. For more details see caption of Fig.~\ref{over_vla1}.
}
\label{Bover_VLA1}
\end{figure*}
\subsubsection{VLA\,1}
\label{mpVLA1}
We note that the highest values of $P_{\rm{l}}$ were measured in epochs 2014.46 (15.6\%) and 2018.44 
(10.6\%) and the corresponding maser features are located far southwest of the continuum emission at
Q-band (2014.46) and at west of the bright core of this continuum emission (2018.44). Whereas the 
lowest values of $P_{\rm{l}}$ are all measured in epoch 2020.82 (see Table~\ref{para}).\\
\indent As described in Sect.~\ref{obssect}, we can estimate some intrinsic characteristics of the
maser features by modeling the linearly polarized emission with the \code. The averaged intrinsic 
linewidth is larger in epoch 2014.46 (\mdvi$=3.0^{+0.2}_{-0.3}$~\kms) and then consistently 
decreases, to reach a minimum value of $2.6^{+0.1}_{-0.3}$~\kms ~in epoch 2020.82. This implies that 
the gas temperature of the region $T$ has also decreased from epoch 2014.46 to epoch 2020.82. Indeed, 
from the equation $T\approx100 \cdot (\langle \Delta V_{\rm{i}}\rangle/0.5)^2$ \citep{ned92},
that is valid if the broadening of the maser line due to the turbulence is negligible, we have that 
$T_{\rm{2014.46}}=3600^{+496}_{-684}$~K and $T_{\rm{2020.82}}=2704^{+212}_{-588}$~K. If the \water 
~masers are pumped by nondisocciative shocks the gas can reach temperatures around 4000~K and
above this threshold the \water ~molecule is dissociated \citep{kau96}. However, the estimated 
high temperatures indicate that the contribution of the turbulence to \dvi ~is not actually negligible. However, if we assume that this contribution is constant in time and everywhere 
in the source we can still qualitatively compare the estimated temperatures between the different epochs.
The averaged emerging brightness temperature (\mtbo) instead increases from epoch 2014.46 to epoch 
2020.82 (see Table~\ref{para}), this might indicate that the saturation level of the 
maser features increases from one epoch to the other. 
This is related to the efficiency of the pumping mechanism. Indeed the saturation regime is 
reached when the stimulated emission rate becomes larger than the decay rate to the upper level of the 
maser transition, in other words the
pumping mechanism is not able anymore to provide enough population inversion between the maser levels.
In the case of the \water ~maser the pumping mechanism is due to the shocks produced by the outflow
hitting the surrounding matter. Therefore, the shocks should have lost part of their
energy from one epoch to the other, which might also be suggested by the estimated gas temperature.\\
\indent Circular polarization was measured only in epochs 2016.45 and 2018.44 (see 
Appendix~\ref{appA}) 
when the total \water ~maser intensity reached its highest values. From the three circularly polarized
maser features, we measure a magnetic field strength along the line of sight between $-676\pm102$~mG 
(epoch 2016.45; see Table~\ref{VLA1.2_tab}) and $-764\pm22$~mG (epoch 2018.44; see 
Table~\ref{VLA1.3_tab}), which are consistent with what \citetalias{sur112} and \citetalias{sur142}
measured in the previous two VLBA epochs ($-400~\rm{mG}<B_{\rm{||}}^{\rm{2005.89}}<+810$~mG and
$-540~\rm{mG}<B_{\rm{||}}^{\rm{2012.54}}<+18$~mG). We note that the positive and negative signs of 
$B_{\rm{||}}$ indicate that the magnetic field is pointing away and toward the observer, respectively.
\subsubsection{VLA\,2}
\label{mpVLA2}
We note that all the linearly polarized maser features in the four EVN epochs, but VLA2.1.24 
($P_{\rm{l}}=4.6~\%$), show a $P_{\rm{l}}$ in the range between 0.2\% and 2.7\% (see Table~\ref{para}). 
The outputs of the \code ~provides consistent \tbo ~values in all EVN epochs, only four and one maser 
features show \tbo ~on the order of $10^6$ and $10^{10}$~K~sr respectively, and consistent \dvi ~in 
three of the four EVN epochs. In epoch 2014.46 we have \dvi$<2$~\kms ~while in the other three EVN
epochs \dvi$>2$~\kms ~with a few exceptions (VLA2.2.22, VLA2.3.21, and VLA2.3.23). As we already made 
previously for VLA\,1 (Sect.~\ref{mpVLA1}), and keeping in mind that the values do not actually
indicate the actual temperature of the gas, we can estimate the mean temperature from the \dvi 
~values and we then have $T_{\rm{2014.46}}=1024^{+272}_{-240}$~K, 
$T_{\rm{2016.45}}=3364^{+480}_{-448}$~K, $T_{\rm{2018.44}}=1764^{+1600}_{-608}$~K, and 
$T_{\rm{2020.82}}=3364^{+236}_{-864}$~K. We note that all the linearly polarized maser features in 
epoch 2014.46 are located northeast, where the expanding gas is supposed to encounter a denser 
medium. \\
\indent We were also able to measure $B_{\rm{||}}$ from the circularly polarized maser emission of a 
total of five maser features in the three EVN epochs 2016.45 ($-1498\pm225$~mG and $-2426\pm364$~mG, 
Table~\ref{VLA2.2_tab}), 2018.44 ($-355\pm69$~mG and $+439\pm66$~mG, Table~\ref{VLA2.3_tab}), and 
2020.82 ($-452\pm68$~mG, Table~\ref{VLA2.4_tab}), all of them are located north - northeast. These values are larger than those measured in epochs 
2005.89 ($-186~\rm{mG}<B_{\rm{||}}^{\rm{2005.89}}<+957$~mG; \citetalias{sur112}) and 2012.54 
($-152~\rm{mG}<B_{\rm{||}}^{\rm{2005.89}}<-103$~mG; \citetalias{sur142}) with the VLBA.
\section{Magnetic Field}
\label{magn}
We discuss the magnetic field around VLA\,1 and VLA\,2 in Sects.~\ref{vla1mag} and ~\ref{vla2mag},
respectively. Here, we are able to estimate and compare the orientation of the magnetic field from
the linear polarization vectors measured in the four EVN epochs. 
We determine for each epoch the error-weighted orientation of the magnetic field
($\langle\Phi_{\rm{B}}\rangle$), which are listed in Table~\ref{para}. We note that the position
angle of a magnetic field vector on the plane of the sky has three values contemporary: $\Phi_{\rm{B}}$
and $\Phi_{\rm{B}}\pm180$\d. Therefore, we can state that the magnetic field vectors have a sort of 
\textit{periodicity} of 180\d ~that exists only as a consequence of how $\Phi_{\rm{B}}$ is 
defined on the plane of the sky: positive if measured counterclockwise from north and negative if 
measured clockwise from north.
\subsection{VLA\,1}
\label{vla1mag}
We plot $\langle\Phi_{\rm{B}}\rangle$ as function of time in the left panel of Fig.~\ref{Bover_VLA1}, as
measured in the two previous VLBA epochs (2005.89 and 2012.54) and in the four EVN epochs. Here, we also
show a 180\d-\textit{periodicity} linear fit (dashed gray lines) that has a slope of
$m_{\rm{VLA\,1}}=(-15^{\circ}\!.38\pm1^{\circ}\!.80)~\rm{yr^{-1}}$.
However, the apparent rotation is not due to calibration uncertainties (see 
Table~\ref{Obs}) or to maser polarization variability, which can influence the relation between the
magnetic field orientation and the polarization vectors only if the \water ~maser features are highly
saturated (this is not our case), but it is simply the consequence of averaging the $\Phi_{\rm{B}}$
angles that are estimated from linear polarization vectors measured in different locations along VLA\,1,
where the magnetic field is actually differently oriented. Therefore, a punctual comparison 
between the magnetic field vectors measured in the different epochs is necessary.
However, this cannot be done with common polarized maser features detected in consecutive epochs 
because of the maser variability on short timescales. Instead, what we can do is a comparison of all the 
magnetic field vectors estimated in all the epochs. Indeed, we plot each single magnetic field vector 
estimated from all the linearly
polarized maser features detected in the four EVN epochs in the right panel of Fig.~\ref{Bover_VLA1}.
Here, we can see that the magnetic field vectors estimated in similar location, but in different
epochs, seem to represent a quasi-static magnetic field. We can therefore consider the magnetic 
field vectors estimated from the \water ~maser features in one epoch as representative of the magnetic
field in those locations rather than in that time. Consequently, we can gather the magnetic field 
vectors of all the EVN epochs and consider them as measurements done at the same time. 
This allows us to compare the magnetic field with the continuum emission at Q-band (see 
Fig.~\ref{Bover_VLA1}).
The magnetic field vectors seem to follow the morphology of the continuum emission, in particular the
internal vectors are in accordance with the radio continuum contours (right panel of
Fig.~\ref{Bover_VLA1}). This suggests that the magnetic field is along the thermal jet and it bends 
toward south at the southwest end of the thermal jet, and toward north at the northeast end.
In addition, we note that the magnetic field orientation in the northeast and in the far southwest
coincides with that of the large-scale magnetic field vectors reported by \cite[][their Fig.2]{pal21} who showed the 
1.3~mm polarized continuum emission of W75N(B) as observed by Alves et al. (\textit{in 
preparation}).\\
\begin{figure*}[h!]
\centering
\includegraphics[width = 8 cm]{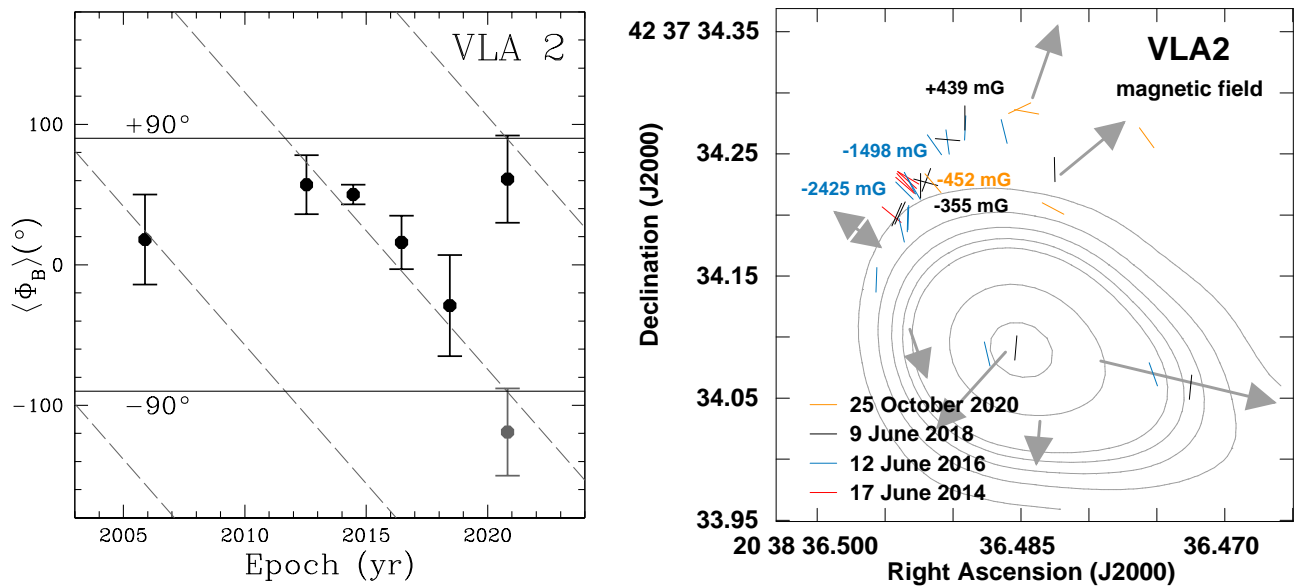}
\includegraphics[width = 9 cm]{VLA2_B_over_K.eps}
\caption{Multi-epoch comparison of $\langle\Phi_{\rm{B}}\rangle$ angles for VLA\,2 (\textit{left panel}). 
The epochs are 2005.89 \citepalias[VLBA,][]{sur112}, 2012.54 \citepalias[VLBA,][]{sur142}, 2014.46
(EVN), 2016.45 (EVN), 2018.44 (EVN), and 2020.82 (EVN). The gray point indicates the 
$\langle\Phi_{\rm{B}}\rangle$ angle measured in 2020.82 after a rotation of 180\d, considering that 
$\langle\Phi_{\rm{B}}\rangle=\langle\Phi_{\rm{B}}\rangle\pm180$\d. The two horizontal solid black lines
indicates the $\pm 90$\d ~angles. The dashed gray lines represents the best linear fit results of the
data considering that the position angle of a magnetic field vector has contemporary multiple values 
equal to $\langle\Phi_{\rm{B}}\rangle=\langle\Phi_{\rm{B}}\rangle\pm180$\d. The slopes of these lines 
are $m_{\rm{VLA\,2}}=(-19^{\circ}\!.73\pm1^{\circ}\!.89)~\rm{ yr^{-1}}$. \textit{Right panel}: 
Magnetic field vectors and strength along the line of sight (next to the corresponding maser 
feature) as estimated from the linearly and circularly polarized maser features detected around VLA\,2 
during the four EVN epochs (2014.46, 2016.45, 2018.44, and 2020.82) and superimposed to the
natural-weighted continuum maps at K-band (central frequency 22~GHz) of the thermal, collimated ionized 
wind emitted by VLA\,2 \citep{car15}. For more details see caption of Fig.~\ref{over_vla2}. The arrows 
represent the direction of the proper motions on the plane of the sky reported in 
Tables~\ref{linearfit}, \ref{polyfit}, and \ref{vzone23}, and the arrows length is proportional to 
their magnitudes. The double arrow in the northeast indicates the uncertainties of the motion 
direction, due to the presence of an inward motion (case A) or of an outward motion (case B) of 
the gas as explained in Sect.~\ref{svVLA2}.}
\label{Bover_VLA2}
\end{figure*}
\indent The three circularly polarized maser features are associated with the ends of the thermal jet
as observed at Q-band (see Fig.~\ref{Bover_VLA1}) and the estimated magnetic field points toward us
(see Sect.~\ref{mpVLA1}).  The 3D magnetic field strength 
can be estimated from $|B|=\langle |B_{||}| \rangle /\rm{cos}\langle \theta \rangle$ (if 
$\theta\neq\pm90$\d), where $\langle |B_{||}| \rangle$ and $\langle \theta \rangle$ are the 
error-weighted mean values of $|B_{||}|$ and $\theta$, respectively.
We must note that the magnetic field derived from the circularly polarized emission of
the \water ~maser features can be very high because these masers probe shocked gas, so it is not
representative of the whole region.
We then have in the two EVN 
epochs $|B|^{\rm{2016.45}}>2.0$~G, which is a lower limit obtained by considering $\theta=71$\d, and
$\langle|B|^{\rm{2018.44}}\rangle=3.6$~G. Both these values are much larger than what was measured on
average in the VLBA epochs ($\langle|B|^{\rm{2005.89}}\rangle\approx0.7$~G and 
$\langle|B|^{\rm{2012.54}}\rangle>0.1$~G; \citetalias{sur112,sur142}). The differences might be due to 
the increment of the magnetic field strength or to the hydrogen number density of the 22~GHz \water 
~maser. In the latter case, by considering the relation $|B|\propto n_{\rm{H_2}}^{0.65}$ \citep{cru10}, 
we can estimate an increment of $n_{\rm{H_2}}^{\rm{H_2O}}$ of one order of magnitude between the VLBA 
epoch 2005.89 and the EVN epoch 2018.44.
\subsection{VLA\,2}
\label{vla2mag}
Similarly to VLA\,1, we observe an apparent clockwise rotation of $\langle\Phi_{\rm{B}}\rangle$ from
the VLBA epoch~2015.89 to the EVN epoch~2020.82 (see left panel of Fig.~\ref{Bover_VLA2}). In this case
we notice that $\langle\Phi_{\rm{B}}\rangle$ measured in the EVN epoch~2020.82 is on a different
180\d-\textit{periodicity} line with respect to those of the previous EVN epochs. The slope
of the 180\d-\textit{periodicity} lines is 
$m_{\rm{VLA\,2}}=(-19^{\circ}\!.73\pm1^{\circ}\!.89)~\rm{yr^{-1}}$, which might
indicate that the rotation in VLA\,2 is faster than in VLA\,1. As explained in the case of VLA\,1, 
the apparent rotation of $\langle\Phi_{\rm{B}}\rangle$ can be affected by the location of each maser
feature for which we are able to estimate the magnetic field vectors and it cannot therefore be 
taken at face value. Indeed, despite our earlier claims \citepalias{sur142}, we now think there is no 
evidence for such a rotation.
A comparison of the magnetic field vectors estimated in the four EVN epochs is reported on the right 
panel of Fig.~\ref{Bover_VLA2}. Two main aspects can be observed here, firstly almost all the magnetic 
field vectors are located northeast and secondly the magnetic field vectors estimated in one 
epoch can be considered representative  of a quasi-static magnetic field in those locations rather 
than in that time. The magnetic field is generally, within the errors, perpendicular to the expansion 
velocities of the gas, which are represented with arrows in Fig.~\ref{Bover_VLA2}, all around VLA\,2 
and it is parallel only in the northeast where the expanding gas encounters the denser matter. The fact 
that the magnetic field vectors are along the shock front is not surprising. Indeed, \water ~masers 
arise behind fast C- or J-type shocks \citep[e.g.,][]{kau96,Hol13} that, while propagating in the 
ambient medium, alter the initial magnetic field configuration. As explained in details 
in \cite{god17}, the magnetic field component perpendicular to the shock velocity is compressed and 
might dominate the parallel component, that remains unaffected, consequently the resulting magnetic 
field probed by the \water ~maser features might be expected to be along the shock front. The sudden 
change of the orientation of the magnetic field vectors in the northeast cannot be explained with the 
typical 90\d-flip because the estimated $\theta$ angles of those \water ~maser features are much 
greater than 55\d ~(see Sect.~\ref{obssect}). Therefore, this can be justified with a variation of the 
ratio between the perpendicular and the parallel components of the magnetic field with respect to the 
expanding velocity. In other words, even though the shock is still able to pump the \water ~molecules 
in order to produce the maser emission, it is not able anymore to compress enough the perpendicular 
component of the magnetic field in order to make it dominating over the parallel component. This can 
be considered a further clue in favor of the presence of a very dense gas in the northeast where the 
magnetic field is oriented northeast-southwest.
\begin {table}[t!]
\caption []{Comparison of the hydrogen number densities of the gas in the northeast (zone 1) of VLA\,2.
} 
\begin{center}
\scriptsize
\begin{tabular}{ l c c c c}
\hline
\hline
\,\,\,\,\,(1)&(2)           & (3)    & (4)                     & (5)               \\
 Epoch       & Maser      & $|B|$  & $n_{\rm{H_2}}$          & $n_{\rm{H_2}}$\tablefootmark{a}  \\
             & feature    &   (G)  &  equation               & ($\rm{cm^{-3}}$)       \\ 
\hline
2005.89\tablefootmark{b}& VLA 2.16, VLA 2.17 & 3.4\tablefootmark{c}    & $n_{1}$                 & $1.2\cdot10^{9}$      \\
                        & VLA 2.22, VLA 2.24 &        &                         & \\
2012.54\tablefootmark{d}& VLA2.44            & 1.5    & $n_{2}=0.3\cdot n_{1}$ & $3.6\cdot10^{8}$  \\
2016.45      & VLA2.2.27    & 27.8   & $n_{3}=89\cdot n_{2}$   & $3.2\cdot10^{10}$  \\
2018.44      & VLA2.3.39    & $0.6$  & $n_{4}=0.003\cdot n_{3}$ &  $10^{8}$  \\
2020.82      & VLA2.4.27    & $0.6$  & $n_{5}=n_{4}$ &  $10^{8}$ \\
 \hline
\end{tabular} 
\end{center}
\tablefoot{
\tablefoottext{a}{The hydrogen number densities $n_{\rm{H_2}}$ is calculated by considering the empirical equation $|B|\propto n_{\rm{H_2}}^{0.65}$ by \citet{cru10}. We assume that $n_{5}$ is the lowest possible hydrogen number density for producing 
22~GHz \water ~maser emission, i.e.  $10^8~\rm{cm^{-3}}$ \citep{eli89}.}
\tablefoottext{b}{\citetalias{sur112}}
\tablefoottext{c}{Average value.}
\tablefoottext{d}{\citetalias{sur142}}
}
\label{Bn}
\end{table}
We also compare the magnetic field vectors with the morphology of the continuum emission at
K-band obtained by \cite{car15} with the VLA in Fig.~\ref{Bover_VLA2}. Here, we see that some 
vectors seems to follow the morphology of the continuum emission suggesting a possible correlation 
between the magnetic field orientation and the shaping of the continuum emission. In addition, the 
orientation of the magnetic field vectors in the northeast of VLA\,2 agree well to the values reported
by \cite{pal21} at large scales ($\sim1.3''$), using submillimeter continuum data, indicating that the magnetic field at small scales might connect with the magnetic field at large scales. \\
\indent To properly compare the magnetic field strength measured in different epochs, we need to 
estimate $|B|$ (see Sect.~\ref{vla1mag}). We have $|B_{\rm{VLA2.2.17}}|=9.5$~G and 
$|B_{\rm{VLA2.2.27}}|=27.8$~G in epoch 2016.45, $|B_{\rm{VLA2.3.14}}|>0.7$~G and 
$|B_{\rm{VLA2.3.39}}|>0.6$~G in epoch 2018.44 (we assume a lower value of $\theta=53$\d, see
Table~\ref{para}), and  $|B_{\rm{VLA2.4.27}}|=0.6$~G in epoch 2020.82. As we already mentioned in
Sect.\ref{vla1mag}, a variation of the magnetic field strength might largely be due to a 
variation of 
$n_{\rm{H_2}}$ of the gas where the \water ~maser features arise. This variation follows the empirical 
equation of \cite{cru10} that is reported in Sect.~\ref{vla1mag}. Considering the measurements 
of $B$ made in the northeast of VLA\,2 (zone~1), we can estimate $n_{\rm{H_2}}$ of the gas in the five 
epochs by assuming that the lowest value must be at least $10^8~\rm{cm^{-3}}$, which is the 
lowest limit for having \water ~maser emission at 22~GHz \citep{eli89} and this can be 
assumed for the epoch with the weakest magnetic field strength that we measured. 
These values are reported in Table~\ref{Bn}, where we name the $n_{\rm{H_2}}$ of the gas as 
$n_{1}-n_{5}$ from epoch 2005.89 to epoch 2020.82, respectively. We note that $n_{\rm{H_2}}$ decreases 
from the VLBA epoch 2005.89 to the VLBA epoch
2012.54, when the morphology of the maser feature distribution changed from quasi-circular to elliptical
(\citealt{kim13}; \citetalias{sur142}). This might be due to the dilution of the gas. Furthermore, in 
this transition we also observe a change of sign of the magnetic field from positive to negative and 
it therefore indicates that the magnetic field changes its pointing direction from away to toward the 
observer. From the VLBA epoch 2012.54 to the EVN epoch 2016.45, when the expansion of the gas in the 
northeast continues, $n_{\rm{H_2}}$ increases to the highest possible value for having \water ~maser 
emission and this might be due to the encounter of the expanding gas with the denser medium. This might 
be a further indication that the absence of \water ~maser emission farther in the northeast is due to 
the presence of a too high density medium (case~A in Sect.~\ref{svVLA2}). Finally, $n_{\rm{H_2}}$ 
shows the lowest possible value in the EVN epochs 2018.44 and 2020.82 when the \water ~masers are 
probing the gas in the same region as they did in the VLBA epoch 2012.54. This suggests that in this 
zone not only the maser features in the EVN epoch 2020.82 are pumped by a different shock, as we 
supposed in Sect.~\ref{svVLA2}, but that likely all the maser features in all the EVN epochs are pumped 
by different shocks in this zone.
\section{Discussion}
\label{discussion}
In this section, we put together all the pieces of information that we have presented in previous 
sections to provide a full picture of both VLA\,1 and VLA\,2.
\subsection{VLA\,1}
\label{vla1}
\indent We can state that the 22~GHz \water ~maser features around VLA\,1 are probing the passage 
throughout the gas where the masers arise of a nondissociative shock. This shock is produced by the 
expansion of a thermal jet from VLA\,1. 
The magnetic field is along the axis of the thermal jet, pointing toward us, and it bends toward south
and north at the southwest and northeast ends of the thermal jet, respectively, following the large-scale
magnetic field morphology \citep{pal21}. From the different maser and gas characteristics observed 
between the VLBA and EVN epochs, we can also conclude that the compression shock probed by the maser 
features in the recent epochs is a different one than that probed in the past. It is also possible 
that the maser features in the six epochs were pumped by a series of shocks rather than a single one.
The presence of more shocks might be justified by episodic variations in the velocity of the jet as
observed for instance in the intermediate-mass protostar in the Serpens star-forming region 
\citep{rod22}.

\subsection{VLA\,2}
\label{vla2}

\indent In VLA\,2 we have an asymmetric expansion of the gas on the plane of the sky that depends 
on the particular direction
of the motion. Indeed, the expansion of the gas is prevented by the presence of a dense gas in the 
northeast, in the northwest the gas is moving outward with a velocity around 26-28~\kms, in the center 
the motion is toward southeast with a velocity around 38~\kms, in the east the motion of the gas is 
toward south with a velocity of 12~\kms, in the south the gas is expanding southward with a velocity
of 8~\kms, and in the southwest the gas is the fastest one with a westward velocity of 78~\kms. If we 
average the magnitude of all the velocities that we estimated we get $|V|\approx33$~\kms ~that is 
similar to the symmetric expanding velocity of 30~\kms ~measured by \citetalias{sur142}, who 
considered the fitted ellipses centered to a common center. Furthermore, the maser features in the
northeast show both blue- and redshifted $V_{\rm{lsr}}$ without an ordered spatial distribution
suggesting that the gas is moving along the walls of the dense core that the gas encounters. In the
northwest and southeast the masers show blueshifted velocities, that is the gas is moving toward us,
while in the center and southwest the masers are all redshifted, that is the gas is moving away from us.
The comparison with the K-band continuum
emission 
\citep[][see Figs.~\ref{over_vla2} and \ref{Bover_VLA2}]{car15} suggests that the gas in the southwest 
can expand without encountering any obstacle, while in the north there might be a very dense medium, 
maybe this is part of the envelope supposed by \cite{car15}, 
that is able to slow down the expanding  gas in the northwest and to stop completely the expansion in 
the northeast. The presence of this denser medium is also responsible to alter the morphology and 
strength of the magnetic field. The magnetic field is generally perpendicular to the proper motions all
around VLA\,2, but in the northeast it becomes parallel and stronger after encountering the denser 
medium. This might be the consequence of the inefficiency of the compression of the gas due to the 
passage of the shock that now it is not able to make the perpendicular component of the magnetic 
field, with respect to the expansion direction, dominating over the parallel component. Furthermore, 
the magnetic field morphology in the northeast region of VLA\,2 agrees with the large-scale magnetic field
reported by \cite{pal21}. Regarding the 3D magnetic field strength, we have that it is higher 
than previously measured only in epoch 2016.45 where the maser features are detected in the northeast high 
density region of VLA\,2. \\
\section{Summary}
\label{summary}
We observed the polarized emission of 22~GHz \water ~maser around the radio sources VLA\,1 and VLA\,2, 
which are located in the HMSFR W75N(B), with the EVN on four epochs, separated by two years from 2014 
to 2020. We detected linearly and circularly polarized emission in all epoch but one (epoch 2014.46) 
around both the radio sources. The comparison of the maser distributions and the magnetic 
fields among the four EVN epochs shows that:
\begin{itemize}
    \item The 22~GHz \water ~maser emission does not probe the magnetic field as it is in a single 
    observing epoch, but it probes a portion of a quasi-static magnetic field in the region where 
    the maser arises. That is, the magnetic field vectors estimated in one epoch must be considered 
    as a measurement of the magnetic field in those locations rather than in that time. Therefore, 
    mapping the magnetic field in different epochs with the 22~GHz \water ~maser allow us to 
    reconstruct the magnetic field in a larger area.\\
    \item In VLA\,1 the 22~GHz \water ~maser features are probing the passage of a nondissociative
    shock produced by the expansion of the thermal radio jet of VLA\,1. The magnetic field is along 
    the axis of the 
    thermal radio jet and bends toward south and north at the southwest and northeast ends of the
    jet. The magnetic field strength along the line of sight measured from the Zeeman splitting of 
    the maser line is in the range $-764~\rm{mG}<B_{\rm{||}}^{\rm{EVN}}<-676$~mG and is consistent
    with previous VLBA measurements.\\
    %However, the 3D magnetic field strength is much larger now 
    %($B>2.0$~G) than in the past, this might be due to an increment of the magnetic field strength 
    %or to an increment of one order of magnitude of the hydrogen number density of the gas where the maser features arise.\\
    %
    \item The 22~GHz \water ~maser features around VLA\,2 are tracing an asymmetric expansion of the gas
    that is actually halted in the northeast where the gas likely encounters a very dense medium. 
    The inferred magnetic field vectors are almost all
    perpendicular to the proper motion, but in the northeast, where the expanding gas encounters the 
    supposed denser medium, the vectors become parallel to the expansion direction.  
    The magnetic field strengths along the line of sight are 
    all measured in the north-northeast and their values are in the range 
    $-355~\rm{mG}<B_{\rm{||}}^{\rm{EVN}}<-2426$~mG, that is larger than previously measured with the 
    VLBA. 
\end{itemize}
In conclusion, W75N(B) is one of the best HMSF laboratory that, thanks to the presence of 
multiple YSOs at different evolutionary stages, can play a crucial role in sheding light on the formation
process of high-mass stars. In addition, the importance of VLBI monitoring 
observations of polarized maser emission in reconstructing the morphology of magnetic field close to 
massive YSOs was demonstrated. \\

\noindent \small{\textit{Acknowledgments.} 
We wish to thank the anonymous referee for the useful suggestions that have improved the paper. 
The European VLBI Network is a joint facility of independent European, African, Asian, and North 
American radio astronomy institutes. Scientific results from data presented in this publication are
derived from the following EVN project code(s): ES074. G.S. thanks Jorge Cant\'{o} for the useful 
discussion and suggestions. J.M.T. acknowledges partial support from the
PID2020-117710GB-100 grant funded by MCIN/AEI/10.13039/501100011033, and by the program Unidad de
Excelencia Mar\'{i}a de Maeztu CEX2020-001058-M. J.F.G. acknowledges support from grants 
PID2020-114461GB-I00 and CEX2021-001131-S, funded by MCIN/ AEI /10.13039/50110001103. 
S.C. acknowledges support from UNAM, 
and CONACyT, M\'{e}xico. This work was partially supported by FAPESP (Funda\c{c}\~{a}o de Amparo \`{a} 
Pesquisa do Estado de S\~{a}o Paulo) under grant 2021/01183-8.}

% bibliography data in biblio.bib
\bibliographystyle{aa}
\bibliography{biblio}
%\end{thebibliograghy}
%

\begin{appendix}
\normalsize
\section{Detailed Results.}
\label{appA}
We report here the detailed results obtained for each EVN epoch, from epoch 2014.46 (Section \ref{res14})
to epoch 2020.82 (Section \ref{res20}). In Fig.~\ref{posplot} we show the 22~GHz \water ~maser distributions 
around VLA\,1 and VLA\,2 for the four epochs. In Tables~\ref{VLA1.1_tab}--\ref{VLA2.4_tab} we list 
all the \water ~maser features detected towards VLA\,1 and VLA\,2 in the four EVN epochs. The Tables are organized 
as follows. The name of the feature is reported in Col.~1. The position offset with respect to the reference maser 
feature are reported in Cols.~2 and 3, and in Col.~4, when it is the case, we report the zone to which the maser feature 
belongs to. The peak intensity, the LSR velocity, and the FWHM of the total
intensity spectra of the maser features, that are obtained using a Gaussian fit, are reported in Cols.~5, 6, and 7, 
respectively. The linear polarization fraction and the linear polarization angles are instead 
reported in Cols.~8 and 9, respectively. The best-fitting results obtained by using the \code ~are reported in Cols.~10
(the intrinsic linewidth), 11 (the emerging brightness temperature), and 14 (the angle between the magnetic field 
and the maser propagation direction). The later is used to solve the 90\d ~ambiguity of the magnetic
field orientation with respect to the linear polarization vector (see Sect.~\ref{obssect}). The circular polarization
fraction and the estimated magnetic field strength along the line of sight are finally reported in Cols.~12 and 13, 
respectively.
\subsection{Epoch 2014.46}
\label{res14}
We detected 28 and 43 \water ~maser features around VLA\,1 and VLA\,2, respectively (see
Tables~\ref{VLA1.1_tab} and \ref{VLA2.1_tab}). The maser
features around VLA\,1 have $V_{\rm{lsr}}$ ranging from 
+7~\kms ~and +20~\kms, with peak intensities in the range $0.4$~\jyb$<I\lesssim1200$~\jyb. The 
distribution is still linear from southwest to northeast as previously observed 
\citepalias{sur112,sur142}, although we did not detect any maser of groups A and C as defined by
\citetalias{sur112} and indicated in Fig.~\ref{posplot}. The overall maser distribution around
VLA\,2, with \water ~maser features with $-12$~\kms$<V_{\rm{lsr}}<+22$~\kms ~and
$0.2$~\jyb$<I<32$~\jyb, is identical to the distribution detected by \cite{kim18} with the 
KVN \& VERA Array (KaVA) and it is similar to the elliptical distribution observed in 2012.54 by
\citetalias{sur142}, even though no maser feature was detected in the west-southwest. 
Actually, we note that most of the maser features (72\%, 31/43) are detected along the arc
structure in the north, which is less rounded than that observed in 2012.54 \citepalias{sur142}. \\
\indent We detected linearly polarized emission toward twelve and five maser features in VLA\,1 
($P_{\rm{l}}=0.4\%-15.6\%$) and VLA\,2 ($P_{\rm{l}}=0.9\%-4.6\%$), respectively. The maser feature
VLA1.1.03 shows the highest mean linear polarization fraction measured among the four epochs, i.e.
$P_{\rm{l}}=15.6\%$, but it is much lower than the highest $P_{\rm{l}}$ 
ever measured toward W75N(B) ($P_{\rm{l}}=25.7\%$, \citetalias{sur112}). The error-weighted linear
polarization angles are $\langle\chi\rangle_{\rm{VLA\,1}}=-47$\d$\pm30$\d ~and
$\langle\chi\rangle_{\rm{VLA\,2}}=-40$\d$\pm7$\d. The \code ~was able to properly fit eight and
four \water ~maser features in VLA\,1 and VLA\,2, respectively, and for all of them it provided
$\theta$ angles much greater than 55\d ~indicating that the magnetic field is perpendicular to the
linear polarization vectors. No circular polarization was detected at 3$\sigma_{\rm{s.-n.}}$, which
implies an upper limit for the brightest maser feature VLA1.1.15 of $P_{\rm{V}}<0.01\%$ (see 
Table~\ref{Obs}).
\subsection{Epoch 2016.45}
\label{res16}
We list in Tables~\ref{VLA1.2_tab} and \ref{VLA2.2_tab} the 20 ($0.2$~\jyb$<I<1620$~\jyb) and 37 
($0.2$~\jyb$<I<91$~\jyb) \water ~maser features detected toward VLA\,1 and VLA\,2, respectively.
All the maser features detected around VLA\,1, but one, have LSR velocities between +7.8~\kms ~and
+16.4~\kms. The maser feature VLA1.2.16 ($I=0.24$~\jyb) is the most blueshifted \water ~maser
feature ever detected toward VLA\,1, with $V_{\rm{lsr}}=+2.68$~\kms. The maser
distribution around VLA\,1 is still linear and only VLA1.2.20 is located about 100~mas northeast
of an area of $70~\rm{mas}\times60$~mas where all the other maser features arose. The elliptical 
distribution of the \water ~maser features around VLA\,2 ($-16$~\kms$<V_{\rm{lsr}}<+21$~\kms) is
also confirmed. In this epoch, we detected six maser
features (VLA2.2.01--VLA2.2.06) in the southwest, these are all redshifted ($V_{\rm{lsr}}>15$~\kms,
see Fig.~\ref{posplot}). One of the maser feature (VLA2.2.37) was actually partially detected
because the 2-MHz bandwidth did not cover the entire line emission, and we are therefore able to
provide only a lower limit of the peak intensity ($I>30$~\jyb). This detection led us to
double the observed bandwidth in the next two EVN epochs to recover the maser emission at
$V_{\rm{lsr}}<-15.7$~\kms ~(see Sect.~\ref{obssect}).\\
\indent Seven and 13 \water ~maser features associated with VLA\,1 ($P_{\rm{l}}=0.7\%-2.2\%$) and
VLA\,2 ($P_{\rm{l}}=0.6\%-2.7\%$) showed linearly polarized emission, respectively. The
error-weighted linear polarization angles are $\langle\chi\rangle_{\rm{VLA\,1}}=-60$\d$\pm16$\d 
~and $\langle\chi\rangle_{\rm{VLA\,2}}=-74$\d$\pm19$\d. We were able to properly fit with the \code
~about 40\% (3 out of 7) of the linearly polarized maser features in VLA\,1 and $\sim$70\% (9 out of
13) of those in VLA\,2. We got $\theta>55$\d ~for all of them, and the magnetic field is 
therefore perpendicular to the linear polarization vectors. In this epoch we were also able to measure 
circular polarization toward three maser features, one in VLA\,1 (VLA1.2.01, $P_{\rm{V}}=3.5\%$)
and two in VLA\,2 (VLA2.2.17 and VLA2.2.27, $P_{\rm{V}}=4.9\%$ and 7.8\%, respectively). Their
spectra are shown in Fig.~\ref{circpol}, where the best-fit model obtained from the \code ~are
overplotted as thick red lines. We note that $P_{\rm{V}}$ of VLA2.2.27 is the highest ever measured
toward both sources VLA\,1 and VLA\,2, while the other two maser features have $P_{\rm{V}}$ 
consistent with previous detections \citepalias{sur112,sur142}.
\subsection{Epoch 2018.44}
\label{res18}
Toward VLA\,1 we detected 20 \water ~maser features in the velocity range
$+7.7$~\kms$\leq V_{\rm{lsr}}\leq+15.0$~\kms ~and with peak intensities between 2~\jyb ~and 
450~\jyb ~(see Table~\ref{VLA1.3_tab}). These maser features are linearly distributed from southwest
to northeast ($\sim270$~au) and are associated only with group~B of \citetalias{sur112}. We
detected 44 \water ~maser features ($0.04$~\jyb$\leq I\leq128$~\jyb) elliptically distributed
around VLA\,2 and with a wide range of velocities ($-16.0$~\kms$< V_{\rm{lsr}}<+28.3$~\kms;
see Table~\ref{VLA2.3_tab}).\\
\indent The \code ~was able to properly fit all the maser features that showed linearly polarized 
emission, these are 12 in VLA\,1 ($P_{\rm{l}}=0.7\%-10.6\%$) and 10 in VLA\,2
($P_{\rm{l}}=1.0\%-2.3\%$). The error-weighted linear polarization angles are 
$\langle\chi\rangle_{\rm{VLA\,1}}=-87$\d$\pm5$\d ~and
$\langle\chi\rangle_{\rm{VLA\,2}}=+76$\d$\pm12$\d. Differently from the previous two EVN epochs,
the magnetic field is not always perpendicular to the linear polarization vectors. Indeed, for two
maser features in VLA\,1 (VLA1.3.06 and VLA1.3.07) and two in VLA\,2 (VLA2.3.23 and VLA2.3.28)
the \code ~provided such $\theta$ angle values that the magnetic field is more likely parallel than
perpendicular. To determine the relative orientation of the magnetic field, we consider the
associated errors $\varepsilon^{\rm{\pm}}$ of $\theta$, where the plus and minus signs indicate 
the positive and negative errors respectively, and in particular the quantities
$\theta^{\rm{\pm}}=\theta+\varepsilon^{\rm{\pm}}$. If
$|\theta^{\rm{+}}-55$\d$|<|\theta^{\rm{-}}-55$\d$|$ the magnetic field is more likely parallel
to the linear polarization vectors, while if $|\theta^{\rm{+}}-55$\d$|>|\theta^{\rm{-}}-55$\d$|$ 
the magnetic field is more likely perpendicular \citep[e.g.,][]{sur15}. We also 
detected circularly polarized emission toward two maser features in VLA\,1 (VLA.1.3.04 and 
VLA.1.3.16) and two maser features in VLA\,2 (VLA2.3.14 and VLA2.3.39). Whereas we were able to
fit the $V$ spectra of VLA.1.3.04 and VLA.1.3.16 with the model obtained by using the outputs of
the \code, we had to find the best models for VLA2.3.14 and VLA2.3.39 by considering the values of 
\tbo ~and \dvi ~that best fit their total intensity spectra. This was necessary because both 
VLA2.3.14 and VLA2.3.39 did not show any linearly polarized emission and therefore we could not use
the \code ~to determine them. In particular, we found that the best estimates of \tbo ~and \dvi
~are equal to $10^6$~K~sr and 1.4~\kms ~for VLA2.3.14, respectively, and \tbo$~=3.2\times10^9$~K~sr
and \dvi$~=2.0$~\kms ~for VLA2.3.39. The results of the fit can be seen in Fig.~\ref{circpol}.
\subsection{Epoch 2020.82}
\label{res20}
\begin{figure*}[]
\centering
\includegraphics[width = 6.2 cm]{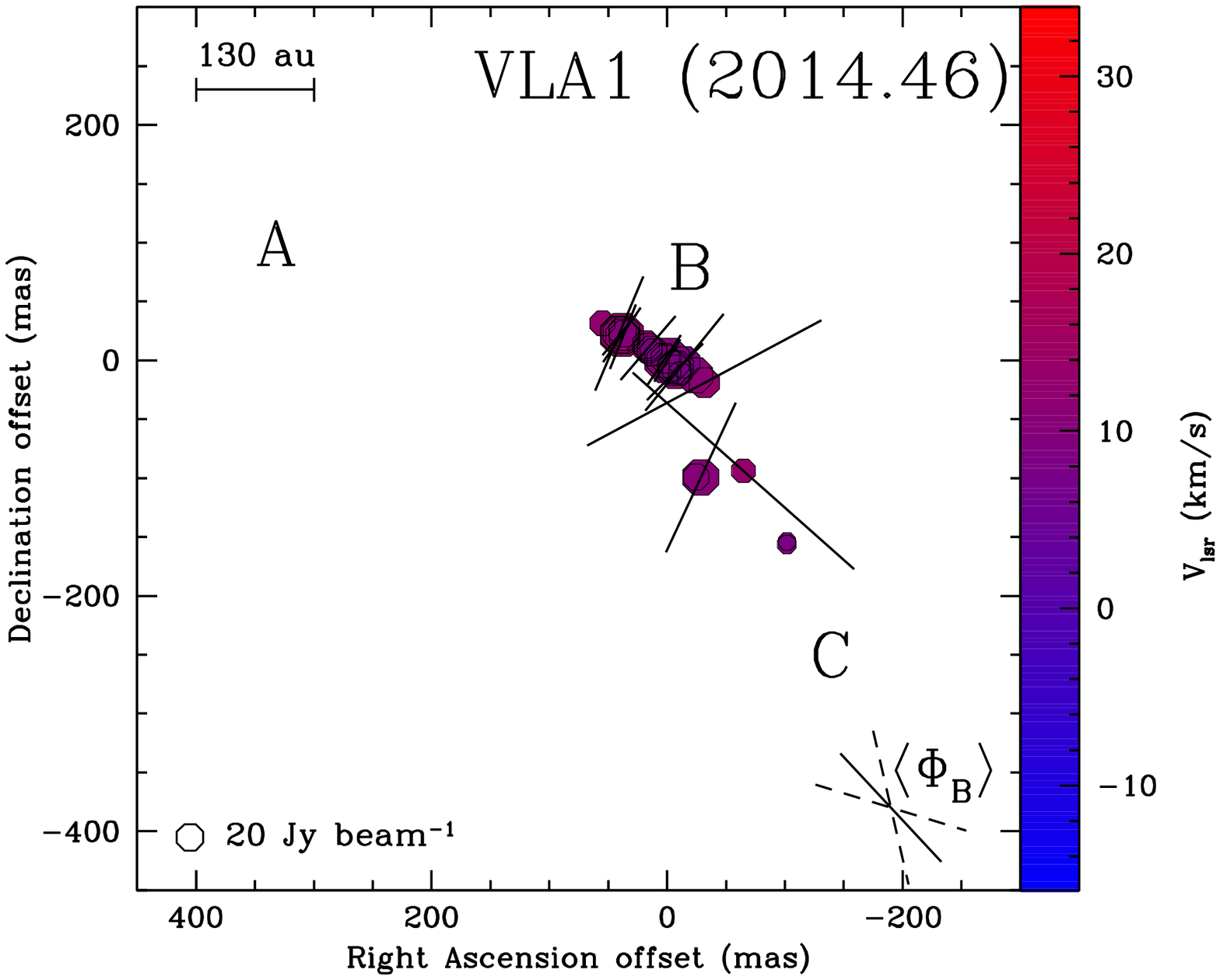}
\includegraphics[width = 6.2 cm]{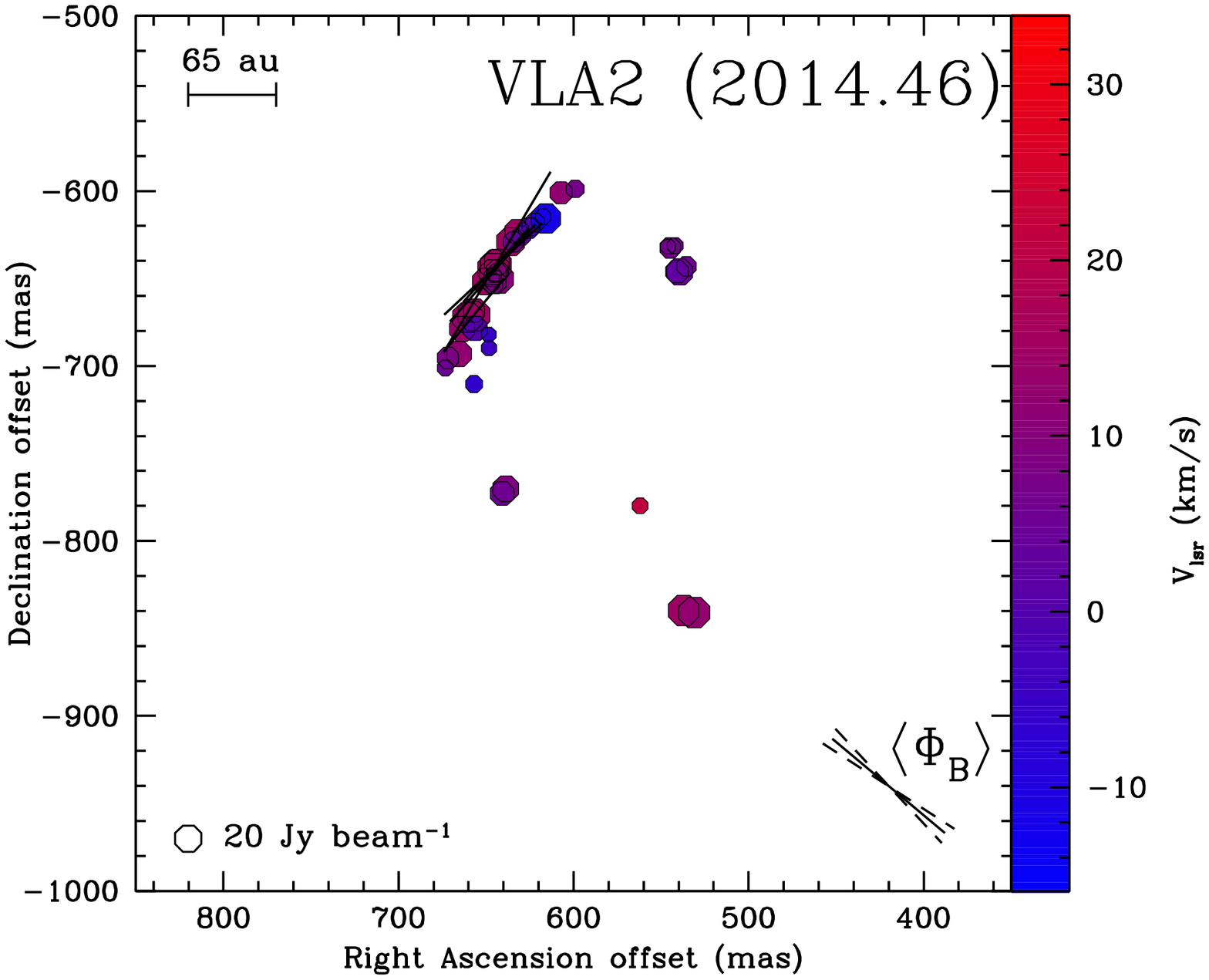}
\includegraphics[width = 6.2 cm]{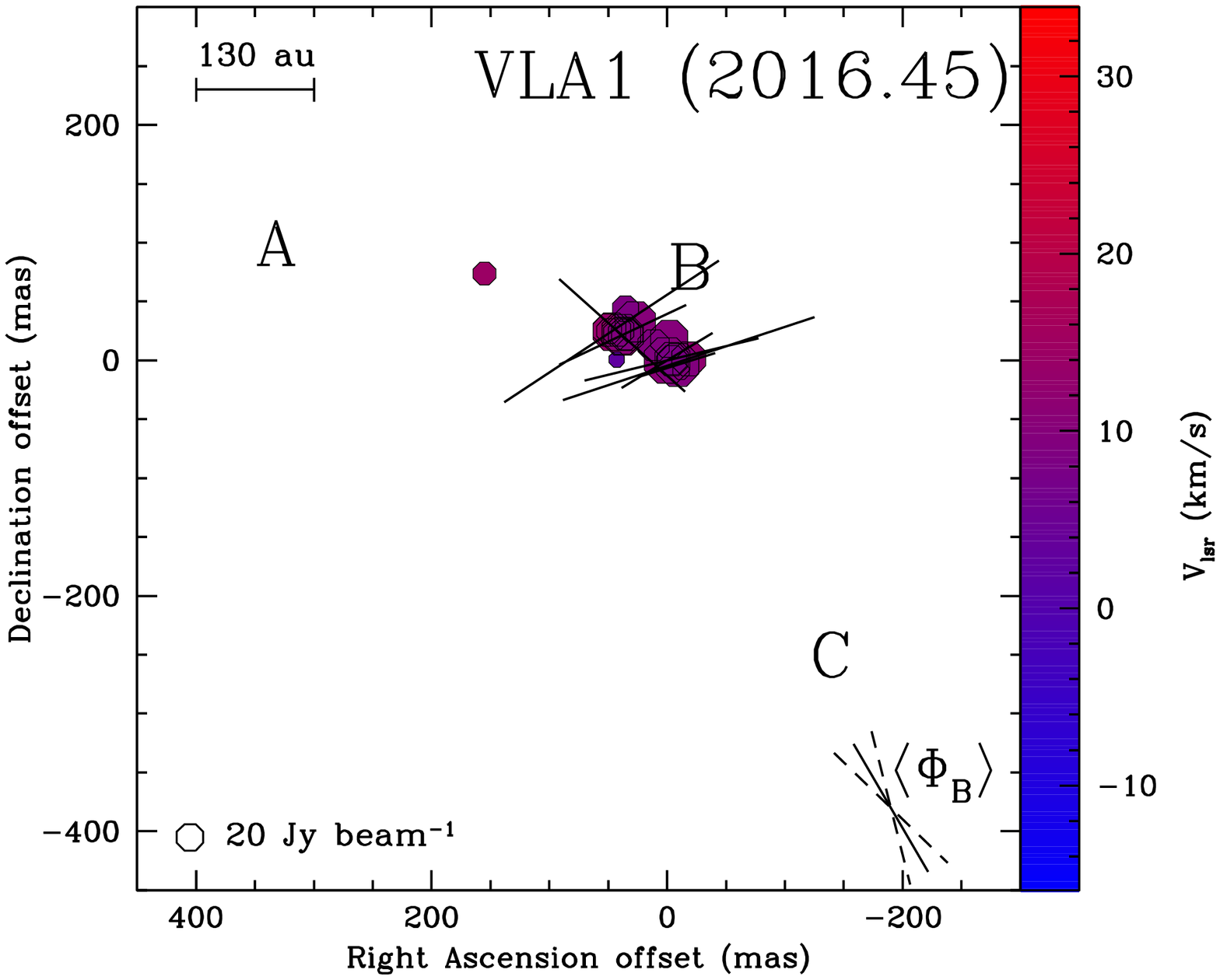}
\includegraphics[width = 6.2 cm]{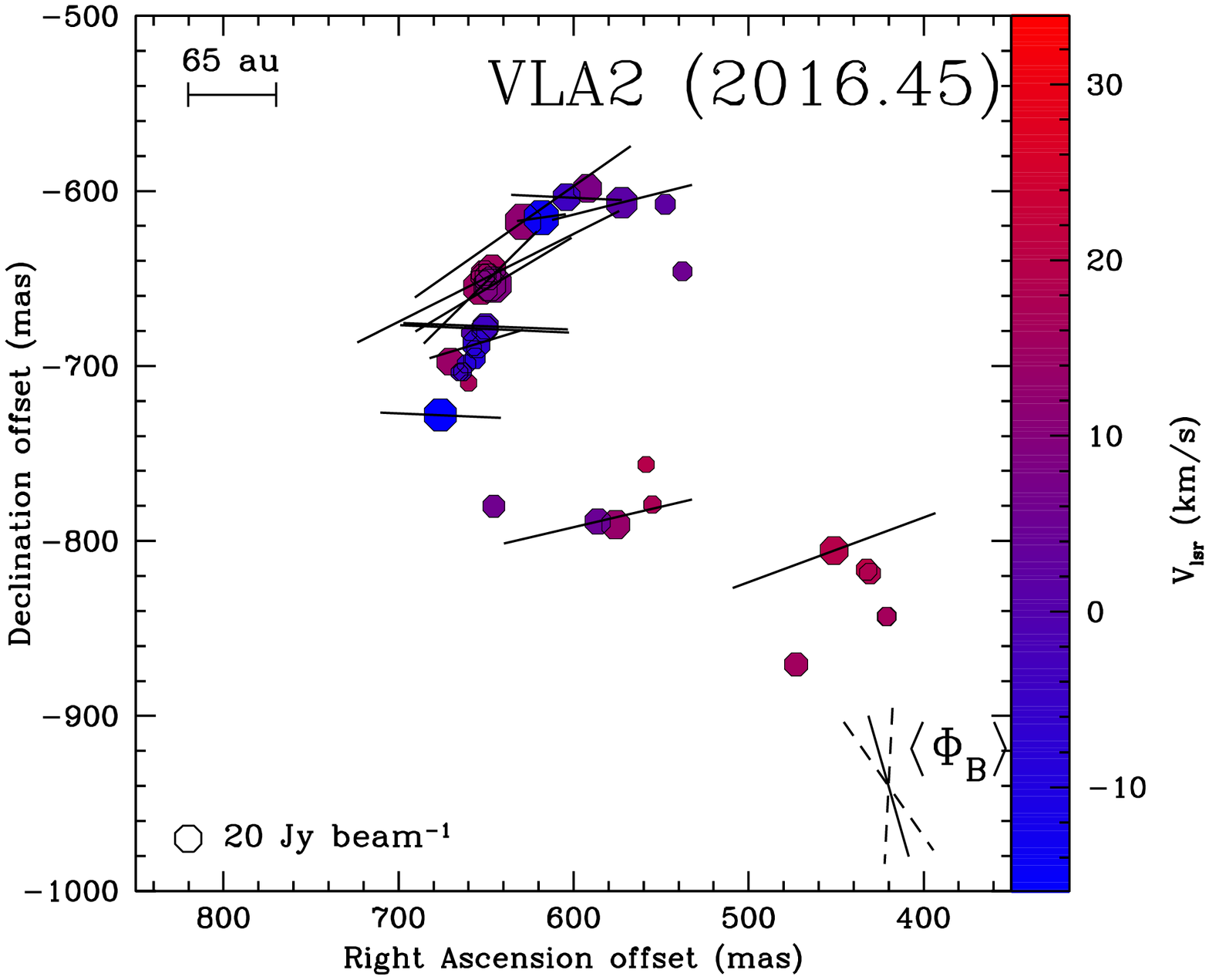}
\includegraphics[width = 6.2 cm]{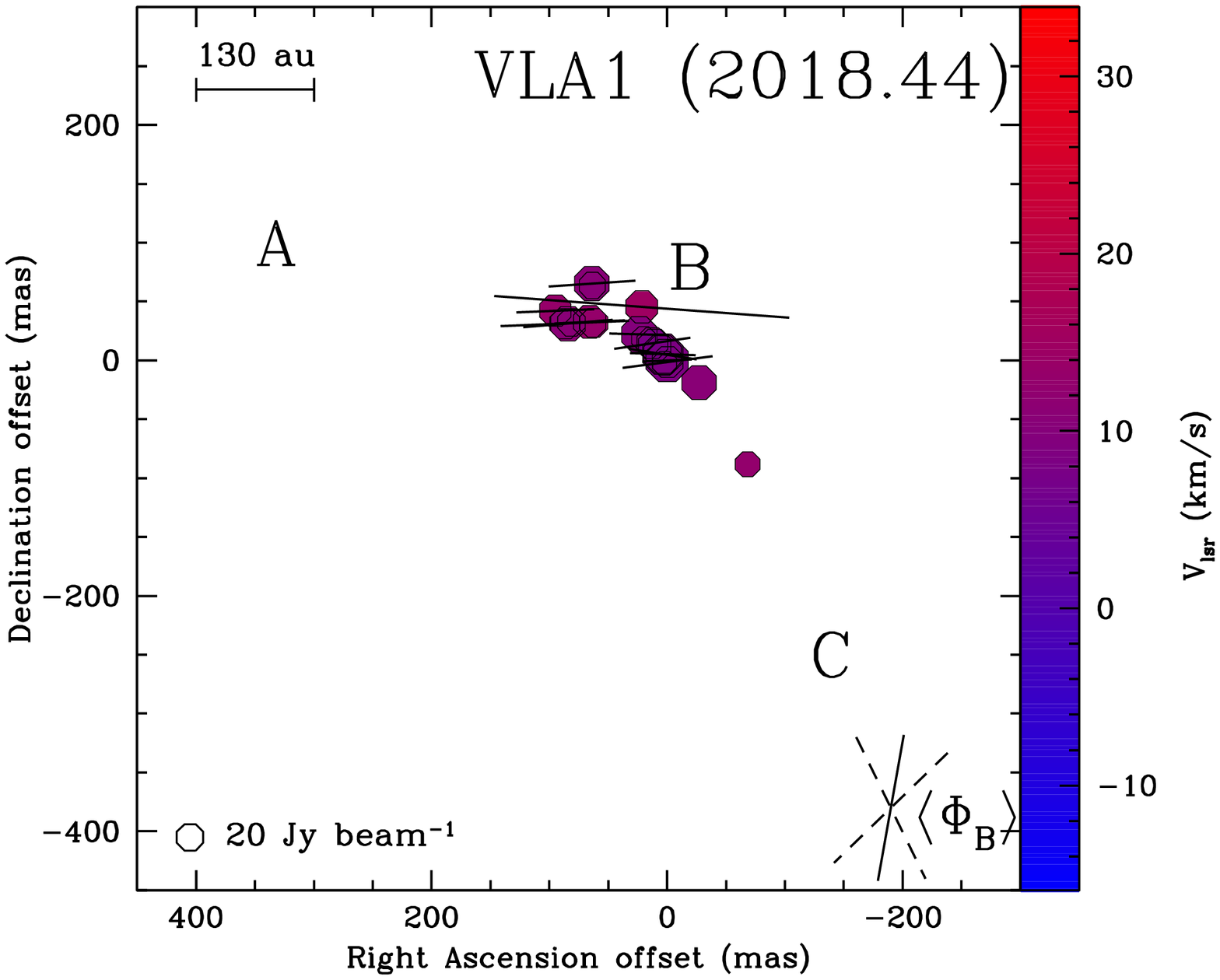}
\includegraphics[width = 6.2 cm]{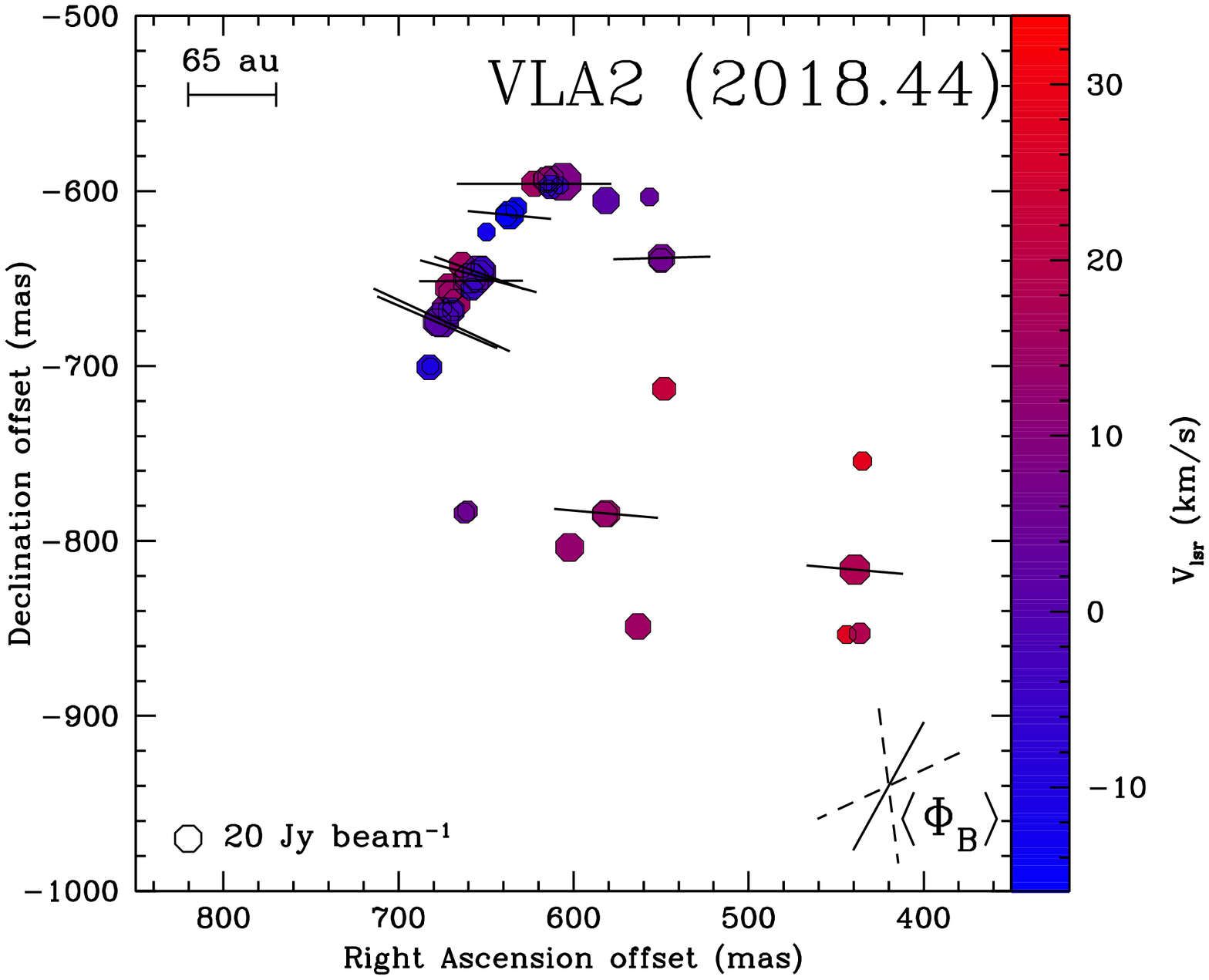}
\includegraphics[width = 6.2 cm]{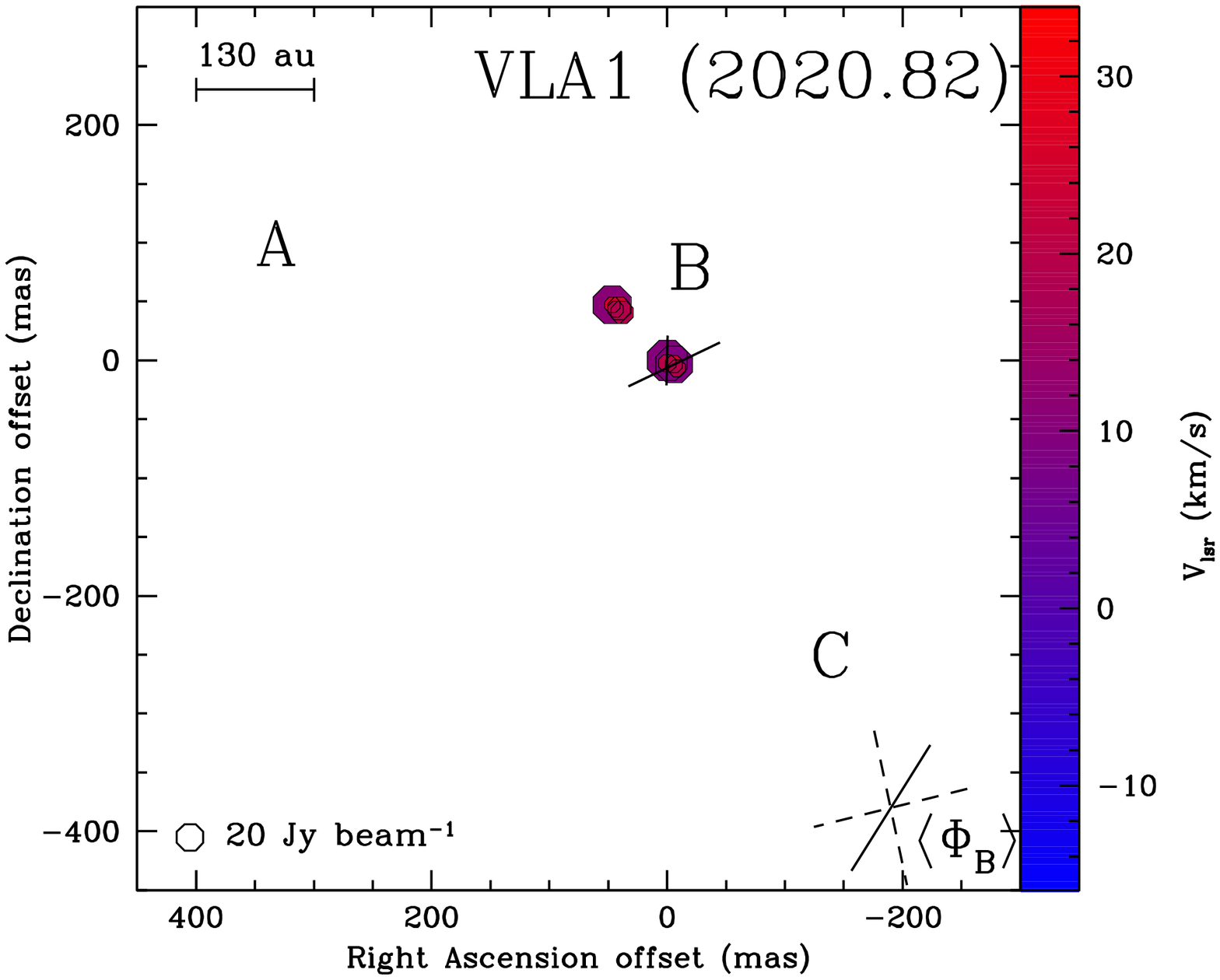}
\includegraphics[width = 6.2 cm]{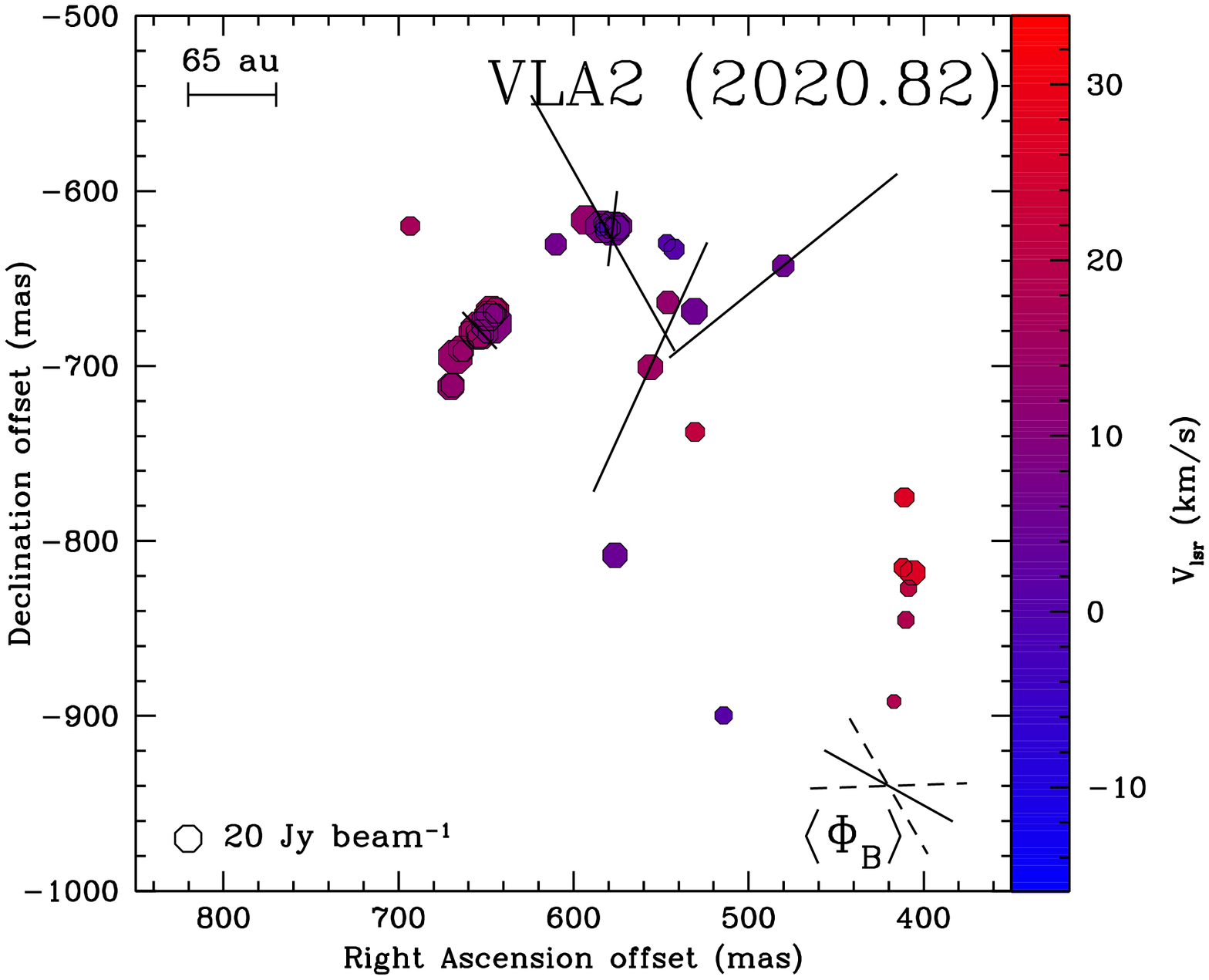}
\caption{A close-up view of the 22~GHz \water ~maser features detected around the radio sources
VLA\,1 (\textit{left panels}) and VLA\,2 (\textit{right panels}) in epochs 2014.46 (top panels), 2016.45, 2018.44, 
and 2020.82 (bottom panels). The reference positions are reported in Table~\ref{abspos}. The
letters A, B, and C indicate the position of the three maser groups identified around VLA\,1 
by \citetalias{sur112} and they are reported here only for reference.
The octagonal symbols are the identified maser features scaled logarithmically
according to their peak intensity (Tables~\ref{VLA1.1_tab}-\ref{VLA2.4_tab}). Maser
LSR radial velocities are indicated by color (the assumed systemic velocity of the
region is $V_{\rm{lsr}}=+10.0$~\kms; \citealt{she03}). The linear
polarization vectors, scaled logarithmically according to the polarization fraction $P_{\rm{l}}$, are
overplotted. In the bottom-right corner of all panels the error-weighted orientation of the
magnetic field ($\langle\Phi_{\rm{B}}\rangle$) is also reported; the two dashed segments indicate 
the uncertainties.}
\label{posplot}
\end{figure*}

\begin{figure*}[t!]
\centering
\includegraphics[width = 5 cm]{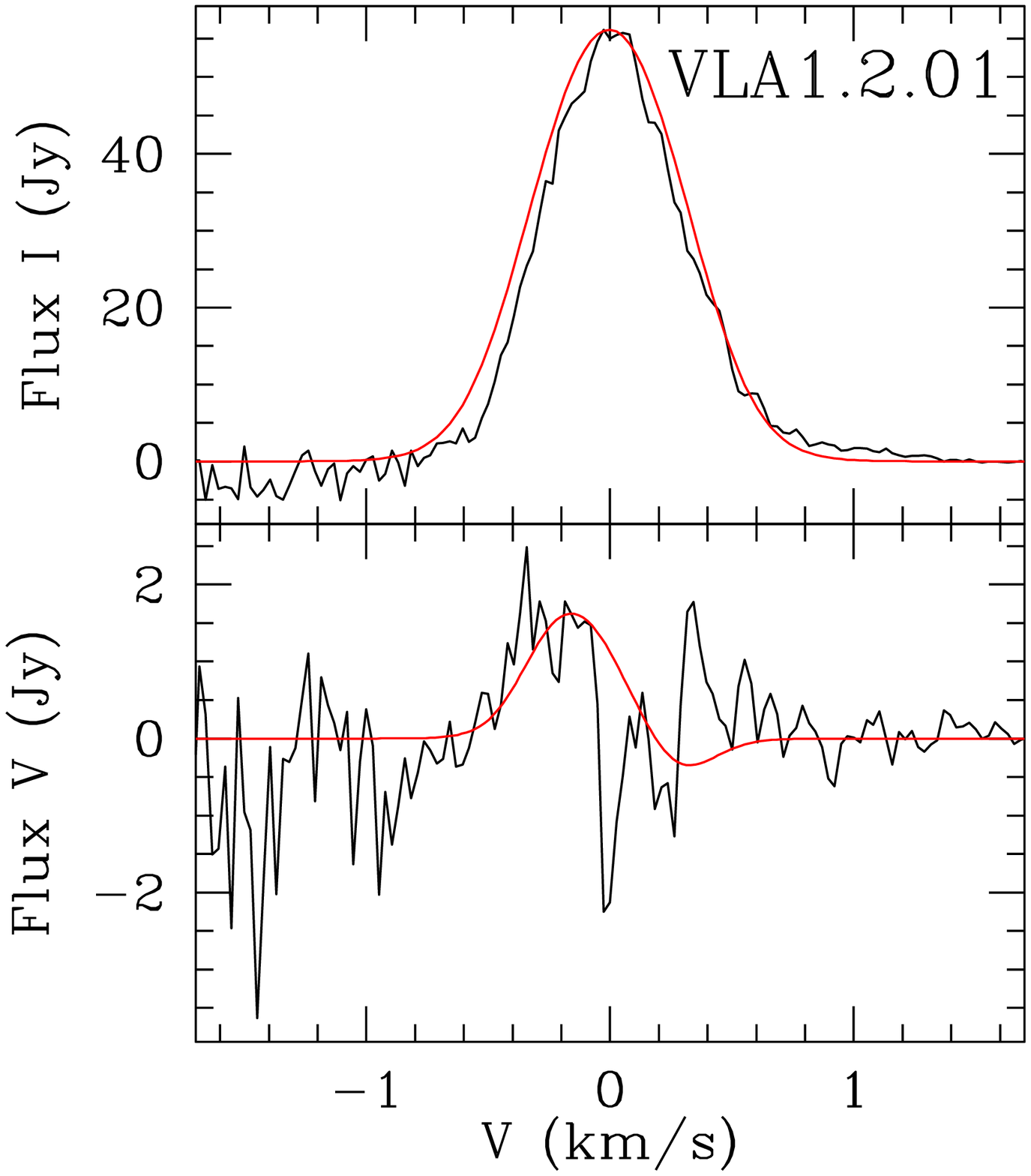}
\includegraphics[width = 5 cm]{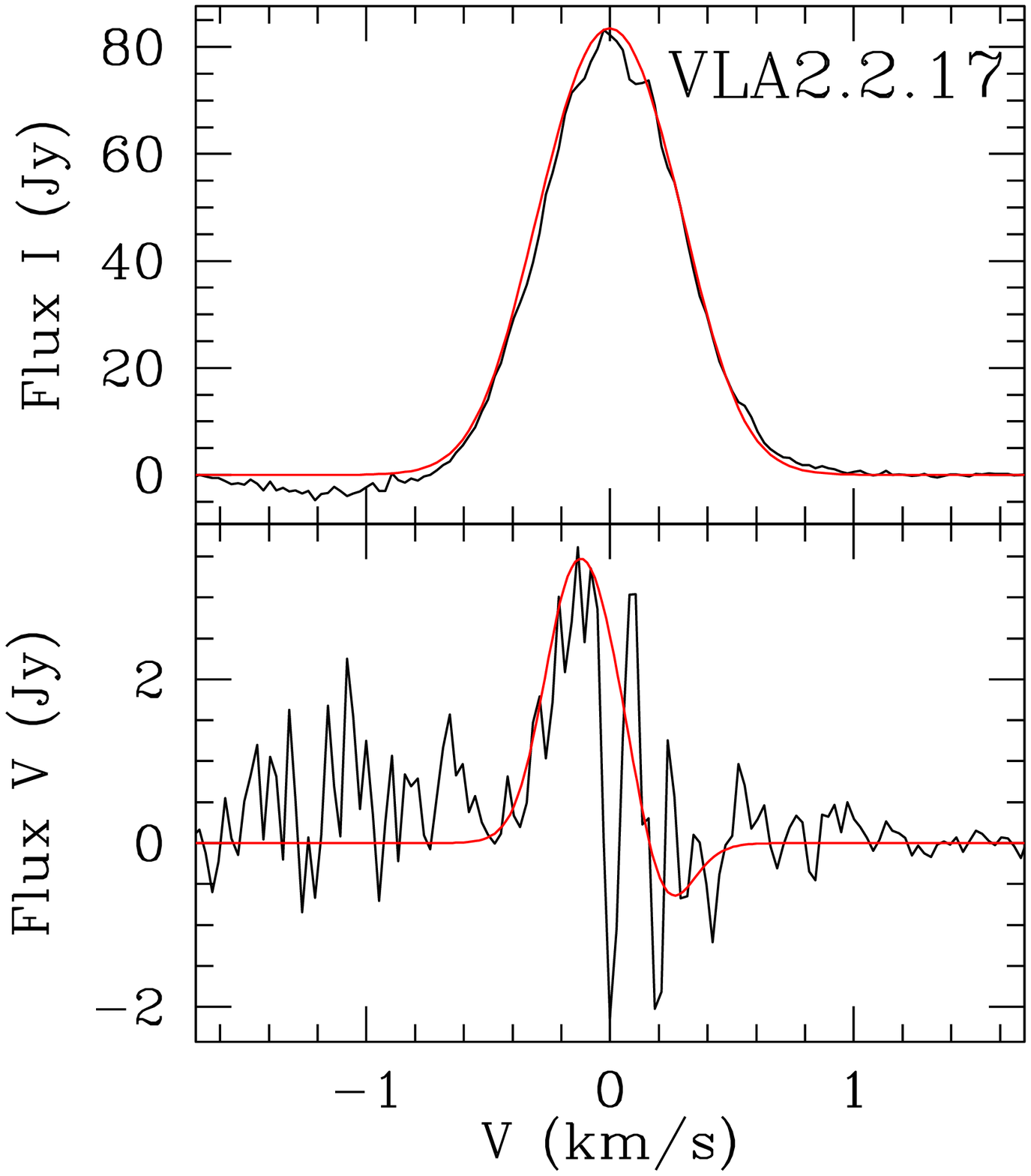}
\includegraphics[width = 5 cm]{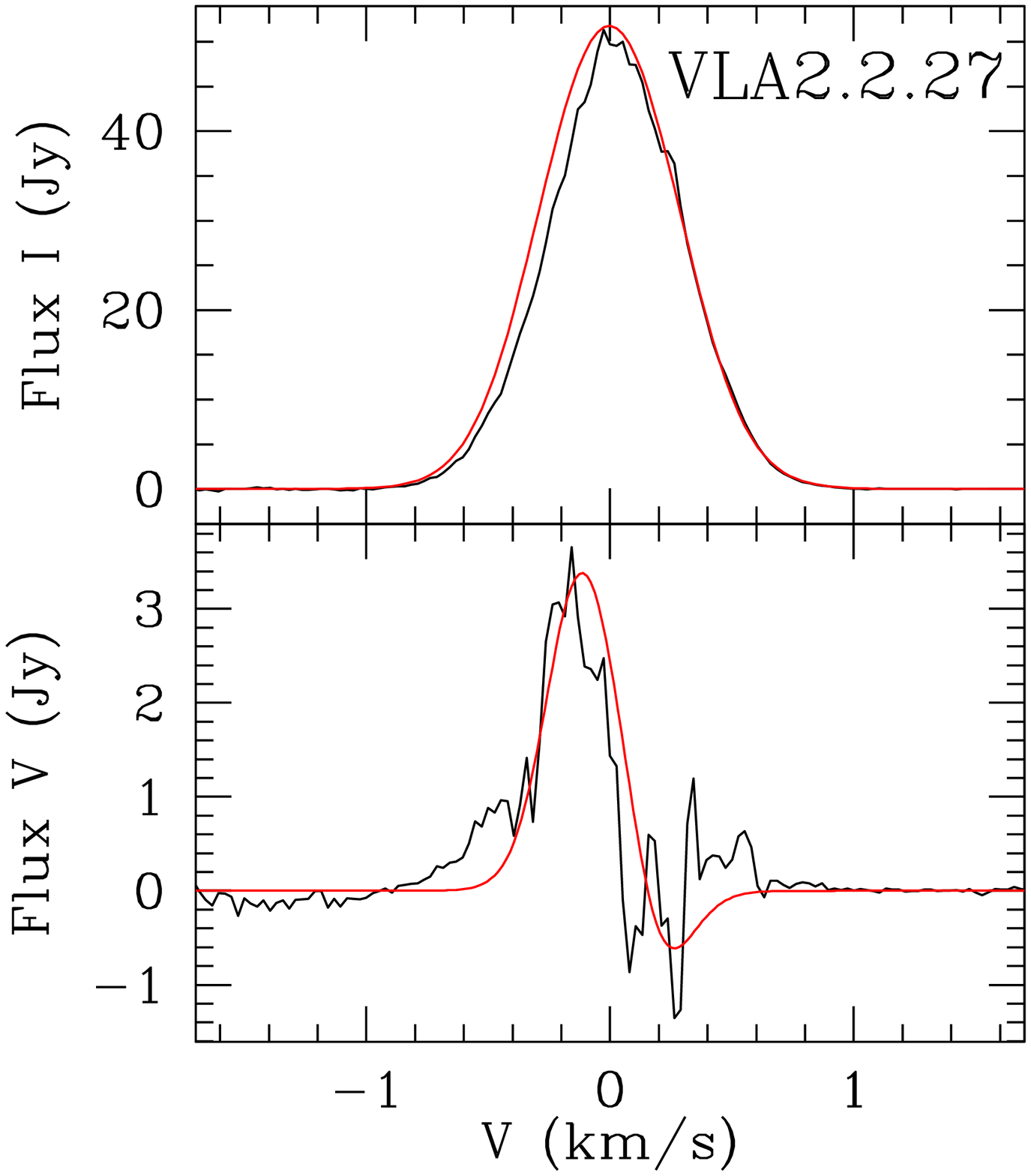}
\includegraphics[width = 5 cm]{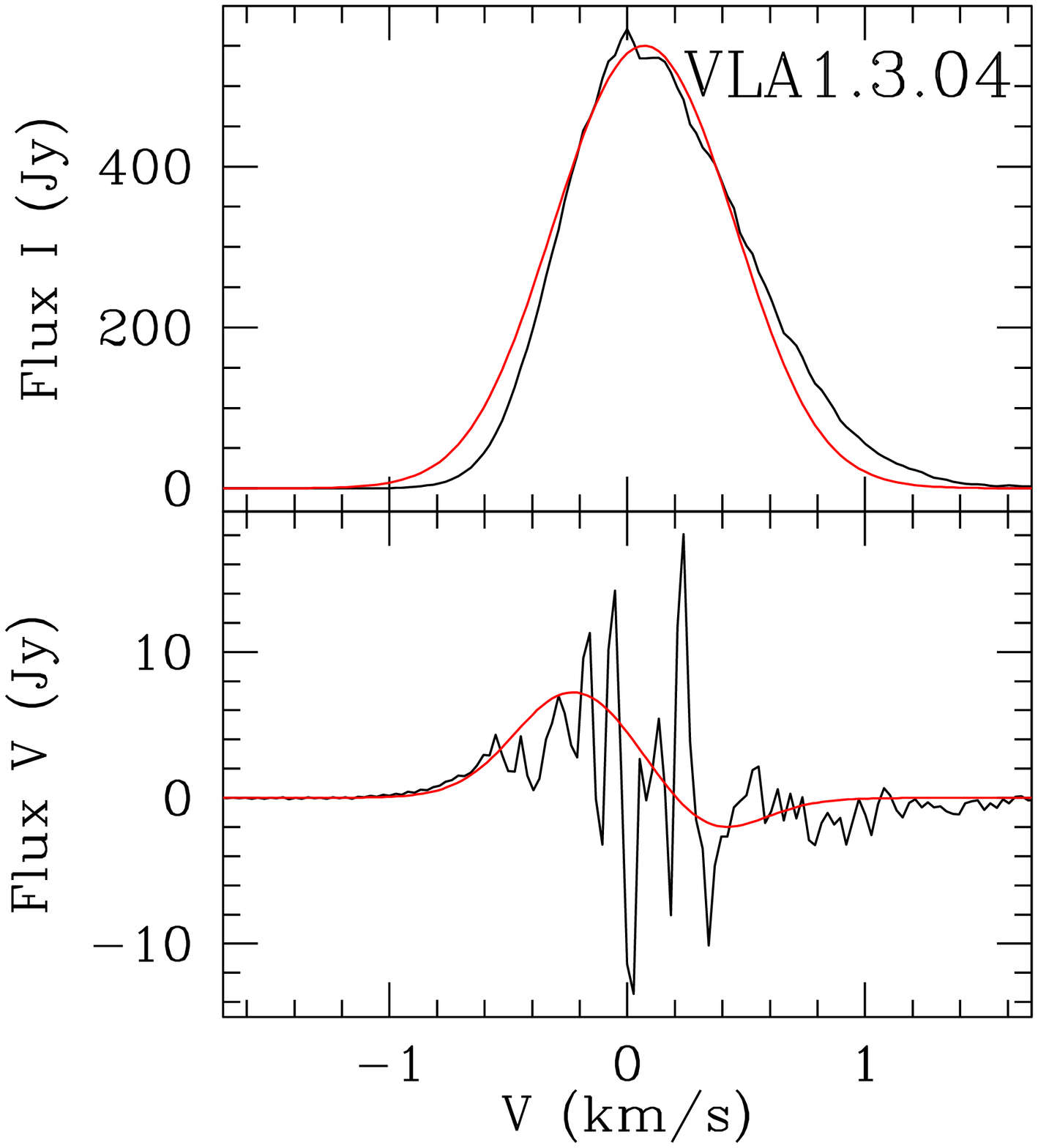}
\includegraphics[width = 5 cm]{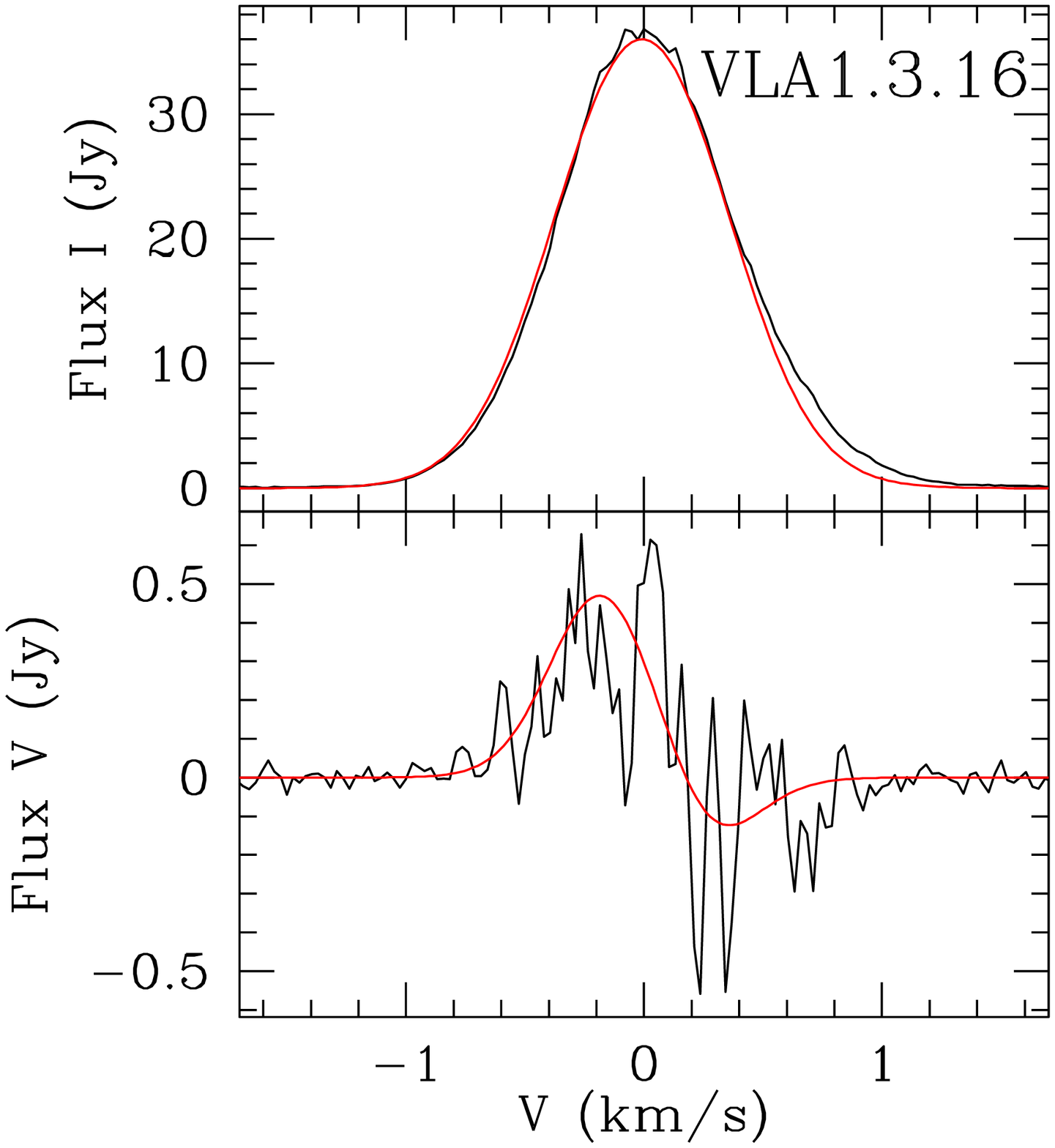}
\includegraphics[width = 5 cm]{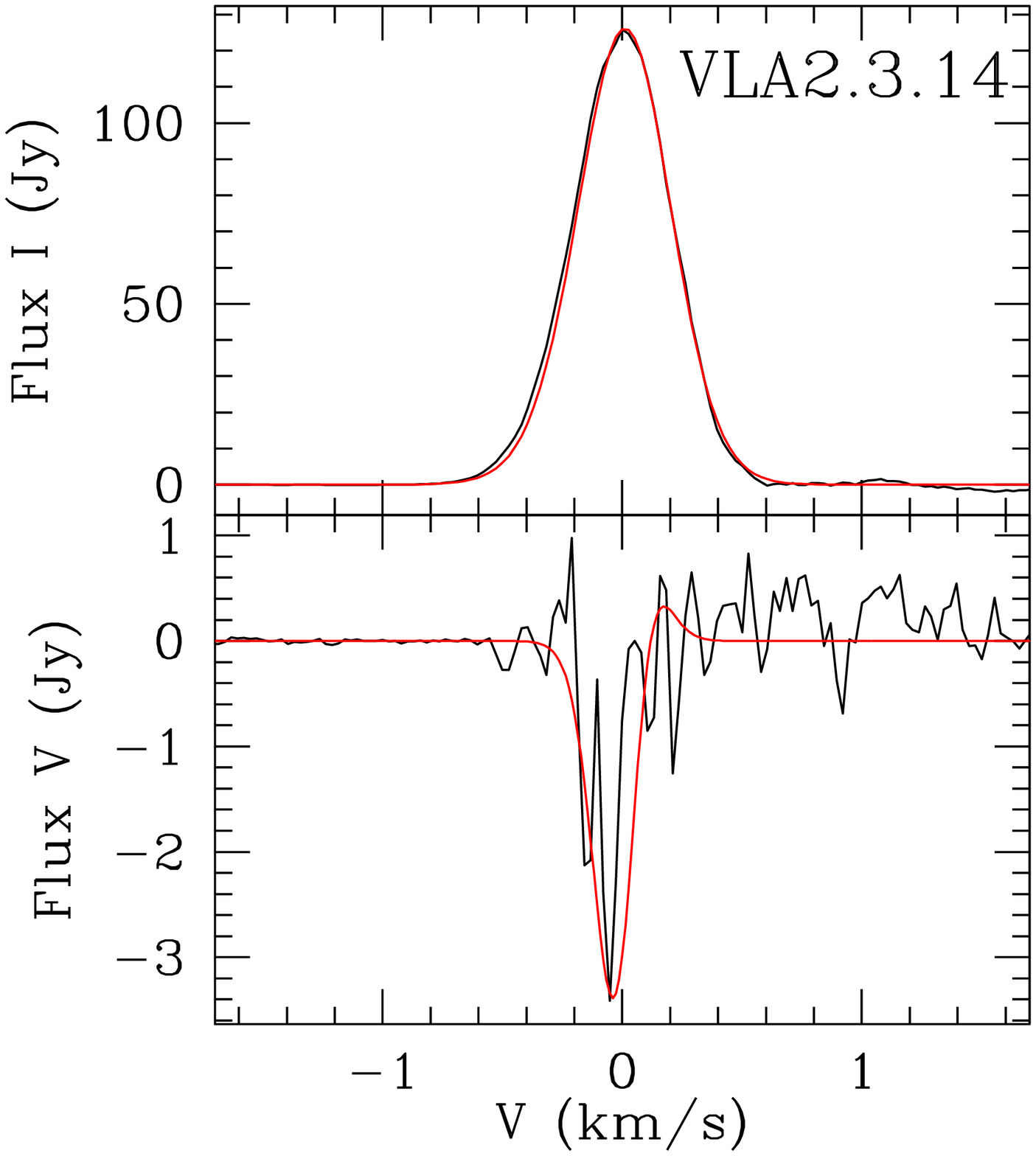}
\includegraphics[width = 5 cm]{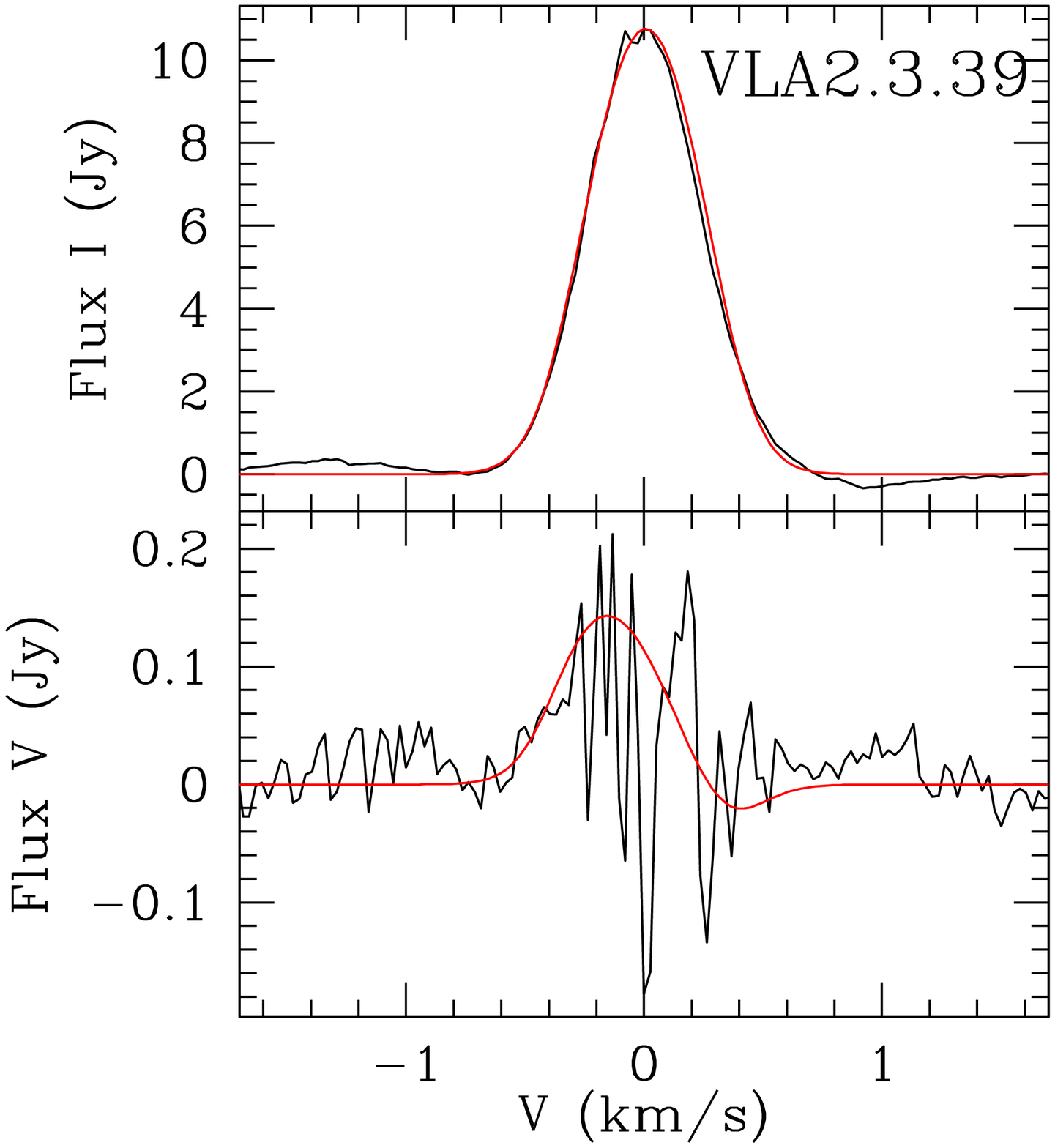}
\includegraphics[width = 5 cm]{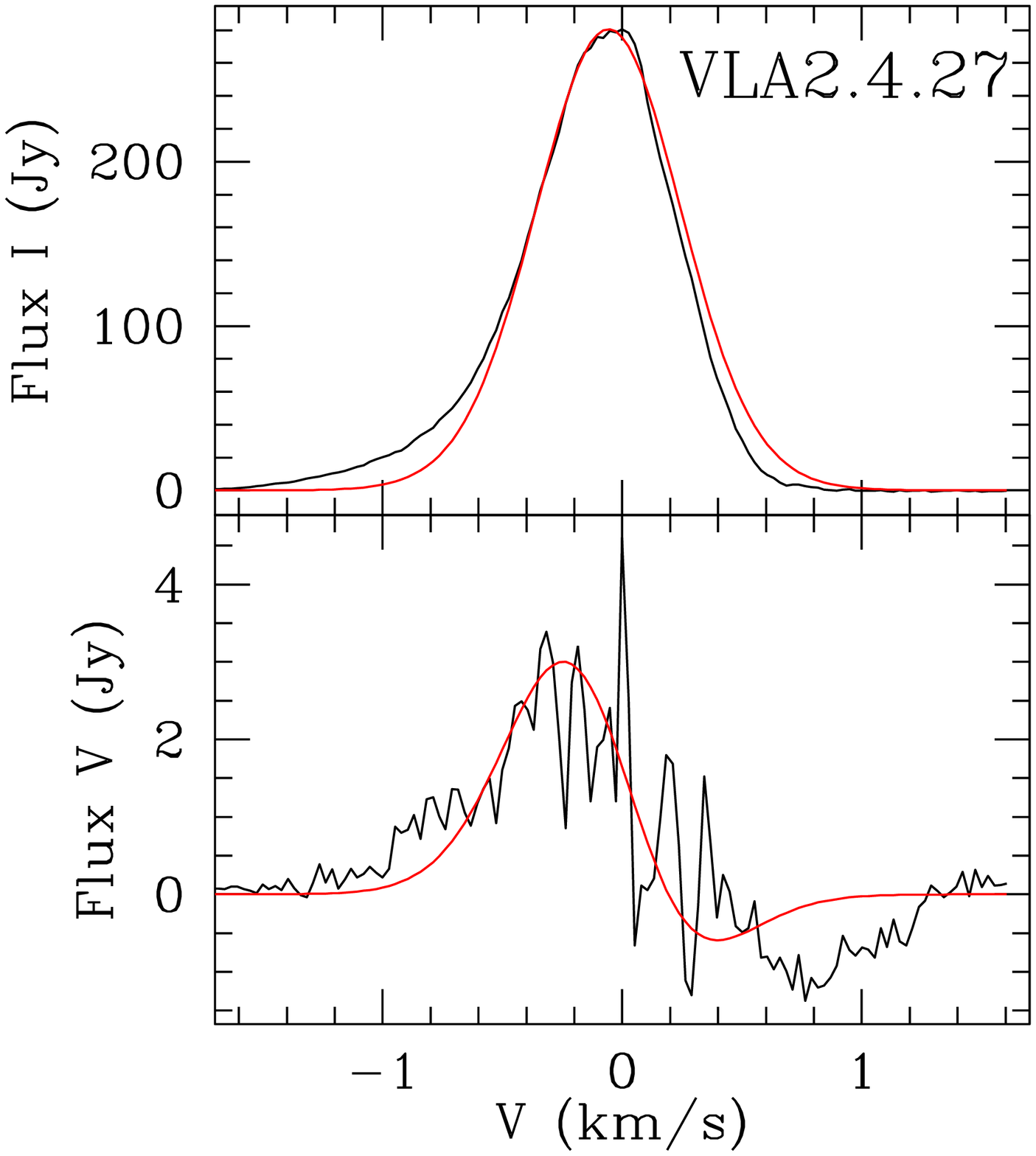}
\caption{Total intensity spectra (\textit{I}, \textit{upper panel}) and circular polarization
intensity spectra (\textit{V}, \textit{lower panel}) for the \water ~masers VLA1.2.01, VLA2.2.17,
VLA2.2.27, VLA1.3.04, VLA1.3.16, VLA2.3.14, VLA2.3.39, and VLA2.4.27 (see Tables~\ref{VLA1.2_tab},
\ref{VLA2.2_tab}, \ref{VLA1.3_tab}, \ref{VLA2.3_tab}, and \ref{VLA2.4_tab}).
The thick red lines are the best-fit models of \textit{I} and \textit{V} emission obtained using the 
outputs of the \code ~for 22~GHz \water ~masers (see Sect.~\ref{obssect}), in particular these
models are used to fit the spectra (for more details see \citetalias{sur112}). The maser features were centered on zero velocity.}
\label{circpol}
\end{figure*}

In the fourth and last EVN epoch, we detected 10 and 39 \water ~maser features toward VLA\,1 and 
VLA\,2 (see Tables~\ref{VLA1.4_tab} and \ref{VLA2.4_tab}), respectively. The maser features in
VLA\,1 ($0.1$~\jyb$< I<303$~\jyb) are compactly located around the two brightest maser features 
VLA1.4.05 ($V_{\rm{lsr}}=9.87$~\kms) and VLA1.4.10 ($V_{\rm{lsr}}=10.87$~\kms), which are about
47~mas apart ($\sim60$~au, see Fig.~\ref{posplot}) and are associated with group~B of
\citetalias{sur112}.
The 70\% of the \water ~maser features detected along VLA\,1 have velocities in the range
$+18.9$~\kms$< V_{\rm{lsr}}<+26.4$~\kms. The last time that \water ~maser emission with
$V_{\rm{lsr}}>+19$~\kms ~was detected toward VLA\,1 was in 2005 \citepalias{sur112}, but those
maser features were associated only with group~A of \citetalias{sur112} and they were at $\sim500$~mas 
($\sim650$~au) northeast of group~B. The \water ~maser features elliptically distributed around
VLA\,2 covers the smallest range of velocities ($-1.0$~\kms$< V_{\rm{lsr}}<+27.2$~\kms) and the
largest range of peak intensities ($0.05$~\jyb$\leq I<286$~\jyb) among the four EVN epochs. \\
\indent The highest linear polarization fraction of epoch 2020.82 was measured in VLA\,2 and its
value is $P_{\rm{l}}=1.9\%$. In particular, we detected linearly polarized emission toward a total
of 7 maser features, two in VLA\,1 and five in VLA\,2 (see Tables~\ref{VLA1.4_tab} and
\ref{VLA2.4_tab}). The \code ~properly fit all of them, providing $\theta$ angles for VLA1.4.02 
and VLA1.4.05 lower than 55\d ~which implies a magnetic field parallel to the linear
polarization vectors. In the case of VLA2.4.31 we have that the magnetic field is more likely 
parallel, indeed for this feature we have $|\theta^{\rm{+}}-55$\d$|<|\theta^{\rm{-}}-55$\d$|$, 
while for all the other four linearly polarized maser features in VLA\,2 the magnetic field is 
perpendicular to the linear polarization vectors (see Table~\ref{VLA2.4_tab}). Only one maser
feature shows circular polarization (VLA2.4.27, $P_{\rm{V}}=1.3\%$), but because it does not show
any linear polarization we had to determine the best estimates of \tbo ~and \dvi ~for modeling its
$V$ spectra (see Fig.~\ref{circpol}). These values are $10^8$~K~sr and $3.0$~\kms, respectively.
\begin {table*}[th!]
\caption []{Parameters of all the 22 GHz \water ~maser features detected around VLA1 (epoch 2014.46).} 
\begin{center}
\scriptsize
\begin{tabular}{ l c c c c c c c c c c c c c c c}
\hline
\hline
\,\,\,\,\,(1)&(2)   & (3)      & (4)            & (5)       & (6)              & (7)         & (8)        & (9)                     & (10)                    & (11)                        & (12)         &(13)               &(14)            \\
Maser     & RA\tablefootmark{a} & Dec\tablefootmark{a}  & Zone & Peak  & $V_{\rm{lsr}}$& $\Delta v\rm{_{L}}$ &$P_{\rm{l}}$ &$\chi$& $\Delta V_{\rm{i}}$\tablefootmark{b}& $T_{\rm{b}}\Delta\Omega$\tablefootmark{b}& $P_{\rm{V}}$ & $B_{||}$  &$\theta$\tablefootmark{c}\\
feature          &  offset &  offset  & & Intensity (I)     &           &                  &             &	          &                         &                         &              &                      &      \\ 
          &  (mas)  &  (mas)   & & (Jy/beam)      &  (km/s)   &      (km/s)      & (\%)        &   (\d)     & (km/s)                  & (log K sr)              &   ($\%$)     &  (mG)               &(\d)       \\ 
\hline
VLA1.1.01      & -101.803& -153.618 & - & $0.43\pm0.03$  &  8.37     &      $0.38$      & $-$         &   $-$     &  $-$                    & $-$                     & $-$	    & $-$            	   &$-$ \\ 
VLA1.1.02      & -101.761& -156.307 & - & $0.57\pm0.04$  &  8.48     &      $0.44$      & $-$         &   $-$     &  $-$                    & $-$                     & $-$	    & $-$            	   &$-$ \\ 
VLA1.1.03      & -64.879 & -93.826  & - & $2.59\pm0.17$  &  12.77    &      $0.52$      & $15.6\pm2.0$& $+48\pm3$ &  $<0.4$                 & $8.5^{+0.6}_{-1.1}$     & $-$	    & $-$            	   &$90^{+7}_{-7}$ \\
VLA1.1.04      & -31.619 & -19.016  & - & $15.85\pm0.15$ &  12.61    &      $2.88$      & $10.1\pm2.5$& $-62\pm1$ &  $2.7^{+0.4}_{-0.3}$    & $7.0^{+1.3}_{-1.1}$     & $-$	    & $-$            	   &$90^{+13}_{-13}$ \\ 
VLA1.1.05      & -28.672 & -99.377  & - & $86.37\pm1.06$ &  10.61    &      $2.31$      & $2.3\pm0.6$ & $-25\pm4$ &  $0.5$                  & $9.4^{+0.2}_{-0.1}$     & $-$	    & $-$            	   &$90^{+17}_{-17}$ \\ 
VLA1.1.06      & -24.588 & -99.083  & - & $8.37\pm0.34$  &  11.35    &      $1.45$      & $-$         &   $-$     &  $-$                    & $-$                     & $-$	    & $-$            	   &$-$ \\ 
VLA1.1.07      & -23.703 & -12.516  & - & $74.21\pm0.79$ &  10.98    &      $0.58$      & $-$         &   $-$     &  $-$                    & $-$                     & $-$ 	    & $-$ 		   &$-$ \\ 
VLA1.1.08      & -14.778 & -1.484   & - & $23.15\pm0.04$ &  8.63     &      $0.57$      & $1.3\pm0.1$ & $-39\pm3$ &  $3.2^{+0.2}_{-0.4}$    & $9.1^{+0.6}_{-0.1}$     & $-$	    & $-$            	   &$75^{+14}_{-36}$ \\ 
VLA1.1.09      & -10.610 & -10.700  & - & $1.71\pm0.02$  &  7.90     &      $2.81$      & $-$         &   $-$     &  $-$                    & $-$                     & $-$	    & $-$            	   &$-$ \\ 
VLA1.1.10      & -7.621  & -6.420   & - & $61.34\pm0.35$ &  9.77     &      $2.10$      & $-$         &   $-$     &  $-$                    & $-$                     & $-$	    & $-$            	   &$-$ \\ 
VLA1.1.11      & -6.821  & -9.647   & - & $44.38\pm0.13$ &  12.48    &      $1.22$      & $0.7\pm0.1$ & $-45\pm5$ &  $-$                    & $-$                     & $-$	    & $-$            	   &$-$ \\ 
VLA1.1.12      & -5.810  & -6.111   & - & $34.60\pm0.21$ &  9.56     &      $0.58$      & $-$         &   $-$     &  $-$                    & $-$                     & $-$	    & $-$            	   &$-$ \\ 
VLA1.1.13      & -4.084  & -9.914   & - & $0.54\pm0.01$  &  16.77    &      $3.61$      & $-$         &   $-$     &  $-$                    & $-$                     & $-$	    & $-$            	   &$-$ \\ 
VLA1.1.14      &  1.684  &  1.221   & - & $9.01\pm0.20$  &  9.53     &      $0.87$      & $-$         &   $-$     &  $-$                    & $-$                     & $-$	    & $-$            	   &$-$ \\ 
VLA1.1.15      & 0       & 0        & - &$1205.52\pm1.05$&  10.45    &      $0.92$      & $0.4\pm0.1$ & $-32\pm5$ &  $-$                    & $-$                     & $-$	    & $-$            	   &$-$ \\ 
VLA1.1.16      &  2.779  & -0.122   & - & $49.67\pm0.21$ &  13.03    &      $1.55$      & $0.5\pm0.1$ & $-34\pm5$ &  $2.9^{+0.1}_{-0.2}$    & $8.7^{+0.5}_{-0.1}$     & $-$	    & $-$            	   &$80^{+10}_{-37}$ \\ 
VLA1.1.17      &  11.662 &  7.095   & - & $5.72\pm0.04$  &  8.90     &      $1.65$      & $-$         &   $-$     &  $-$                    & $-$                     & $-$	    & $-$            	   &$-$ \\ 
VLA1.1.18      &  12.673 &  8.083   & - & $14.22\pm0.10$ &  9.29     &      $1.08$      & $-$         &   $-$     &  $-$                    & $-$                     & $-$	    & $-$            	   &$-$ \\ 
VLA1.1.19      &  16.251 &  11.299  & - & $9.16\pm0.26$  &  11.37    &      $0.97$      & $-$         &   $-$     &  $-$                    & $-$                     & $-$	    & $-$            	   &$-$ \\ 
VLA1.1.20      &  16.378 &  10.674  & - & $13.52\pm0.02$ &  19.67    &      $1.87$      & $0.7\pm0.1$ & $-41\pm5$ &  $-$                    & $-$                     & $-$	    & $-$            	   &$-$ \\ 
VLA1.1.21      &  17.767 &  13.378  & - & $8.08\pm0.25$  &  11.37    &      $0.77$      & $-$         &   $-$     &  $-$                    & $-$                     & $-$	    & $-$            	   &$-$ \\ 
VLA1.1.22      &  36.250 &  24.662  & - & $21.39\pm0.37$ &  9.79     &      $1.38$      & $-$         &   $-$     &  $-$                    & $-$                     & $-$	    & $-$            	   &$-$ \\ 
VLA1.1.23      &  36.418 &  20.989  & - & $15.00\pm0.10$ &  12.27    &      $2.05$      & $-$         &   $-$     &  $-$                    & $-$                     & $-$	    & $-$            	   &$-$ \\ 
VLA1.1.24      &  37.639 &  20.233  & - & $62.51\pm0.14$ &  12.58    &      $0.72$      & $0.6\pm0.1$ & $-22\pm4$ &  $3.8^{+0.1}_{-0.3}$    & $7.6^{+0.5}_{-0.2}$     & $-$	    & $-$            	   &$80^{+10}_{-31}$ \\ 
VLA1.1.25      &  38.397 &  21.591  & - & $680.26\pm0.61$&  14.14    &      $1.36$      & $0.6\pm0.1$ & $-35\pm2$ &  $-$                    & $-$                     & $-$	    & $-$            	   &$-$ \\ 
VLA1.1.26      &  40.418 &  21.824  & - & $134.29\pm0.25$&  13.19    &      $0.68$      & $0.5\pm0.1$ & $-40\pm4$ &  $4.2^{+0.1}_{-0.2}$    & $8.7^{+0.3}_{-0.2}$     & $-$	    & $-$            	   &$86^{+4}_{-12}$ \\  
VLA1.1.27      &  40.460 &  22.739  & - & $42.89\pm0.24$ &  13.19    &      $0.83$      & $1.3\pm0.1$ & $-23\pm2$ &  $4.0^{+0.3}_{-0.3}$    & $9.1^{+1.0}_{-3.1}$     & $-$	    & $-$            	   &$87^{+3}_{-23}$ \\ 
VLA1.1.28      &  54.943 &  31.521  & - & $3.67\pm0.13$  &  12.50    &      $0.88$      & $-$         &   $-$     &  $-$                    & $-$                     & $-$	    & $-$            	   &$-$ \\ 
\hline
\end{tabular} \end{center}
\tablefoot{
\tablefoottext{a}{The reference position is $\alpha_{2000}=20^{\rm{h}}38^{\rm{m}}36^{\rm{s}}\!.43399\pm0^{\rm{s}}\!.00008$ and 
$\delta_{2000}=+42^{\circ}37'34''\!\!.8710\pm0''\!\!.0009$.}
\tablefoottext{b}{The best-fitting results obtained by using a model based on the radiative transfer theory of \water ~masers 
for $\Gamma+\Gamma_{\nu}=1~\rm{s^{-1}}$ \citetalias{sur112}. The errors were determined by analyzing the full probability distribution 
function. For $T\sim3600$~K ($\Gamma_{\nu}=19~\rm{s}^{-1}$) \tbo ~has to be adjusted by adding $+1.3$~log~K~sr \citep{and93}.}
\tablefoottext{c}{The angle between the magnetic field and the maser propagation direction is determined by using the observed $P_{\rm{l}}$ 
and the fitted emerging brightness temperature. The errors were determined by analyzing the full probability distribution function.}
}
\label{VLA1.1_tab}
\end{table*}
\begin {table*}[th!]
\caption []{Parameters of all the 22 GHz \water ~maser features detected around VLA2 (epoch 2014.46).} 
\begin{center}
\scriptsize
\begin{tabular}{ l c c c c c c c c c c c c c c c}
\hline
\hline
\,\,\,\,\,(1)&(2)   & (3)      & (4)            & (5)       & (6)              & (7)         & (8)        & (9)                     & (10)                    & (11)                        & (12)         &(13)        & (14)                   \\
Maser     & RA\tablefootmark{a} & Dec\tablefootmark{a}  & Zone & Peak  & $V_{\rm{lsr}}$& $\Delta v\rm{_{L}}$ &$P_{\rm{l}}$ &$\chi$& $\Delta V_{\rm{i}}$\tablefootmark{b}& $T_{\rm{b}}\Delta\Omega$\tablefootmark{b}& $P_{\rm{V}}$ & $B_{||}$  &$\theta$\tablefootmark{c}\\
feature          &  offset &  offset  &  &Intensity (I)     &           &                  &             &	          &                         &                         &              &                      &      \\ 
          &  (mas)  &  (mas)   & & (Jy/beam)      &  (km/s)   &      (km/s)      & (\%)        &   (\d)     & (km/s)                  & (log K sr)              &   ($\%$)     &  (mG)               &(\d)       \\ 
\hline
VLA2.1.01      & 531.033 & -840.988 & 3 & $21.02\pm0.13$ & 12.48     &      $0.66$      & $-$         &   $-$     &  $-$                    & $-$                     & $-$	    & $-$              	   &$-$ \\ 
VLA2.1.02      & 535.453 & -643.040 & 2 & $0.89\pm0.01$  & 5.87      &      $1.00$      & $-$         &   $-$     &  $-$                    & $-$                     & $-$	    & $-$              	   &$-$ \\ 
VLA2.1.03      & 537.137 & -839.603 & 3 & $21.13\pm0.25$ & 14.90     &      $0.65$      & $-$         &   $-$     &  $-$                    & $-$                     & $-$	    & $-$              	   &$-$ \\ 
VLA2.1.04      & 539.369 & -646.263 & 2 & $5.87\pm0.02$  & 4.26      &      $2.12$      & $-$         &   $-$     &  $-$                    & $-$                     & $-$	    & $-$              	   &$-$ \\ 
VLA2.1.05      & 540.927 & -646.011 & 2 & $2.11\pm0.01$  & 3.13      &      $0.66$      & $-$         &   $-$     &  $-$                    & $-$                     & $-$	    & $-$              	   &$-$ \\ 
VLA2.1.06      & 541.979 & -631.249 & 2 & $0.27\pm0.01$  & 6.29      &      $0.53$      & $-$         &   $-$     &  $-$                    & $-$                     & $-$	    & $-$              	   &$-$ \\ 
VLA2.1.07      & 544.758 & -632.530 & 2 & $1.13\pm0.02$  & 7.37      &      $0.56$      & $-$         &   $-$     &  $-$                    & $-$                     & $-$	    & $-$              	   &$-$ \\ 
VLA2.1.08      & 545.895 & -631.870 & 2 & $0.22\pm0.01$  & 6.21      &      $0.45$      & $-$         &   $-$     &  $-$                    & $-$                     & $-$	    & $-$              	   &$-$ \\ 
VLA2.1.09      & 561.935 & -779.900 & 3 & $0.27\pm0.01$  & 21.19     &      $0.66$      & $-$         &   $-$     &  $-$                    & $-$                     & $-$	    & $-$              	   &$-$ \\ 
VLA2.1.10      & 598.985 & -598.778 & 1 & $0.49\pm0.02$  & 7.21      &      $0.62$      & $-$         &   $-$     &  $-$                    & $-$                     & $-$	    & $-$              	   &$-$ \\ 
VLA2.1.11      & 607.153 & -600.987 & 1 & $1.85\pm0.09$  & 12.27     &      $0.59$      & $-$         &   $-$     &  $-$                    & $-$                     & $-$	    & $-$              	   &$-$ \\ 
VLA2.1.12      & 615.784 & -615.742 & 1 & $14.16\pm0.02$ & -11.59    &      $0.55$      & $-$         &   $-$     &  $-$                    & $-$                     & $-$	    & $-$              	   &$-$ \\ 
VLA2.1.13      & 617.426 & -614.739 & 1 & $0.26\pm0.01$  & -6.03     &      $0.79$      & $-$         &   $-$     &  $-$                    & $-$                     & $-$	    & $-$              	   &$-$ \\ 
VLA2.1.14      & 622.226 & -617.950 & 1 & $0.72\pm0.01$  & 1.52      &      $1.10$      & $-$         &   $-$     &  $-$                    & $-$                     & $-$	    & $-$              	   &$-$ \\ 
VLA2.1.15      & 624.246 & -620.457 & 1 & $0.55\pm0.01$  & -0.74     &      $0.92$      & $-$         &   $-$     &  $-$                    & $-$                     & $-$	    & $-$              	   &$-$ \\ 
VLA2.1.16      & 625.552 & -621.185 & 1 & $1.16\pm0.01$  & 0.37      &      $0.78$      & $-$         &   $-$     &  $-$                    & $-$                     & $-$	    & $-$              	   &$-$ \\ 
VLA2.1.17      & 629.846 & -625.595 & 1 & $0.75\pm0.01$  & 2.34      &      $0.66$      & $-$         &   $-$     &  $-$                    & $-$                     & $-$	    & $-$              	   &$-$ \\ 
VLA2.1.18      & 632.414 & -623.024 & 1 & $2.76\pm0.02$  & 14.98     &      $0.45$      & $-$         &   $-$     &  $-$                    & $-$                     & $-$	    & $-$              	   &$-$ \\ 
VLA2.1.19      & 634.351 & -629.128 & 1 & $1.02\pm0.01$  & 6.11      &      $0.69$      & $-$         &   $-$     &  $-$                    & $-$                     & $-$	    & $-$              	   &$-$ \\ 
VLA2.1.20      & 635.951 & -628.952 & 1 & $10.24\pm0.02$ & 15.24     &      $0.43$      & $-$         &   $-$     &  $-$                    & $-$                     & $-$	    & $-$              	   &$-$ \\ 
VLA2.1.21      & 638.561 & -770.130 & 3 & $5.25\pm0.05$  & 8.90      &      $0.55$      & $-$         &   $-$     &  $-$                    & $-$                     & $-$	    & $-$              	   &$-$ \\ 
VLA2.1.22      & 640.793 & -772.827 & 3 & $2.81\pm0.01$  & 5.61      &      $0.53$      & $-$         &   $-$     &  $-$                    & $-$                     & $-$	    & $-$              	   &$-$ \\ 
VLA2.1.23      & 642.687 & -650.074 & 1 & $14.99\pm0.71$ & 10.08     &      $1.85$      & $-$         &   $-$     &  $-$                    & $-$                     & $-$	    & $-$              	   &$-$ \\ 
VLA2.1.24      & 643.529 & -640.629 & 1 & $4.68\pm0.15$  & 15.24     &      $0.42$      & $4.6\pm1.2$ & $-31\pm5$ &  $-$                    & $-$                     & $-$	    & $-$              	   &$-$ \\ 
VLA2.1.25      & 643.698 & -645.699 & 1 & $1.62\pm0.01$  & 16.85     &      $0.52$      & $-$         &   $-$     &  $-$                    & $-$                     & $-$	    & $-$              	   &$-$ \\ 
VLA2.1.26      & 644.245 & -642.471 & 1 & $25.15\pm0.14$ & 15.27     &      $0.45$      & $0.9\pm0.2$ & $-42\pm6$ &  $1.6^{+0.1}_{-0.2}$    & $8.9^{+0.3}_{-1.2}$     & $-$	    & $-$              	   &$79^{+11}_{-17}$ \\ 
VLA2.1.27      & 645.213 & -652.153 & 1 & $2.48\pm0.07$  & 9.19      &      $0.57$      & $-$         &   $-$     &  $-$                    & $-$                     & $-$	    & $-$              	   &$-$ \\ 
VLA2.1.28      & 645.466 & -647.984 & 1 & $0.22\pm0.01$  & 16.67     &      $0.44$      & $-$         &   $-$     &  $-$                    & $-$                     & $-$	    & $-$              	   &$-$ \\ 
VLA2.1.29      & 645.887 & -644.447 & 1 & $25.57\pm0.11$ & 15.40     &      $0.46$      & $1.5\pm0.7$ & $-47\pm4$ &  $1.6^{+0.1}_{-0.3}$    & $9.2^{+0.3}_{-2.1}$     & $-$	    & $-$              	   &$79^{+11}_{-44}$ \\ 
VLA2.1.30      & 645.971 & -653.091 & 1 & $0.99\pm0.04$  & 8.56      &      $0.73$      & $-$         &   $-$     &  $-$                    & $-$                     & $-$	    & $-$              	   &$-$ \\ 
VLA2.1.31      & 646.097 & -646.828 & 1 & $6.83\pm0.22$  & 15.88     &      $0.48$      & $1.4\pm0.2$ & $-41\pm6$ &  $1.6^{+0.2}_{-0.2}$    & $9.2^{+0.3}_{-1.2}$     & $-$	    & $-$              	   &$86^{+3}_{-13}$ \\ 
VLA2.1.32      & 648.371 & -689.751 & 1 & $0.23\pm0.13$  & -3.85     &      $0.76$      & $-$         &   $-$     &  $-$                    & $-$                     & $-$	    & $-$              	   &$-$ \\ 
VLA2.1.33      & 648.539 & -682.083 & 1 & $0.21\pm0.01$  & -5.51     &      $0.55$      & $-$         &   $-$     &  $-$                    & $-$                     & $-$	    & $-$              	   &$-$ \\ 
VLA2.1.34      & 650.139 & -651.642 & 1 & $6.93\pm0.06$  & 15.59     &      $0.46$      & $-$         &   $-$     &  $-$                    & $-$                     & $-$	    & $-$              	   &$-$ \\ 
VLA2.1.35      & 656.202 & -678.604 & 1 & $3.28\pm0.01$  & 0.68      &      $0.86$      & $-$         &   $-$     &  $-$                    & $-$                     & $-$          & $-$                  &$-$ \\ 
VLA2.1.36      & 656.875 & -710.316 & 1 & $0.40\pm0.01$  & -6.14     &      $0.49$      & $-$         &   $-$     &  $-$                    & $-$                     & $-$	    & $-$              	   &$-$ \\ 
VLA2.1.37      & 656.918 & -668.846 & 1 & $1.61\pm0.04$  & 15.45     &      $0.50$      & $-$         &   $-$     &  $-$                    & $-$                     & $-$	    & $-$              	   &$-$ \\ 
VLA2.1.38      & 656.918 & -670.769 & 1 & $31.70\pm0.20$ & 15.09     &      $2.12$      & $0.9\pm0.1$ & $-39\pm9$ &  $1.9^{+0.1}_{-0.3}$    & $8.9^{+0.5}_{-0.3}$     & $-$	    & $-$              	   &$79^{+12}_{-36}$ \\ 
VLA2.1.39      & 660.580 & -672.241 & 1 & $17.93\pm0.45$ & 14.56     &      $0.39$      & $-$         &   $-$     &  $-$                    & $-$                     & $-$	    & $-$              	   &$-$ \\ 
VLA2.1.40      & 663.654 & -678.890 & 1 & $4.59\pm0.35$  & 13.43     &      $0.38$      & $-$         &   $-$     &  $-$                    & $-$                     & $-$	    & $-$              	   &$-$ \\ 
VLA2.1.41      & 665.380 & -693.222 & 1 & $4.19\pm0.13$  & 12.48     &      $0.68$      & $-$         &   $-$     &  $-$                    & $-$                     & $-$	    & $-$              	   &$-$ \\ 
VLA2.1.42      & 671.653 & -695.343 & 1 & $1.35\pm0.02$  & 7.98      &      $0.62$      & $-$         &   $-$     &  $-$                    & $-$                     & $-$	    & $-$              	   &$-$ \\ 
VLA2.1.43      & 673.295 & -701.130 & 1 & $0.30\pm0.01$  & 5.05      &      $0.65$      & $-$         &   $-$     &  $-$                    & $-$                     & $-$	    & $-$              	   &$-$ \\ 
\hline
\end{tabular} \end{center}
\tablefoot{
\tablefoottext{a}{The reference position is $\alpha_{2000}=20^{\rm{h}}38^{\rm{m}}36^{\rm{s}}\!.43399\pm0^{\rm{s}}\!.00008$ and 
$\delta_{2000}=+42^{\circ}37'34''\!\!.8710\pm0''\!\!.0009$.}
\tablefoottext{b}{The best-fitting results obtained by using a model based on the radiative transfer theory of \water ~masers 
for $\Gamma+\Gamma_{\nu}=1~\rm{s^{-1}}$ \citetalias{sur112}. The errors were determined by analyzing the full probability distribution 
function. For $T\sim1024$~K ($\Gamma_{\nu}=6~\rm{s}^{-1}$) \tbo ~has to be adjusted 
by adding $+0.8$~log~K~sr \citep{and93}.}
\tablefoottext{c}{The angle between the magnetic field and the maser propagation direction is determined by using the observed $P_{\rm{l}}$ 
and the fitted emerging brightness temperature. The errors were determined by analyzing the full probability distribution function.}
}
\label{VLA2.1_tab}
\end{table*}
\begin {table*}[th!]
\caption []{Parameters of all the 22 GHz \water ~maser features detected around VLA1 (epoch 2016.45).} 
\begin{center}
\scriptsize
\begin{tabular}{ l c c c c c c c c c c c c c c c}
\hline
\hline
\,\,\,\,\,(1)&(2)   & (3)      & (4)            & (5)       & (6)              & (7)         & (8)        & (9)                     & (10)                    & (11)                        & (12)         &(13)      & (14)                     \\
Maser     & RA\tablefootmark{a} & Dec\tablefootmark{a}  & Zone & Peak  & $V_{\rm{lsr}}$& $\Delta v\rm{_{L}}$ &$P_{\rm{l}}$ &$\chi$& $\Delta V_{\rm{i}}$\tablefootmark{b}& $T_{\rm{b}}\Delta\Omega$\tablefootmark{b}& $P_{\rm{V}}$ & $B_{||}$  &$\theta$\tablefootmark{c}\\
feature          &  offset &  offset  & & Intensity (I)     &           &                  &             &	          &                         &                         &              &                      &      \\ 
          &  (mas)  &  (mas)   & & (Jy/beam)      &  (km/s)   &      (km/s)      & (\%)        &   (\d)     & (km/s)                  & (log K sr)              &   ($\%$)     &  (mG)               &(\d)       \\ 
\hline
VLA1.2.01 & -18.441 &  1.366   & - & $57.07\pm1.42$ &  12.29    & 0.66             & $2.2\pm0.2$ & $-72\pm10$ & $3.8^{+0.1}_{-0.5}$     & $9.3^{+0.5}_{-1.6}$     & $3.5$        & $-676\pm102$        &$90^{+11}_{-11}$ \\ 
VLA1.2.02 & -11.915 & -9.304   & - & $0.29\pm0.02$  &  16.40    & 0.76             & $-$         & $-$        & $-$                     & $-$                     & $-$	     & $-$            	   &$-$ \\ 
VLA1.2.03 & -9.052  & -3.773   & - & $659.23\pm2.17$&  9.48     & 0.66             & $0.7\pm0.1$ & $-73\pm4$  & $3.9^{+0.2}_{-0.7}$     & $9.2^{+0.4}_{-1.2}$     & $-$	     & $-$            	   &$-$ \\
VLA1.2.04 & -4.210  &  0.237   & - & $13.25\pm0.42$ &  7.82     & 1.99             & $-$         & $-$        & $-$                     & $-$                     & $-$	     & $-$            	   &$-$ \\ 
VLA1.2.05 & -3.873  &  0.980   & - & $47.27\pm0.64$ &  8.98     & 0.75             & $1.3\pm0.1$ & $-76\pm15$ & $4.3^{+0.1}_{-0.6}$     & $6.8^{+0.2}_{-0.7}$     & $-$ 	     & $-$            	   &$90^{+11}_{-11}$ \\ 
VLA1.2.06 & -1.810  &  18.555  & - & $134.08\pm1.61$&  10.14    & 0.70             & $-$         & $-$        & $-$                     & $-$                     & $-$	     & $-$            	   &$-$ \\ 
VLA1.2.07 &  0      &  0       & - &$1619.87\pm4.42$&  10.92    & 1.19             & $0.8\pm0.1$ & $-59\pm3$  & $-$                     & $-$                     & $-$ 	     & $-$ 		   &$-$ \\ 
VLA1.2.08 &  11.452 &  12.955  & - & $25.57\pm0.86$ &  9.27     & 1.14             & $-$         & $-$        & $-$                     & $-$                     & $-$	     & $-$            	   &$-$ \\ 
VLA1.2.09 &  26.945 &  32.818  & - & $330.99\pm2.28$&  10.35    & 0.58             & $-$         & $-$        & $-$                     & $-$                     & $-$	     & $-$            	   &$-$ \\ 
VLA1.2.10 &  33.892 &  21.893  & - & $24.05\pm0.72$ &  13.43    & 3.68             & $-$         & $-$        & $-$                     & $-$                     & $-$	     & $-$            	   &$-$ \\ 
VLA1.2.11 &  35.366 &  44.006  & - & $10.46\pm0.23$ &  9.00     & 1.21             & $-$         & $-$        & $-$                     & $-$                     & $-$	     & $-$            	   &$-$ \\ 
VLA1.2.12 &  36.124 &  22.461  & - & $35.42\pm1.19$ &  9.29     & 0.81             & $-$         & $-$        & $-$                     & $-$                     & $-$ 	     & $-$            	   &$-$ \\ 
VLA1.2.13 &  37.723 &  21.629  & - & $412.53\pm1.48$&  12.69    & 1.10             & $1.0\pm0.1$ & $-65\pm16$ & $-$                     & $-$                     & $-$	     & $-$	  	   &$-$ \\ 
VLA1.2.14 &  38.144 &  21.221  & - & $98.05\pm4.33$ &  10.82    & 1.31             & $1.2\pm0.4$ & $-11\pm14$ & $-$                     & $-$                     & $-$	     & $-$            	   &$-$ \\ 
VLA1.2.15 &  41.765 &  25.604  & - & $28.31\pm0.73$ &  13.40    & 0.80             & $-$         & $-$        & $-$                     & $-$                     & $-$	     & $-$            	   &$-$ \\ 
VLA1.2.16 &  42.691 &  0.790   & - & $0.24\pm0.02$  &  2.68     & 0.49             & $-$         & $-$        & $-$                     & $-$                     & $-$	     & $-$            	   &$-$ \\ 
VLA1.2.17 &  47.112 &  24.559  & - & $150.64\pm0.68$&  13.51    & 1.05             & $2.1\pm0.5$ & $-56\pm3$  & $-$                     & $-$                     & $-$ 	     & $-$            	   &$-$ \\ 
VLA1.2.18 &  48.923 &  25.745  & - & $35.99\pm0.84$ &  13.56    & 0.91             & $-$         & $-$        & $-$                     & $-$                     & $-$	     & $-$            	   &$-$ \\ 
VLA1.2.19 &  49.301 &  24.933  & - & $4.04\pm0.64$  &  15.72    & 1.05             & $-$         & $-$        & $-$                     & $-$                     & $-$	     & $-$            	   &$-$ \\ 
VLA1.2.20 & 155.146 &  73.784  & - & $3.72\pm0.21$  &  13.64    & 1.09             & $-$         & $-$        & $-$                     & $-$                     & $-$	     & $-$            	   &$-$ \\ 
\hline
\end{tabular} \end{center}
\tablefoot{
\tablefoottext{a}{The reference position is $\alpha_{2000}=20^{\rm{h}}38^{\rm{m}}36^{\rm{s}}\!.43403\pm0^{\rm{s}}\!.00011$ and 
$\delta_{2000}=+42^{\circ}37'34''\!\!.8667\pm0''\!\!.0010$.}
\tablefoottext{b}{The best-fitting results obtained by using a model based on the radiative transfer theory of \water ~masers 
for $\Gamma+\Gamma_{\nu}=1~\rm{s^{-1}}$ \citetalias{sur112}. The errors were determined by analyzing the full probability distribution 
function. For $T\sim3136$~K ($\Gamma_{\nu}=17~\rm{s}^{-1}$) \tbo ~has to be adjusted by adding $+1.3$~log~K~sr \citep{and93}.}
\tablefoottext{c}{The angle between the magnetic field and the maser propagation direction is determined by using the observed $P_{\rm{l}}$ 
and the fitted emerging brightness temperature. The errors were determined by analyzing the full probability distribution function.}}
\label{VLA1.2_tab}
\end{table*}
\begin {table*}[th!]
\caption []{Parameters of all the 22 GHz \water ~maser features detected around VLA2 (epoch 2016.45).} 
\begin{center}
\scriptsize
\begin{tabular}{ l c c c c c c c c c c c c c c c}
\hline
\hline
\,\,\,\,\,(1)&(2)   & (3)      & (4)            & (5)       & (6)              & (7)         & (8)        & (9)                     & (10)                    & (11)                        & (12)         &(13)    &(14)                       \\
Maser     & RA\tablefootmark{a} & Dec\tablefootmark{a}  & Zone & Peak  & $V_{\rm{lsr}}$& $\Delta v\rm{_{L}}$ &$P_{\rm{l}}$ &$\chi$& $\Delta V_{\rm{i}}$\tablefootmark{b}& $T_{\rm{b}}\Delta\Omega$\tablefootmark{b}& $P_{\rm{V}}$ & $B_{||}$  &$\theta$\tablefootmark{c}\\
feature          &  offset &  offset  & & Intensity (I)     &           &                  &             &	          &                         &                         &              &                      &      \\ 
          &  (mas)  &  (mas)   & & (Jy/beam)      &  (km/s)   &      (km/s)      & (\%)        &   (\d)     & (km/s)                  & (log K sr)              &   ($\%$)     &  (mG)               &(\d)       \\ 
\hline
VLA2.2.01 & 420.978 & -843.044 & 4 & $0.77\pm0.02$  &  16.40    &      0.51        & $-$         & $-$       & $-$                     & $-$                     & $-$	    & $-$              	   &$-$ \\ 
VLA2.2.02 & 421.399 & -843.182 & 4 & $0.76\pm0.02$  &  16.50    &      0.55        & $-$         & $-$       & $-$                     & $-$                     & $-$	    & $-$              	   &$-$ \\ 
VLA2.2.03 & 430.619 & -818.573 & 4 & $1.21\pm0.02$  &  18.59    &      0.82        & $-$         & $-$       & $-$                     & $-$                     & $-$	    & $-$              	   &$-$ \\ 
VLA2.2.04 & 432.598 & -816.361 & 4 & $1.17\pm0.02$  &  20.67    &      0.71        & $-$         & $-$       & $-$                     & $-$                     & $-$	    & $-$              	   &$-$ \\ 
VLA2.2.05 & 451.207 & -805.458 & 4 & $9.17\pm0.04$  &  19.40    &      0.63        & $1.6\pm0.3$ & $-70\pm8$ & $3.9^{+0.2}_{-0.3}$     & $9.3^{+0.1}_{-2.3}$     & $-$	    & $-$              	   &$82^{+6}_{-12}$ \\ 
VLA2.2.06 & 472.848 & -870.491 & 3 & $2.55\pm0.17$  &  15.51    &      0.60        & $-$         & $-$       & $-$                     & $-$                     & $-$	    & $-$              	   &$-$ \\ 
VLA2.2.07 & 537.853 & -645.912 & 2 & $0.68\pm0.02$  &  5.63     &      0.80        & $-$         & $-$       & $-$                     & $-$                     & $-$	    & $-$              	   &$-$ \\ 
VLA2.2.08 & 547.705 & -607.678 & 2 & $1.01\pm0.02$  &  2.29     &      0.43        & $-$         & $-$       & $-$                     & $-$                     & $-$	    & $-$              	   &$-$ \\ 
VLA2.2.09 & 555.073 & -779.221 & 3 & $0.44\pm0.02$  &  17.27    &      0.65        & $-$         & $-$       & $-$                     & $-$                     & $-$	    & $-$              	   &$-$ \\ 
VLA2.2.10 & 558.609 & -756.298 & 3 & $0.31\pm0.03$  &  19.51    &      0.49        & $-$         & $-$       & $-$                     & $-$                     & $-$	    & $-$              	   &$-$ \\ 
VLA2.2.11 & 572.419 & -606.586 & 2 & $21.72\pm0.08$ &  1.71     &      0.54        & $1.1\pm0.2$ & $-76\pm10$& $3.6^{+0.1}_{-0.5}$     & $9.0^{+0.8}_{-0.3}$     & $-$	    & $-$              	   &$76^{+15}_{-34}$ \\ 
VLA2.2.12 & 575.913 & -790.714 & 3 & $10.59\pm0.52$ &  13.45    &      0.61        & $-$         & $-$       & $-$                     & $-$                     & $-$	    & $-$              	   &$-$ \\ 
VLA2.2.13 & 586.144 & -788.815 & 3 & $4.67\pm0.02$  &  4.87     &      0.71        & $1.4\pm0.2$ & $-77\pm9$ & $-$                     & $-$                     & $-$	    & $-$              	   &$-$ \\ 
VLA2.2.14 & 592.038 & -598.564 & 1 & $10.02\pm0.25$  &  8.08     &      0.61        & $-$         & $-$       & $-$                     & $-$                     & $-$	    & $-$              	   &$-$ \\ 
VLA2.2.15 & 604.037 & -603.630 & 1 & $7.59\pm0.04$  & -3.85     &      0.81        & $0.9\pm0.1$ & $+87\pm7$ & $4.0^{+0.1}_{-0.2}$     & $6.1^{+1.6}_{-0.1}$     & $-$	    & $-$              	   &$90^{+13}_{-13}$ \\ 
VLA2.2.16 & 618.310 & -615.192 & 1 & $73.11\pm0.28$ & -13.88    &      0.68        & $0.6\pm0.1$ & $-82\pm5$ & $4.2^{+0.1}_{-0.6}$     & $8.7^{+0.1}_{-1.5}$     & $-$	    & $-$              	   &$90^{+30}_{-30}$ \\ 
VLA2.2.17 & 629.004 & -617.607 & 1 & $84.57\pm0.94$ &  12.11    &      0.65        & $2.2\pm0.4$ & $-55\pm5$ & $3.5^{+0.1}_{-0.6}$     & $9.4^{+0.4}_{-2.0}$     & $4.9$	    & $-1498\pm225$   	   &$81^{+9}_{-10}$ \\ 
VLA2.2.18 & 645.508 & -780.159 & 3 & $1.61\pm0.02$  &  5.87     &      0.66        & $-$         & $-$       & $-$                     & $-$                     & $-$	    & $-$              	   &$-$ \\ 
VLA2.2.19 & 645.718 & -653.431 & 1 & $90.99\pm0.40$ &  8.71     &      0.53        & $1.3\pm0.6$ & $-59\pm7$ & $-$                     & $-$                     & $-$	    & $-$              	   &$-$ \\ 
VLA2.2.20 & 646.560 & -644.295 & 1 & $7.34\pm0.14$  &  15.56    &      0.47        & $-$         & $-$       & $-$                     & $-$                     & $-$	    & $-$              	   &$-$ \\ 
VLA2.2.21 & 647.908 & -653.736 & 1 & $30.24\pm1.18$ &  9.69     &      0.55        & $-$         & $-$       & $-$                     & $-$                     & $-$	    & $-$              	   &$-$ \\ 
VLA2.2.22 & 648.834 & -648.983 & 1 & $5.75\pm0.07$  &  15.74    &      0.51        & $2.7\pm0.7$ & $-63\pm19$& $<0.4$                  & $8.5^{+0.5}_{-1.0}$     & $-$	    & $-$              	   &$90^{+19}_{-19}$ \\ 
VLA2.2.23 & 650.181 & -677.208 & 1 & $4.29\pm0.02$  & -2.16     &      0.64        & $1.2\pm0.3$ & $+88\pm7$ & $3.6^{+0.1}_{-0.6}$     & $9.1^{+0.1}_{-2.1}$     & $-$	    & $-$              	   &$90^{+15}_{-15}$ \\ 
VLA2.2.24 & 650.771 & -678.692 & 1 & $4.99\pm0.04$  & -3.85     &      0.63        & $1.2\pm0.1$ & $+88\pm7$ & $3.6^{+0.4}_{-0.1}$     & $9.1^{+0.3}_{-0.1}$     & $-$	    & $-$              	   &$85^{+4}_{-9}$ \\ 
VLA2.2.25 & 651.360 & -646.961 & 1 & $3.63\pm0.12$  & 14.48     &      0.48        & $-$         & $-$       & $-$                     & $-$                     & $-$	    & $-$              	   &$-$ \\ 
VLA2.2.26 & 651.907 & -648.590 & 1 & $2.83\pm0.15$  & 14.40     &      0.79        & $-$         & $-$       & $-$                     & $-$                     & $-$	    & $-$              	   &$-$ \\ 
VLA2.2.27 & 653.255 & -655.125 & 1 & $52.19\pm0.23$ & 15.19     &      0.63        & $1.2\pm0.1$ & $-45\pm3$ & $3.7^{+0.2}_{-0.2}$     & $9.1^{+0.3}_{-0.1}$     & $7.8$	    & $-2426\pm364$   	   &$85^{+6}_{-9}$ \\ 
VLA2.2.28 & 655.570 & -687.443 & 1 & $7.77\pm0.04$  & -6.90     &      1.14        & $0.8\pm0.2$ & $-77\pm9$ & $-$                     & $-$                     & $-$	    & $-$              	   &$-$ \\ 
VLA2.2.29 & 656.202 & -695.667 & 1 & $1.08\pm0.02$  & -8.88     &      0.59        & $-$         & $-$       & $-$                     & $-$                     & $-$	    & $-$              	   &$-$ \\ 
VLA2.2.30 & 656.875 & -689.842 & 1 & $0.23\pm0.02$  & -5.11     &      0.86        & $-$         & $-$       & $-$                     & $-$                     & $-$	    & $-$              	   &$-$ \\ 
VLA2.2.31 & 659.865 & -709.724 & 1 & $0.35\pm0.02$  & 16.85     &      0.49        & $-$         & $-$       & $-$                     & $-$                     & $-$	    & $-$              	   &$-$ \\ 
VLA2.2.32 & 659.612 & -681.156 & 1 & $0.23\pm0.02$  & 0.71      &      0.83        & $-$         & $-$       & $-$                     & $-$                     & $-$	    & $-$              	   &$-$ \\ 
VLA2.2.33 & 661.212 & -698.853 & 1 & $0.58\pm0.02$  & -4.61     &      0.88        & $-$         & $-$       & $-$                     & $-$                     & $-$	    & $-$              	   &$-$ \\ 
VLA2.2.34 & 663.485 & -703.415 & 1 & $0.54\pm0.02$  & -4.59     &      0.76        & $-$         & $-$       & $-$                     & $-$                     & $-$	    & $-$              	   &$-$ \\ 
VLA2.2.35 & 665.085 & -703.720 & 1 & $0.38\pm0.03$  & -3.61     &      0.62        & $-$         & $-$       & $-$                     & $-$                     & $-$          & $-$                  & $-$ \\
VLA2.2.36 & 670.348 & -697.502 & 1 & $7.46\pm0.30$  & 13.95     &      0.49        & $-$         & $-$       & $-$                     & $-$                     & $-$	    & $-$              	   &$-$ \\ 
VLA2.2.37\tablefootmark{d} & 676.032 & -728.066 & 1 & $>30$ & -15.62    &      2.70        & $0.9\pm0.1$ & $+88\pm3$ & $-$                     & $-$                     & $-$	    & $-$              	   &$-$ \\ 
\hline
\end{tabular} \end{center}
\tablefoot{
\tablefoottext{a}{The reference position is $\alpha_{2000}=20^{\rm{h}}38^{\rm{m}}36^{\rm{s}}\!.43403\pm0^{\rm{s}}\!.00011$ and 
$\delta_{2000}=+42^{\circ}37'34''\!\!.8667\pm0''\!\!.0010$.}
\tablefoottext{b}{The best-fitting results obtained by using a model based on the radiative transfer theory of \water ~masers 
for $\Gamma+\Gamma_{\nu}=1~\rm{s^{-1}}$ \citetalias{sur112}. The errors were determined by analyzing the full probability distribution 
function. For $T\sim3364$~K ($\Gamma_{\nu}=17~\rm{s}^{-1}$) \tbo ~has to be adjusted by adding $+1.3$~log~K~sr \citep{and93}.}
\tablefoottext{c}{The angle between the magnetic field and the maser propagation direction is determined by using the observed $P_{\rm{l}}$ 
and the fitted emerging brightness temperature. The errors were determined by analyzing the full probability distribution function.}
\tablefoottext{d}{The values reported here are referred only to the available channels (see Sect.~\ref{res16})}.
}
\label{VLA2.2_tab}
\end{table*}
\begin {table*}[th!]
\caption []{Parameters of all the 22 GHz \water ~maser features detected around VLA1 (epoch 2018.44).} 
\begin{center}
\scriptsize
\begin{tabular}{ l c c c c c c c c c c c c c c c}
\hline
\hline
\,\,\,\,\,(1)&(2)   & (3)      & (4)            & (5)       & (6)              & (7)         & (8)        & (9)                     & (10)                    & (11)                        & (12)         &(13)      & (14)                     \\
Maser     & RA\tablefootmark{a} & Dec\tablefootmark{a}  & Zone & Peak  & $V_{\rm{lsr}}$& $\Delta v\rm{_{L}}$ &$P_{\rm{l}}$ &$\chi$& $\Delta V_{\rm{i}}$\tablefootmark{b}& $T_{\rm{b}}\Delta\Omega$\tablefootmark{b}& $P_{\rm{V}}$ & $B_{||}$  &$\theta$\tablefootmark{c}\\
feature          &  offset &  offset  & & Intensity (I)     &           &                  &             &	          &                         &                         &              &                      &      \\ 
          &  (mas)  &  (mas)   & & (Jy/beam)      &  (km/s)   &      (km/s)      & (\%)        &   (\d)     & (km/s)                  & (log K sr)              &   ($\%$)     &  (mG)               &(\d)       \\ 
\hline
VLA1.3.01 & -68.500 & -88.299  & - & $2.13\pm0.14$   &  12.90    &  0.82            & $-$         & $-$        & $-$                     & $-$                 & $-$             & $-$                 &$-$ \\ 
VLA1.3.02 & -27.156 & -18.986  & - & $36.76\pm0.53$  &  10.74    &  0.92            & $-$         & $-$        & $-$                     & $-$                 & $-$	     & $-$            	   &$-$ \\ 
VLA1.3.03 & -0.463  & -1.347   & - & $17.56\pm0.12$  &  11.82    &  1.29            & $1.1\pm0.1$ & $-83\pm2$  & $<0.4$                  & $9.0^{+0.3}_{-0.3}$ & $-$ 	     & $-$            	   &$90^{+7}_{-7}$ \\ 
VLA1.3.04 & 0       &  0       & - & $446.25\pm4.18$ &  8.63     &  1.10            & $0.7\pm0.1$ & $-88\pm6$  & $3.6^{+0.1}_{-0.6}$     & $9.1^{+0.4}_{-0.3}$ & $1.6$ 	     & $-764\pm115$    	   &$68^{+18}_{-33}$ \\  
VLA1.3.05 & 2.147   & -2.552   & - & $2.08\pm0.10$   &  11.98    &  1.45            & $-$         & $-$        & $-$                     & $-$                 & $-$	     & $-$            	   &$-$ \\ 
VLA1.3.06 & 3.705   &  5.379   & - & $283.86\pm1.56$ &  9.58     &  1.14            & $0.8\pm0.1$ & $+88\pm5$  & $3.9^{+0.2}_{-0.3}$     & $9.1^{+0.8}_{-0.2}$ & $-$ 	     & $-$            	   &$\mathbf{63^{+11}_{-40}}$ \\  
VLA1.3.07 & 3.705   &  5.184   & - & $265.98\pm1.73$ &  9.56     &  1.15            & $0.9\pm0.1$ & $+81\pm6$  & $3.8^{+0.1}_{-0.7}$     & $9.4^{+0.4}_{-0.7}$ & $-$ 	     & $-$            	   &$\mathbf{62^{+3}_{-43}}$ \\   
VLA1.3.08 & 10.526  &  12.749  & - & $21.66\pm0.18$  &  7.77     &  1.16            & $-$         & $-$        & $-$                     & $-$                 & $-$	     & $-$            	   &$-$ \\ 
VLA1.3.09 & 12.673  &  14.587  & - & $11.87\pm0.31$  &  11.45    &  0.80            & $1.0\pm0.1$ & $-82\pm3$  & $1.5^{+0.1}_{-0.3}$     & $9.0^{+0.3}_{-0.1}$ & $-$ 	     & $-$            	   &$85^{+5}_{-12}$ \\  
VLA1.3.10 & 19.872  &  18.551  & - & $8.22\pm0.14$   &  10.77    &  1.60            & $-$         & $-$        & $-$                     & $-$                 & $-$	     & $-$            	   &$-$ \\ 
VLA1.3.11 & 21.514  &  45.429  & - & $21.83\pm0.16$  &  14.96    &  0.92            & $10.6\pm0.2$& $+86\pm6$  & $1.1^{+0.3}_{-0.3}$     & $10.5^{+0.1}_{-0.3}$& $-$	     & $-$            	   &$90^{+6}_{-6}$ \\ 
VLA1.3.12 & 23.703  &  22.160  & - & $67.31\pm1.06$  &  9.98     &  1.01            & $0.8\pm0.3$ & $+88\pm7$  & $3.2^{+0.1}_{-0.5}$     & $8.9^{+0.2}_{-1.9}$ & $-$ 	     & $-$            	   &$90^{+41}_{-41}$ \\  
VLA1.3.13 & 62.479  &  31.822  & - & $7.38\pm0.18$   &  13.59    &  1.88            & $-$         & $-$        & $-$                     & $-$                 & $-$	     & $-$            	   &$-$ \\ 
VLA1.3.14 & 63.532  &  63.618  & - & $4.04\pm0.31$   &  11.32    &  0.93            & $-$         & $-$        & $-$                     & $-$                 & $-$	     & $-$            	   &$-$ \\ 
VLA1.3.15 & 63.827  &  65.319  & - & $48.05\pm0.57$  &  10.66    &  0.94            & $1.1\pm0.1$ & $-86\pm7$  & $2.7^{+0.2}_{-0.2}$     & $9.0^{+0.5}_{-0.3}$ & $-$ 	     & $-$            	   &$82^{+7}_{-13}$ \\  
VLA1.3.16 & 65.721  &  32.665  & - & $29.64\pm0.23$  &  13.30    &  1.10            & $0.9\pm0.1$ & $-88\pm4$  & $3.6^{+0.1}_{-0.5}$     & $8.9^{+0.3}_{-0.2}$ & $1.6$ 	     & $-733\pm113$    	   &$82^{+7}_{-13}$ \\  
VLA1.3.17 & 81.467  &  32.475  & - & $4.71\pm0.39$   &  11.09    &  0.82            & $-$         & $-$        & $-$                     & $-$                 & $-$	     & $-$            	   &$-$ \\ 
VLA1.3.18 & 84.120  &  31.231  & - & $33.10\pm0.42$  &  11.11    &  0.80            & $1.1\pm0.1$ & $-85\pm4$  & $1.6^{+0.1}_{-0.4}$     & $9.0^{+0.4}_{-0.1}$ & $-$ 	     & $-$            	   &$82^{+8}_{-31}$ \\  
VLA1.3.19 & 86.856  &  31.273  & - & $11.23\pm0.55$  &  10.80    &  0.87            & $1.7\pm0.1$ & $-88\pm7$  & $1.8^{+0.1}_{-0.3}$     & $9.3^{+0.3}_{-0.1}$ & $-$ 	     & $-$            	   &$78^{+1}_{-7}$ \\  
VLA1.3.20 & 94.814  &  42.259  & - & $18.33\pm0.22$  &  13.38    &  0.99            & $1.0\pm0.1$ & $-88\pm6$  & $2.8^{+0.2}_{-0.2}$     & $9.0^{+0.2}_{-0.2}$ & $-$             & $-$            	   &$84^{+6}_{-11}$ \\  
\hline
\end{tabular} \end{center}
\tablefoot{
\tablefoottext{a}{The reference position is $\alpha_{2000}=20^{\rm{h}}38^{\rm{m}}36^{\rm{s}}\!.43201\pm0^{\rm{s}}\!.00008$ and 
$\delta_{2000}=42^{\circ}37'34''\!\!.34.8588\pm0''\!\!.0009$.}
\tablefoottext{b}{The best-fitting results obtained by using a model based on the radiative transfer theory of \water ~masers 
for $\Gamma+\Gamma_{\nu}=1~\rm{s^{-1}}$ \citetalias{sur112}. The errors were determined 
by analyzing the full probability distribution function. For $T\sim 2916$~K ($\Gamma_{\nu}=15~\rm{s}^{-1}$) \tbo ~has to be adjusted 
by adding $+1.2$~log~K~sr \citep{and93}.}
\tablefoottext{c}{The angle between the magnetic field and the maser propagation direction is determined by using the observed $P_{\rm{l}}$ 
and the fitted emerging brightness temperature. The errors were determined by analyzing the full probability distribution function. The boldface indicates that $|\theta^{\rm{+}}-55$\d$|<|\theta^{\rm{-}}-55$\d$|$, i.e., the magnetic field is parallel to the linear polarization vector (see Sect.~\ref{obssect}).}
}
\label{VLA1.3_tab}
\end{table*}
\begin {table*}[th!]
\caption []{Parameters of all the 22 GHz \water ~maser features detected around VLA2 (epoch 2018.44).} 
\begin{center}
\scriptsize
\begin{tabular}{ l c c c c c c c c c c c c c c c}
\hline
\hline
\,\,\,\,\,(1)&(2)   & (3)      & (4)            & (5)       & (6)              & (7)         & (8)        & (9)                     & (10)                    & (11)                        & (12)         &(13)    & (14)                       \\
Maser     & RA\tablefootmark{a} & Dec\tablefootmark{a}  & Zone & Peak  & $V_{\rm{lsr}}$& $\Delta v\rm{_{L}}$ &$P_{\rm{l}}$ &$\chi$& $\Delta V_{\rm{i}}$\tablefootmark{b}& $T_{\rm{b}}\Delta\Omega$\tablefootmark{b}& $P_{\rm{V}}$ & $B_{||}$  &$\theta$\tablefootmark{c}\\
feature          &  offset &  offset  & & Intensity (I)     &           &                  &             &	          &                         &                         &              &                      &      \\ 
          &  (mas)  &  (mas)   & & (Jy/beam)      &  (km/s)   &      (km/s)      & (\%)        &   (\d)     & (km/s)                  & (log K sr)              &   ($\%$)     &  (mG)               &(\d)       \\ 
\hline
VLA2.3.01 & 434.956 & -754.345 & 4 & $0.04\pm0.01$   &  28.23    & 0.55             & $-$         & $-$       & $-$                     & $-$                     & $-$	    & $-$              	   &$-$ \\ 
VLA2.3.02 & 436.429 & -852.776 & 4 & $1.06\pm0.02$   &  18.59    & 0.73             & $-$         & $-$       & $-$                     & $-$                     & $-$	    & $-$              	   &$-$ \\ 
VLA2.3.03 & 439.334 & -816.387 & 4 & $15.90\pm0.06$  &  17.88    & 0.75             & $1.2\pm01$  & $+85\pm3$ & $2.7^{+0.2}_{-0.5}$     & $9.4^{+0.5}_{-0.6}$     & $-$	    & $-$              	   &$70^{+15}_{-36}$ \\ 
VLA2.3.04 & 443.923 & -853.413 & 4 & $0.04\pm0.01$   &  27.59    & 0.77             & $-$         & $-$       & $-$                     & $-$                     & $-$	    & $-$              	   &$-$ \\ 
VLA2.3.05 & 548.252 & -713.097 & 2 & $2.18\pm0.02$   &  22.25    & 0.60             & $-$         & $-$       & $-$                     & $-$                     & $-$	    & $-$              	   &$-$ \\
VLA2.3.06 & 549.642 & -638.168 & 2 & $5.95\pm0.03$   &  6.27     & 0.62             & $1.2\pm0.1$ & $-89\pm4$ & $2.5^{+0.2}_{-0.2}$     & $8.7^{+0.5}_{-1.0}$     & $-$	    & $-$              	   &$90^{+13}_{-13}$ \\ 
VLA2.3.07 & 549.894 & -639.736 & 2 & $2.55\pm0.03$   &  5.92     & 0.62             & $-$         & $-$       & $-$                     & $-$                     & $-$	    & $-$              	   &$-$ \\
VLA2.3.08 & 556.546 & -603.367 & 2 & $0.58\pm0.01$   &  3.53     & 0.50             & $-$         & $-$       & $-$                     & $-$                     & $-$	    & $-$              	   &$-$ \\
VLA2.3.09 & 563.156 & -848.869 & 3 & $4.85\pm0.13$   &  14.69    & 0.56             & $-$         & $-$       & $-$                     & $-$                     & $-$	    & $-$              	   &$-$ \\
VLA2.3.10 & 581.302 & -784.256 & 3 & $6.26\pm0.15$   &  14.80    & 0.94             & $1.3\pm0.2$ & $+85\pm11$& $3.4^{+0.2}_{-0.3}$     & $6.0^{+2.0}_{-0.2}$     & $-$	    & $-$              	   &$90^{+11}_{-11}$ \\ 
VLA2.3.11 & 581.429 & -605.461 & 2 & $6.23\pm0.11$   &  1.79     & 0.49             & $-$         & $-$       & $-$                     & $-$                     & $-$	    & $-$              	   &$-$ \\
VLA2.3.12 & 582.186 & -784.851 & 3 & $3.86\pm0.17$   &  13.43    & 0.73             & $-$         & $-$       & $-$                     & $-$                     & $-$	    & $-$              	   &$-$ \\
VLA2.3.13 & 602.311 & -803.642 & 3 & $9.25\pm0.10$   &  12.80    & 0.63             & $-$         & $-$       & $-$                     & $-$                     & $-$	    & $-$              	   &$-$ \\
VLA2.3.14 & 606.185 & -594.807 & 1 & $128.02\pm1.00$  &  8.19     & 0.47             & $-$         & $-$       & $-$                     & $-$                     & $2.9\tablefootmark{d}$ & $+439\pm66\tablefootmark{d}$  &$-$ \\ 
VLA2.3.15 & 607.827 & -596.924 & 1 & $0.40\pm0.01$   & -3.37     & 0.69             & $-$         & $-$       & $-$                     & $-$                     & $-$	    & $-$              	   &$-$ \\
VLA2.3.16 & 613.005 & -597.683 & 1 & $2.62\pm0.02$   & -4.87     & 0.70             & $-$         & $-$       & $-$                     & $-$                     & $-$	    & $-$              	   &$-$ \\
VLA2.3.17 & 613.089 & -592.804 & 1 & $4.30\pm0.06$   &  12.38    & 0.53             & $-$         & $-$       & $-$                     & $-$                     & $-$	    & $-$              	   &$-$ \\
VLA2.3.18 & 614.226 & -598.274 & 1 & $0.35\pm0.02$   & -5.66     & 1.08             & $-$         & $-$       & $-$                     & $-$                     & $-$	    & $-$                  & $-$ \\
VLA2.3.19 & 615.868 & -593.662 & 1 & $4.33\pm0.07$   &  12.03    & 0.86             & $-$         & $-$       & $-$                     & $-$                     & $-$	    & $-$              	   &$-$ \\
VLA2.3.20 & 616.036 & -593.086 & 1 & $2.23\pm0.10$   &  13.72    & 0.65             & $-$         & $-$       & $-$                     & $-$                     & $-$	    & $-$                  & $-$ \\
VLA2.3.21 & 622.562 & -595.924 & 1 & $4.46\pm0.12$   &  15.19    & 0.51             & $2.3\pm0.5$ & $+90\pm7$ & $1.6^{+0.1}_{-0.4}$     & $6.0^{+2.3}_{-0.2}$     & $-$	    & $-$              	   &$90^{+17}_{-17}$ \\  
VLA2.3.22 & 632.835 & -609.512 & 1 & $1.13\pm0.01$   & -13.88    & 0.79             & $-$         & $-$       & $-$                     & $-$                     & $-$	    & $-$              	   &$-$ \\
VLA2.3.23 & 636.540 & -613.735 & 1 & $8.40\pm0.04$   & -15.85    & 0.58             & $1.0\pm0.1$ & $+85\pm4$ & $0.8^{+0.1}_{-0.1}$     & $10.5^{+0.1}_{-0.2}$    & $-$	    & $-$              	   &$\mathbf{14^{+37}_{-1}}$ \\  
VLA2.3.24 & 638.561 & -614.048 & 1 & $1.34\pm0.02$   & -11.77    & 1.02             & $-$         & $-$       & $-$                     & $-$                     & $-$	    & $-$ 		   & $-$\\
VLA2.3.25 & 649.760 & -623.409 & 1 & $0.48\pm0.01$   & -12.09    & 0.62             & $-$         & $-$       & $-$                     & $-$                     & $-$	    & $-$ 		   & $-$\\
VLA2.3.26 & 652.076 & -644.253 & 1 & $2.34\pm0.02$   & -4.24     & 0.67             & $-$         & $-$       & $-$                     & $-$                     & $-$	    & $-$		   & $-$ \\
VLA2.3.27 & 653.970 & -646.782 & 1 & $48.41\pm0.19$  &  2.74     & 0.62             & $1.1\pm0.5$ & $+70\pm6$ & $2.9^{+0.1}_{-0.5}$     & $8.2^{+0.4}_{-1.7}$     & $-$	    & $-$              	   &$90^{+54}_{-54}$ \\  
VLA2.3.28 & 654.391 & -648.689 & 1 & $21.67\pm0.12$  &  1.74     & 0.67             & $1.6\pm0.2$ & $+75\pm6$ & $3.4^{+0.3}_{-0.5}$     & $8.5^{+0.1}_{-2.28}$    & $-$	    & $-$              	   &$\mathbf{58^{+32}_{-46}}$ \\ 
VLA2.3.29 & 658.138 & -656.162 & 1 & $1.68\pm0.02$   & -4.63     & 1.18             & $-$         & $-$       & $-$                     & $-$                     & $-$	    & $-$              	   &$-$ \\
VLA2.3.30 & 658.560 & -651.417 & 1 & $61.99\pm1.22$  &  9.74     & 0.67             & $1.3\pm1.2$ & $+90\pm27$& $3.6^{+0.1}_{-0.5}$     & $6.2^{+0.7}_{-0.2}$     & $-$	    & $-$              	   &$90^{+62}_{-62}$ \\
VLA2.3.31 & 659.023 & -649.757 & 1 & $15.90\pm0.42$  &  10.56    & 0.47             & $-$         & $-$       & $-$                     & $-$                     & $-$	    & $-$              	   &$-$ \\
VLA2.3.32 & 660.580 & -782.860 & 3 & $0.91\pm0.02$   &  5.90     & 1.04             & $-$         & $-$       & $-$                     & $-$                     & $-$	    & $-$              	   &$-$ \\
VLA2.3.33 & 662.349 & -784.054 & 3 & $0.98\pm0.02$   &  5.21     & 0.67             & $-$         & $-$       & $-$                     & $-$                     & $-$	    & $-$              	   &$-$ \\
VLA2.3.34 & 663.780 & -642.242 & 1 & $3.84\pm0.11$   &  15.54    & 0.49             & $-$         & $-$       & $-$                     & $-$                     & $-$	    & $-$              	   &$-$ \\
VLA2.3.35 & 666.938 & -664.009 & 1 & $7.33\pm0.13$   &  7.21     & 0.60             & $-$         & $-$       & $-$                     & $-$                     & $-$         & $-$                  &$-$ \\
VLA2.3.36 & 669.043 & -668.186 & 1 & $0.94\pm0.02$   &  0.13     & 0.50             & $-$         & $-$       & $-$                     & $-$                     & $-$	    & $-$              	   &$-$ \\
VLA2.3.37 & 669.885 & -668.453 & 1 & $4.71\pm0.03$   & -2.08     & 0.62             & $-$         & $-$       & $-$                     & $-$                     & $-$	    & $-$              	   &$-$ \\
VLA2.3.38 & 670.390 & -657.791 & 1 & $1.94\pm0.04$   &  14.24    & 0.40             & $-$         & $-$       & $-$                     & $-$                     & $-$	    & $-$              	   &$-$ \\
VLA2.3.39 & 670.769 & -655.510 & 1 & $10.93\pm0.13$  &  14.67    & 0.55             & $-$         & $-$       & $-$                     & $-$                     & $1.5\tablefootmark{e}$ & $-355\pm69\tablefootmark{e}$   &$-$ \\ 
VLA2.3.40 & 675.232 & -666.389 & 1 & $3.59\pm0.04$   &  8.74     & 0.91             & $-$         & $-$       & $-$                     & $-$                     & $-$	    & $-$              	   &$-$ \\
VLA2.3.41 & 675.358 & -673.714 & 1 & $53.39\pm0.40$  &  1.21     & 0.64             & $2.2\pm0.2$ & $+65\pm1$ & $2.4^{+0.2}_{-0.4}$     & $9.5^{+0.7}_{-1.2}$     & $-$         & $-$                  &$73^{+15}_{-34}$ \\
VLA2.3.42 & 677.969 & -675.056 & 1 & $7.74\pm0.22$   &  1.21     & 0.64             & $1.8\pm0.1$ & $+67\pm7$ & $2.6^{+0.2}_{-0.2}$     & $8.7^{+0.7}_{-1.0}$     & $-$	    & $-$              	   &$90^{+17}_{-17}$ \\ 
VLA2.3.43 & 681.758 & -700.119 & 1 & $0.73\pm0.01$   & -7.95     & 3.32             & $-$         & $-$       & $-$                     & $-$                     & $-$	    & $-$              	   &$-$ \\
VLA2.3.44 & 682.431 & -700.844 & 1 & $3.69\pm0.02$   & -7.87     & 1.14             & $-$         & $-$       & $-$                     & $-$                     & $-$	    & $-$              	   &$-$ \\
\hline                                                           
\end{tabular} \end{center}                                       
\tablefoot{
\tablefoottext{a}{The reference position is $\alpha_{2000}=20^{\rm{h}}38^{\rm{m}}36^{\rm{s}}\!.43201\pm0^{\rm{s}}\!.00008$ and 
$\delta_{2000}=42^{\circ}37'34''\!\!.8588\pm0''\!\!.0009$.}
\tablefoottext{b}{The best-fitting results obtained by using a model based on the radiative transfer theory of \water ~masers 
for $\Gamma+\Gamma_{\nu}=1~\rm{s^{-1}}$ \citetalias{sur112}. The errors were determined 
by analyzing the full probability distribution function. For $T\sim1764$~K ($\Gamma_{\nu}=9~\rm{s}^{-1}$) \tbo ~has to be adjusted 
by adding $+1.0$~log~K~sr \citep{and93}).}
\tablefoottext{c}{The angle between the magnetic field and the maser propagation direction is determined by using the observed $P_{\rm{l}}$ 
and the fitted emerging brightness temperature. The errors were determined by analyzing the full probability distribution function. The boldface indicates that $|\theta^{\rm{+}}-55$\d$|<|\theta^{\rm{-}}-55$\d$|$, that is, the magnetic field is parallel to the linear polarization vector (see Sect.~\ref{obssect}).}
\tablefoottext{d}{In the fitting model we include the values \tbo$~=1.0\times10^6$~K~sr and 
\dvi$~=1.4$~\kms ~that best fit the total intensity emission.}
\tablefoottext{e}{In the fitting model we include the values \tbo$~=3.2\times10^9$~K~sr and 
\dvi$~=2.0$~\kms ~that best fit the total intensity emission.}
}
\label{VLA2.3_tab}
\end{table*}
\begin {table*}[t!]
\caption []{Parameters of all the 22 GHz \water ~maser features detected around VLA1 (epoch 2020.82).} 
\begin{center}
\scriptsize
\begin{tabular}{ l c c c c c c c c c c c c c}
\hline
\hline
\,\,\,\,\,(1)&(2)   & (3)      & (4)             & (5)       & (6)              & (7)         & (8)        & (9)                     & (10)                    & (11)                        & (12)         &(13)   & (14)                        \\
Maser     & RA\tablefootmark{a} & Dec\tablefootmark{a}  & Zone & Peak & $V_{\rm{lsr}}$& $\Delta v\rm{_{L}}$ &$P_{\rm{l}}$ &$\chi$& $\Delta V_{\rm{i}}$\tablefootmark{b}& $T_{\rm{b}}\Delta\Omega$\tablefootmark{b}& $P_{\rm{V}}$ & $B_{||}$  &$\theta$\tablefootmark{c}\\
feature          &  offset &  offset  & & Intensity (I)     &           &                  &             &	          &                         &                         &              &                      &      \\ 
          &  (mas)  &  (mas)   & & (Jy/beam)      &  (km/s)   &      (km/s)      & (\%)        &   (\d)     & (km/s)                  & (log K sr)              &   ($\%$)     &  (mG)               &(\d)       \\ 
\hline
VLA1.4.01 & -8.378  & -6.729   & - & $0.37\pm0.02$   & 19.03     & 0.55             & $-$         & $-$        & $-$                     & $-$                     & $-$          & $-$                &$-$ \\ 
VLA1.4.02 & -5.894  & -3.410   & - & $114.86\pm3.00$ & 8.84      & 1.22             & $0.3\pm0.1$ & $-67\pm19$ & $2.6^{+0.1}_{-0.9}$     & $10.6^{+0.1}_{-0.1}$    & $-$	      & $-$            	   &$\mathbf{54^{+2}_{-48}}$ \\ 
VLA1.4.03 & -5.852  & -3.094   & - & $0.31\pm0.03$   & 25.91     & 3.36             & $-$         & $-$        & $-$                     & $-$                     & $-$ 	      & $-$            	   &$-$ \\ 
VLA1.4.04 &  0      & -2.228   & - & $0.44\pm0.01$   & 20.32     & 0.64             & $-$         & $-$        & $-$                     & $-$                     & $-$ 	      & $-$       	   &$-$ \\  
VLA1.4.05 &  0      &  0       & - & $302.47\pm2.59$ & 9.87      & 0.97             & $0.1\pm0.1$ & $-5\pm15$  & $2.6^{+0.1}_{-0.9}$     & $10.6^{+0.1}_{-0.1}$    & $-$	      & $-$            	   &$\mathbf{5^{+51}_{-1}}$ \\ 
VLA1.4.06 & 38.271  &  40.607  & - & $1.79\pm0.02$   & 18.98     & 0.59             & $-$         & $-$        & $-$                     & $-$                     & $-$ 	      & $-$            	   &$-$ \\  
VLA1.4.07 & 40.923  &  44.167  & - & $1.10\pm0.03$   & 26.01     & 3.43             & $-$         & $-$        & $-$                     & $-$                     & $-$ 	      & $-$            	   &$-$ \\   
VLA1.4.08 & 43.786  &  43.541  & - & $0.25\pm0.02$   & 18.98     & 0.58             & $-$         & $-$        & $-$                     & $-$                     & $-$	      & $-$            	   &$-$ \\ 
VLA1.4.09 & 46.523  &  47.241  & - & $0.14\pm0.02$   & 26.35     & 4.20             & $-$         & $-$        & $-$                     & $-$                     & $-$ 	      & $-$            	   &$-$ \\  
VLA1.4.10 & 46.817  &  47.165  & - & $160.64\pm1.83$ & 10.87     & 1.59             & $-$         & $-$        & $-$                     & $-$                     & $-$	      & $-$            	   &$-$ \\ 
\hline
\end{tabular} \end{center}
\tablefoot{
\tablefoottext{a}{The reference position is $\alpha_{2000}=20^{\rm{h}}38^{\rm{m}}36^{\rm{s}}\!.43111\pm0^{\rm{s}}\!.00009$ and 
$\delta_{2000}=42^{\circ}37'34''\!\!.34.8794\pm0''\!\!.0009$.}
\tablefoottext{b}{The best-fitting results obtained by using a model based on the radiative transfer theory of \water ~masers 
for $\Gamma+\Gamma_{\nu}=1~\rm{s^{-1}}$ \citetalias{sur112}. The errors were determined 
by analyzing the full probability distribution function. For $T\sim2704$~K ($\Gamma_{\nu}=14~\rm{s}^{-1}$) \tbo ~has to be adjusted 
by adding $+1.2$~log~K~sr \citep{and93}.}
\tablefoottext{c}{The angle between the magnetic field and the maser propagation direction is determined by using the observed $P_{\rm{l}}$ 
and the fitted emerging brightness temperature. The errors were determined by analyzing the full probability distribution function. The boldface indicates that $|\theta^{\rm{+}}-55$\d$|<|\theta^{\rm{-}}-55$\d$|$, that is, the magnetic field is parallel to the linear polarization vector (see Sect.~\ref{obssect}).}
}
\label{VLA1.4_tab}
\end{table*}
\begin {table*}[t!]
\caption []{Parameters of all the 22 GHz \water ~maser features detected around VLA2 (epoch 2020.82).} 
\begin{center}
\scriptsize
\begin{tabular}{ l c c c c c c c c c c c c c}
\hline
\hline
\,\,\,\,\,(1)&(2)   & (3)      & (4)             & (5)       & (6)              & (7)         & (8)       & (9)                     & (10)                    & (11)                        & (12)         &(13)        &(14)                   \\
Maser     & RA\tablefootmark{a} & Dec\tablefootmark{a}  & Zone & Peak   & $V_{\rm{lsr}}$& $\Delta v\rm{_{L}}$ &$P_{\rm{l}}$ &$\chi$& $\Delta V_{\rm{i}}$\tablefootmark{b}& $T_{\rm{b}}\Delta\Omega$\tablefootmark{b}& $P_{\rm{V}}$ & $B_{||}$  &$\theta$\tablefootmark{c}\\
feature          &  offset &  offset  & & Intensity (I)     &           &                  &             &	          &                         &                         &              &                      &      \\ 
          &  (mas)  &  (mas)   & & (Jy/beam)      &  (km/s)   &      (km/s)      & (\%)        &   (\d)     & (km/s)                  & (log K sr)              &   ($\%$)     &  (mG)               &(\d)       \\ 
\hline
VLA2.4.01 & 406.368 & -818.092 & 4 & $0.42\pm0.04$   & 27.14     & 0.61             & $-$         & $-$        & $-$                     & $-$                 & $-$            & $-$                  &$-$ \\ 
VLA2.4.02 & 408.642 & -826.874 & 4 & $0.35\pm0.02$   & 21.22     & 0.61             & $-$         & $-$        & $-$                     & $-$                 & $-$	    & $-$            	   &$-$ \\ 
VLA2.4.03 & 410.242 & -845.112 & 4 & $0.37\pm0.01$   & 18.43     & 0.65             & $-$         & $-$        & $-$                     & $-$                 & $-$ 	    & $-$            	   &$-$ \\ 
VLA2.4.04 & 411.168 & -775.082 & 4 & $0.08\pm0.01$   & 27.09     & 0.55             & $-$         & $-$        & $-$                     & $-$                 & $-$ 	    & $-$       	   &$-$ \\  
VLA2.4.05 & 411.926 & -815.075 & 4 & $0.05\pm0.01$   & 27.22     & 0.52             & $-$         & $-$        & $-$                     & $-$                 & $-$	    & $-$            	   &$-$ \\ 
VLA2.4.06 & 417.020 & -891.735 & 4 & $0.17\pm0.02$   & 18.66     & 0.56             & $-$         & $-$        & $-$                     & $-$                 & $-$ 	    & $-$            	   &$-$ \\  
VLA2.4.07 & 480.258 & -642.651 & 2 & $1.57\pm0.18$   & 5.63      & 1.60             & $1.9\pm0.6$ & $-54\pm14$ & $2.6^{+0.4}_{-0.2}$     & $8.4^{+0.4}_{-0.8}$ & $-$ 	    & $-$            	   &$90^{+38}_{-38}$ \\   
VLA2.4.08 & 514.318 & -899.658 & 3 & $0.46\pm0.05$   & 1.10      & 1.29             & $-$         & $-$        & $-$                     & $-$                 & $-$	    & $-$            	   &$-$ \\ 
VLA2.4.09 & 530.696 & -737.679 & 2 & $0.68\pm0.02$   & 21.69     & 0.45             & $-$         & $-$        & $-$                     & $-$                 & $-$ 	    & $-$            	   &$-$ \\  
VLA2.4.10 & 531.033 & -668.739 & 2 & $8.68\pm0.31$   & 5.31      & 0.54             & $-$         & $-$        & $-$                     & $-$                 & $-$	    & $-$            	   &$-$ \\ 
VLA2.4.11 & 542.611 & -633.331 & 2 & $0.86\pm0.05$   & 0.84      & 0.48             & $-$         & $-$        & $-$                     & $-$                 & $-$            & $-$                  &$-$ \\ 
VLA2.4.12 & 546.231 & -663.677 & 2 & $1.85\pm0.21$   & 12.77     & 0.55             & $-$         & $-$        & $-$                     & $-$                 & $-$	    & $-$            	   &$-$ \\ 
VLA2.4.13 & 546.779 & -629.688 & 2 & $0.39\pm0.05$   & 0.65      & 0.46             & $-$         & $-$        & $-$                     & $-$                 & $-$ 	    & $-$            	   &$-$ \\ 
VLA2.4.14 & 556.209 & -700.676 & 2 & $10.30\pm0.46$  & 15.40     & 0.55             & $1.5\pm0.5$ & $-28\pm17$ & $2.9^{+0.4}_{-0.2}$     & $8.3^{+0.4}_{-1.3}$ & $-$ 	    & $-$       	   &$90^{+31}_{-31}$ \\  
VLA2.4.15 & 574.355 & -619.995 & 2 & $6.80\pm0.11$   & 9.29      & 2.71             & $-$         & $-$        & $-$                     & $-$                 & $-$	    & $-$            	   &$-$ \\ 
VLA2.4.16 & 575.155 & -621.258 & 2 & $3.53\pm0.09$   & 4.79      & 0.52             & $-$         & $-$        & $-$                     & $-$                 & $-$ 	    & $-$            	   &$-$ \\  
VLA2.4.17 & 576.418 & -808.151 & 3 & $3.66\pm0.25$   & 5.00      & 0.60             & $-$         & $-$        & $-$                     & $-$                 & $-$ 	    & $-$            	   &$-$ \\   
VLA2.4.18 & 577.724 & -621.490 & 2 & $49.34\pm0.51$  & 5.31      & 0.56             & $0.2\pm0.1$ & $-10\pm22$ & $3.6^{+0.1}_{-0.5}$     & $8.3^{+0.6}_{-1.0}$ & $-$	    & $-$            	   &$74^{+14}_{-36}$ \\ 
VLA2.4.19 & 579.281 & -621.262 & 2 & $1.05\pm0.02$   & 6.26      & 0.55             & $-$         & $-$        & $-$                     & $-$                 & $-$ 	    & $-$            	   &$-$ \\  
VLA2.4.20 & 583.281 & -618.427 & 2 & $6.63\pm0.09$   & 5.34      & 0.55             & $1.9\pm0.7$ & $+26\pm27$ & $2.7^{+0.3}_{-0.3}$     & $8.3^{+0.4}_{-1.3}$ & $-$	    & $-$            	   &$90^{+54}_{-54}$ \\ 
VLA2.4.21 & 584.039 & -621.773 & 2 & $0.14\pm0.01$   & -1.03     & 0.45             & $-$         & $-$        & $-$                     & $-$                 & $-$            & $-$                  &$-$ \\ 
VLA2.4.22 & 584.081 & -620.522 & 2 & $50.25\pm1.80$  & 9.21      & 0.46             & $-$         & $-$        & $-$                     & $-$                 & $-$	    & $-$            	   &$-$ \\ 
VLA2.4.23 & 593.259 & -616.524 & 2 & $11.69\pm0.23$  & 12.90     & 0.55             & $-$         & $-$        & $-$                     & $-$                 & $-$ 	    & $-$            	   &$-$ \\ 
VLA2.4.24 & 610.184 & -630.386 & 1 & $1.47\pm0.02$   & 6.47      & 0.61             & $-$         & $-$        & $-$                     & $-$                 & $-$	    & $-$            	   &$-$ \\ 
VLA2.4.25 & 644.329 & -668.327 & 1 & $4.41\pm0.23$   & 16.45     & 0.47             & $-$         & $-$        & $-$                     & $-$                 & $-$            & $-$                  &$-$ \\ 
VLA2.4.26 & 644.792 & -670.151 & 1 & $0.76\pm0.02$   & 16.93     & 0.37             & $-$         & $-$        & $-$                     & $-$                 & $-$	    & $-$            	   &$-$ \\ 
VLA2.4.27 & 646.897 & -675.522 & 1 & $285.16\pm4.13$ & 9.24      & 0.61             & $-$         & $-$        & $-$                     & $-$                 & $1.3\tablefootmark{d}$ & $-452\pm68\tablefootmark{d}$            	   &$-$ \\ 
VLA2.4.28 & 646.939 & -668.846 & 1 & $19.99\pm0.29$  & 16.29     & 0.55             & $-$         & $-$        & $-$                     & $-$                 & $-$ 	    & $-$       	   &$-$ \\  
VLA2.4.29 & 648.623 & -671.463 & 1 & $11.60\pm0.26$  & 16.14     & 0.55             & $-$         & $-$        & $-$                     & $-$                 & $-$	    & $-$            	   &$-$ \\ 
VLA2.4.30 & 653.507 & -681.347 & 1 & $8.14\pm0.26$   & 14.87     & 0.96             & $-$         & $-$        & $-$                     & $-$                 & $-$ 	    & $-$            	   &$-$ \\  
VLA2.4.31 & 653.844 & -679.752 & 1 & $136.56\pm1.49$ & 15.40     & 0.58             &$0.15\pm0.03$& $+40\pm17$ & $3.8^{+0.1}_{-0.3}$     & $8.2^{+0.8}_{-0.3}$ & $-$ 	    & $-$            	   &$\mathbf{66^{+9}_{-43}}$ \\   
VLA2.4.32 & 653.844 & -683.891 & 1 & $2.40\pm0.09$   & 14.19     & 0.42             & $-$         & $-$        & $-$                     & $-$                 & $-$	    & $-$            	   &$-$ \\ 
VLA2.4.33 & 659.949 & -680.241 & 1 & $0.83\pm0.08$   & 14.53     & 0.48             & $-$         & $-$        & $-$                     & $-$                 & $-$ 	    & $-$            	   &$-$ \\  
VLA2.4.34 & 663.022 & -691.490 & 1 & $3.26\pm0.08$   & 10.26     & 2.85             & $-$         & $-$        & $-$                     & $-$                 & $-$	    & $-$            	   &$-$ \\ 
VLA2.4.35 & 664.412 & -690.178 & 1 & $4.98\pm0.07$   & 14.37     & 0.49             & $-$         & $-$        & $-$                     & $-$                 & $-$            & $-$                  &$-$ \\ 
VLA2.4.36 & 667.611 & -694.836 & 1 & $47.57\pm0.61$  & 13.45     & 0.70             & $-$         & $-$        & $-$                     & $-$                 & $-$	    & $-$            	   &$-$ \\ 
VLA2.4.37 & 669.127 & -710.930 & 1 & $2.83\pm0.21$   & 12.58     & 0.54             & $-$         & $-$        & $-$                     & $-$                 & $-$ 	    & $-$            	   &$-$ \\   
VLA2.4.38 & 670.221 & -711.891 & 1 & $5.64\pm0.25$   & 12.55     & 0.38             & $-$         & $-$        & $-$                     & $-$                 & $-$	    & $-$            	   &$-$ \\ 
VLA2.4.39 & 693.378 & -620.022 & 1 & $1.94\pm0.08$   & 16.35     & 1.65             & $-$         & $-$        & $-$                     & $-$                 & $-$ 	    & $-$       	   &$-$ \\
\hline                                                           
\end{tabular} \end{center}                                       
\tablefoot{
\tablefoottext{a}{The reference position is $\alpha_{2000}=20^{\rm{h}}38^{\rm{m}}36^{\rm{s}}\!.43111\pm0^{\rm{s}}\!.00009$ and 
$\delta_{2000}=42^{\circ}37'34''\!\!.8794\pm0''\!\!.0009$.}
\tablefoottext{b}{The best-fitting results obtained by using a model based on the radiative transfer theory of \water ~masers 
for $\Gamma+\Gamma_{\nu}=1~\rm{s^{-1}}$ \citetalias{sur112}. The errors were determined 
by analyzing the full probability distribution function. For $T\sim3364$~K ($\Gamma_{\nu}=17~\rm{s}^{-1}$) \tbo ~has to be adjusted 
by adding $+1.3$~log~K~sr \citep{and93}.}
\tablefoottext{c}{The angle between the magnetic field and the maser propagation direction is determined by using the observed $P_{\rm{l}}$ 
and the fitted emerging brightness temperature. The errors were determined by analyzing the full probability distribution function. The boldface indicates that $|\theta^{\rm{+}}-55$\d$|<|\theta^{\rm{-}}-55$\d$|$, that is, the magnetic field is parallel to the linear polarization vector (see Sect.~\ref{obssect}).}
\tablefoottext{d}{In the fitting model we include the values \tbo$~=1.0\times10^8$~K~sr and 
\dvi$=3.0$~\kms ~that best fit the total intensity emission.}
}
\label{VLA2.4_tab}
\end{table*}
\section{Proper motion estimate of VLA\,1.}
\label{appB}
\begin{figure*}[h!]
\centering
\includegraphics[width = 6 cm]{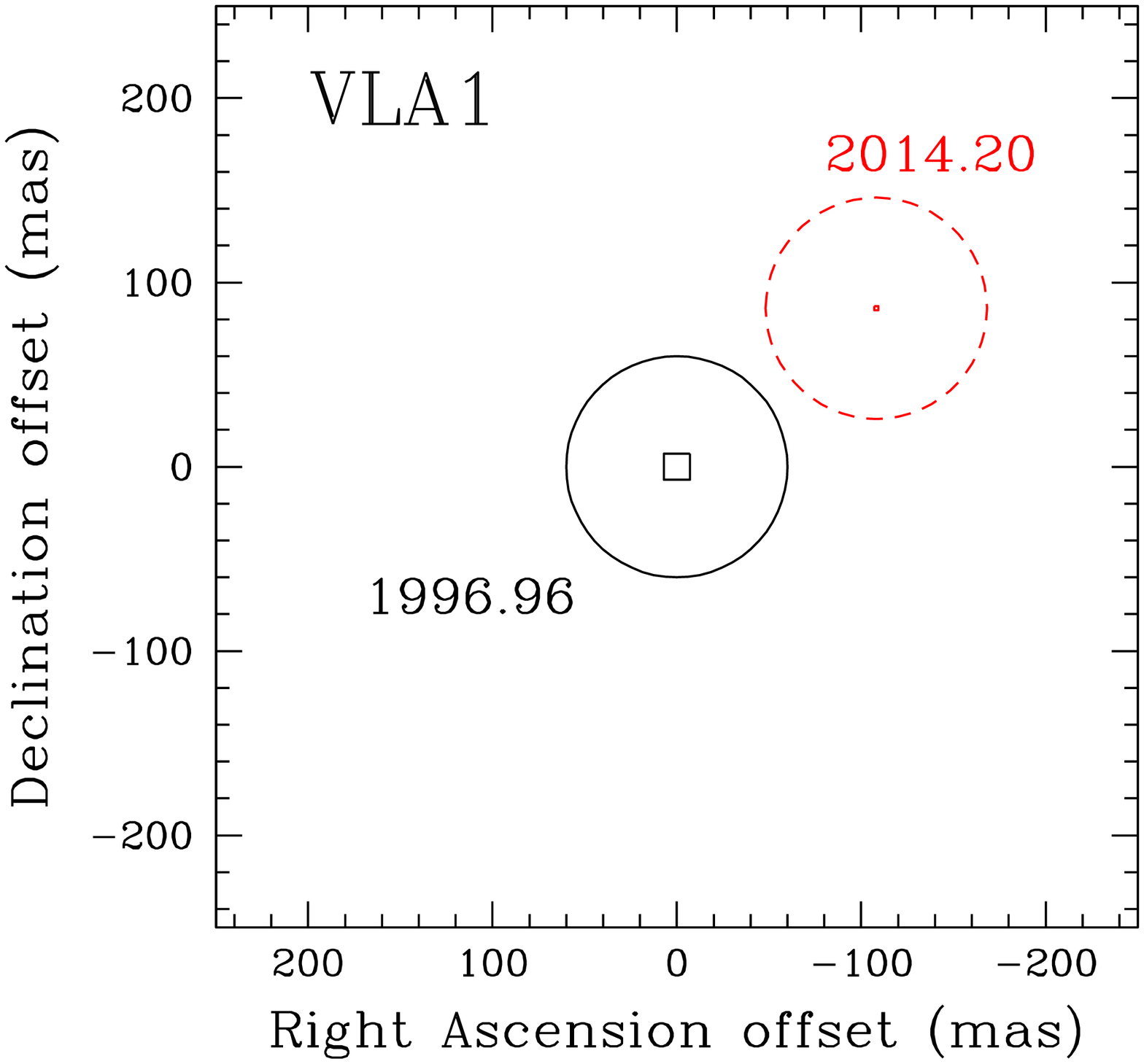}
\includegraphics[width = 6 cm]{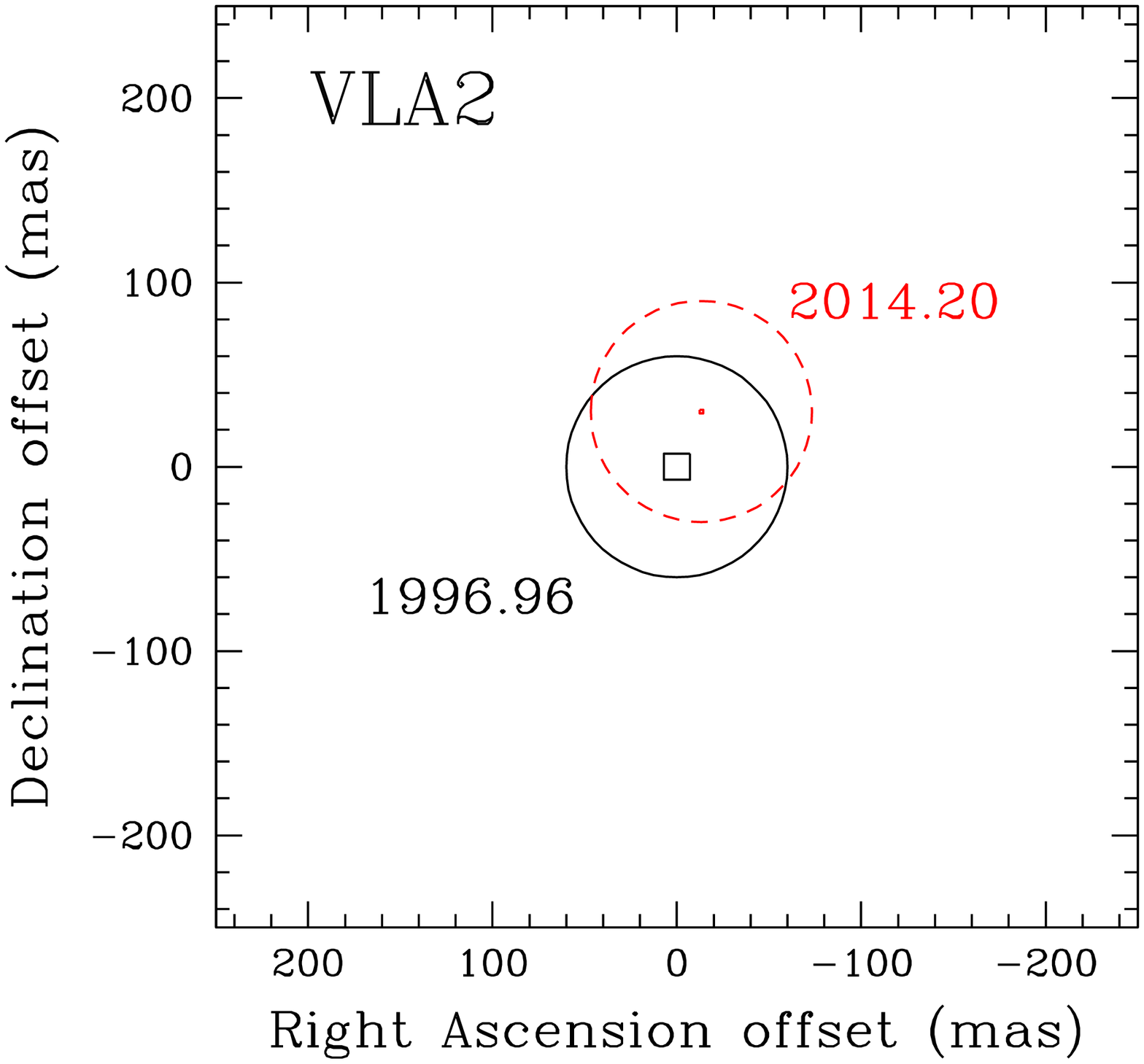}
\includegraphics[width = 6 cm]{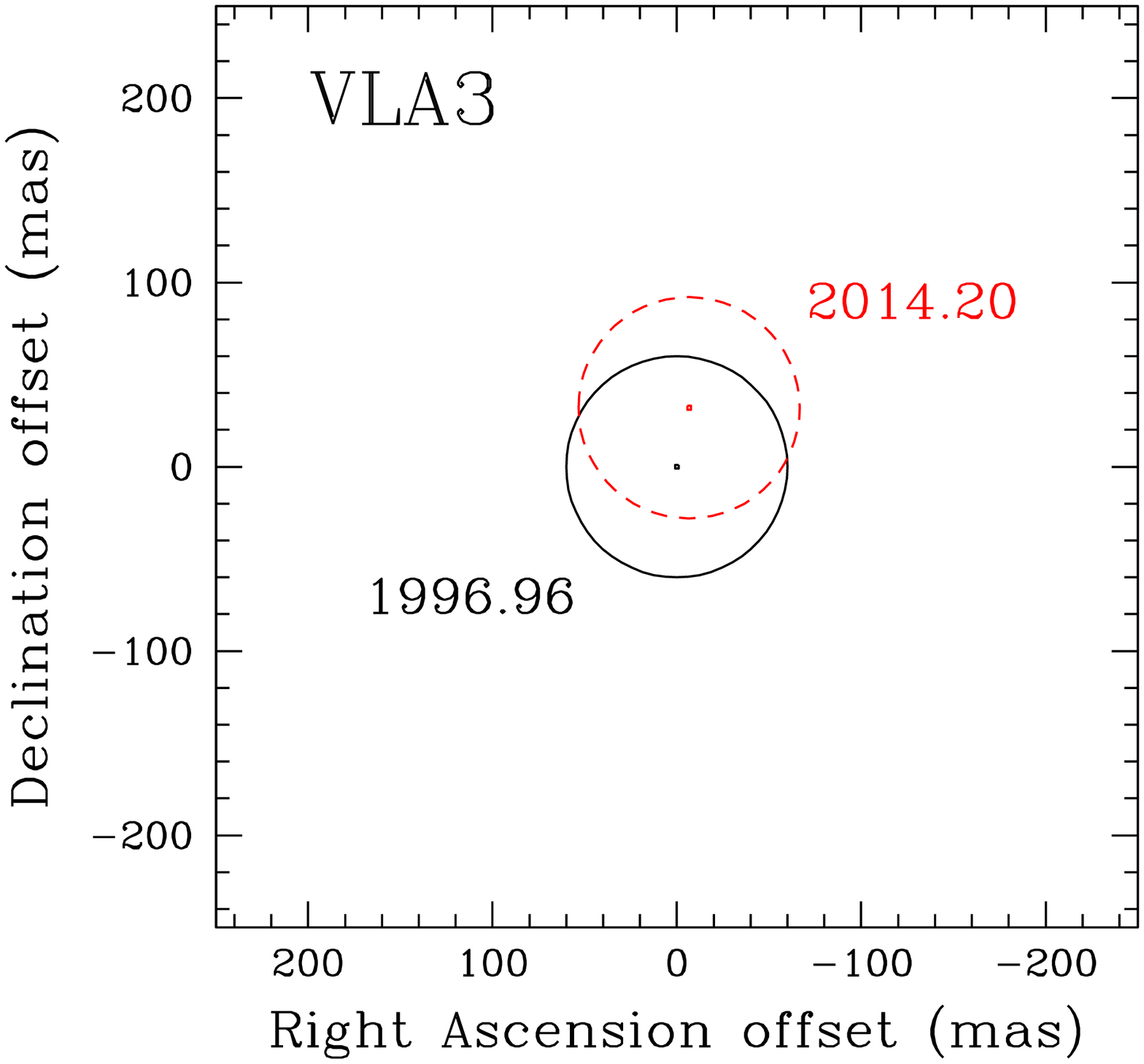}
\caption{Comparison of the positions of the continuum emission at K-band of VLA\,1 (\textit{left panel}), 
VLA\,2 (\textit{central panel}), and VLA\,3 (\textit{right panel}) as observed in epoch 1996.96 
\citep[solid black;][]{tor97} and epoch 2014.20 \citep[dashed red;][]{car15,rod20}. The squares represent the 
uncertainties of the peaks positions (see Table~\ref{absposYSO}) and the circles are the restoring circular 
beam of $0.\!\!''12$ used to properly compare the two epochs. Each panel is centered at the corresponding 
position of the YSO in epoch 1996.96 (see Table~\ref{absposYSO}). The positions of epoch 2014.20 are corrected
for the proper motion of the region W75N(B) as measured by \cite{ryg12}, 
$\langle\mu_{\rm{\alpha}}\rangle=(-1.97\pm0.10)~\rm{mas~yr^{-1}}$
and $\langle\mu_{\rm\delta}\rangle=(-4.16\pm0.15)~\rm{mas~yr^{-1}}$.
}
\label{comp}
\end{figure*}

To properly compare the positions of the \water ~maser features detected in the four EVN epochs with
the continuum emissions of VLA\,1 and VLA\,2 observed in 2014, we must verify if the two 
massive YSOs have any proper motion within the region W75N(B). We report in Table~\ref{absposYSO} the
absolute positions of the three massive YSOs VLA\,1, VLA\,2, and VLA\,3 as measured at K-band with the 
VLA by \cite{tor97}, \cite{car15}, and \cite{rod20} in epochs 1996.96 and 2014.20. 
These positions were obtained by deconvolving the data of the two epochs with the same restoring circular 
beam of $0.\!\!''12$, as already done by \cite{car15} and \cite{rod20}, and by fitting the emission 
with a Gaussian fit. Furthermore, as already done for the maser features of the last three EVN epochs, 
the positions of the three YSOs of epoch 2014.20 are also corrected assuming the proper motion of the 
region W75N(B) equal to the median  proper motion measured for the 6.7 GHz \meth ~maser features by 
\cite{ryg12}, $\langle\mu_{\rm{\alpha}}\rangle=(-1.97\pm0.10)~\rm{mas~yr^{-1}}$
and $\langle\mu_{\rm\delta}\rangle=(-4.16\pm0.15)~\rm{mas~yr^{-1}}$.\\
\indent We plot in the three panels of 
Fig.~\ref{comp} the corrected positions of the three massive YSOs for the two epochs with the relative
errors due to the thermal noise and to the Gaussian fit errors ($\Delta\alpha$, $\Delta\delta$), which are
represented by squares, and with the restoring circular beam centered on each position (circles).
We note from Fig.~\ref{comp} that the positions of VLA\,2 and VLA\,3 in both epochs are consistent within 
the beam and from Table~\ref{shift} that the position shifts between the two epochs are similar suggesting
a possible systematic uncertainty. We cannot verify if this uncertainty is real or not because the
structure of the two sources varied between the two epochs, in particular that of VLA\,2. We can therefore 
assume that VLA\,2 and VLA\,3 did not move within W75N(B) over 17.24~years and that the positions of the 
\water ~maser features in VLA\,2 in the last three EVN epochs do not need further corrections.\\
\indent It is evident from both Fig.~\ref{comp} and Table~\ref{shift} that VLA\,1 actually moved
within W75N(B) between the two epochs. From the shift reported in Table~\ref{shift} we measure a proper 
motion of VLA\,1 within W75N(B)
of $\langle\mu_{\rm{\alpha}}\rangle^{\rm{VLA\,1}}=(-6.3\pm0.4)~\rm{mas~yr^{-1}}$ ($-38.9\pm2.5$~\kms) and
$\langle\mu_{\rm\delta}\rangle^{\rm{VLA\,1}}=(+5.0\pm0.4)~\rm{mas~yr^{-1}}$ ($+30.9\pm2.5$~\kms) along 
right ascension and declination, respectively. This motion might suggest that VLA\,1 is a runaway 
protostar, but to verify this possibility it is necessary to carry-on ad-hoc multi epoch observations.
Nevertheless, we decide to correct the positions of the \water ~maser features detected in the last
three EVN epochs and associated with VLA\,1 accordingly to the estimated proper motion of VLA\,1.
\begin {table*}[h!]
\caption []{Absolute positions of the massive YSOs VLA\,1 and VLA\,2 in epochs 1996.96 and 2014.20.} 
\begin{center}
\scriptsize
\begin{tabular}{ l c c c c c c c}
\hline
\hline
\,\,\,\,\,(1) &(2)         &   (3)             &    (4)           &  (5)              & (6)     & (7)   & (8) \\
epoch\tablefootmark{a}   &  $\alpha_{1950}$ &  $\delta_{1950}$  &  $\alpha_{2000}$ &  $\delta_{2000}$  & $\Delta\alpha$ & $\Delta\delta$ & Ref.\tablefootmark{b}\\ 
        &                  &                   &                  &                   &  \\
        &  ($\rm{^{h}:~^{m}:~^{s}}$)  & ($\rm{^{\circ}:\,':\,''}$) &  ($\rm{^{h}:~^{m}:~^{s}}$)  & ($\rm{^{\circ}:\,':\,''}$) & (mas) & (mas) &\\
\hline      
         \multicolumn{8}{c}{VLA\,1} \\
\hline
1996.96 & 20:36:50.0056    & +42:26:58.507     & 20:38:36.4528    & +42:37:34.850     & 7  & 7  & (1)  \\
2014.20 &                  &                   & 20:38:36.440     & +42:37:34.865     & 0.2 & 0.2 & (2)  \\
2014.20* &                 &                   & 20:38:36.443     & +42:37:34.936     &    &    &  -   \\
\hline      
         \multicolumn{8}{c}{VLA\,2} \\
\hline
1996.96 & 20:36:50.0405    & +42:26:58.783     & 20:38:36.4882    & +42:37:34.128     & 7  & 7  & (1)   \\
2014.20 &                  &                   & 20:38:36.484     & +42:37:34.086     & 1  & 1 & (3)   \\
2014.20*&                  &                   & 20:38:36.487     & +42:37:34.158     &    &    &  -    \\ 
\hline
         \multicolumn{8}{c}{VLA\,3} \\
\hline
1996.96 & 20:36:50.0406    & +42:26:58.095     & 20:38:36.4886    & +42:37:33.440     & 1  & 1  & (1)   \\
2014.20 &                  &                   & 20:38:36.485     & +42:37:33.400     & 0.1 & 0.1 & (3)  \\
2014.20*&                  &                   & 20:38:36.488     & +42:37:33.472     &    &    &  -    \\ 
\hline
\end{tabular}
\end{center}
\tablefoot{
\tablefoottext{a}{Epoch 2014.20* refers to the epoch 2014.20 corrected for the proper motion of the
region W75N(B) as measured by \cite{ryg12}, 
$\langle\mu_{\rm{\alpha}}\rangle=(-1.97\pm0.10)~\rm{mas~yr^{-1}}$
and $\langle\mu_{\rm\delta}\rangle=(-4.16\pm0.15)~\rm{mas~yr^{-1}}$. The reference epoch is epoch 1996.96.}
\tablefoottext{b}{References: (1) \cite{tor97}; (2) \cite{rod20}, (3) \cite{car15}.}
}
\label{absposYSO}
\end{table*}
\begin {table}[th!]
\caption []{Shift of the positions of the three massive YSOs.} 
\begin{center}
\scriptsize
\begin{tabular}{ l c c }
\hline
\hline
\,\,\,\,\,(1) &(2)         &   (3)              \\
YSO   & $\Delta(\alpha^{2014.20^*}-\alpha^{1996.96})$ & $\Delta(\delta^{2014.20^*}-\delta^{1996.96})$   \\ 
      & (mas)      & (mas) \\
\hline     
VLA\,1 & $-108\pm7$ & $+86\pm7$\\
VLA\,2 & $-13\pm7$  & $+30\pm7$\\
VLA\,3 & $-7\pm1$   & $+32\pm1$\\
\hline
\end{tabular}
\end{center}
%\tablefoot{
%}
\label{shift}
\end{table}

%\object{W75N}

\end{appendix}

\end{document}